\documentclass{siamltex}
\usepackage{amsmath,amssymb,mathabx}
\usepackage{graphicx,esint,ulem}

\usepackage[english]{babel}
\usepackage{amsmath}
\usepackage{amsfonts}
\usepackage{fullpage}
\usepackage{graphicx}
\usepackage{setspace}
\usepackage{float}
\usepackage{multirow}
\usepackage{fancyhdr}
\usepackage{time}
\newtheorem{remark}{Remark}[section]

\title{On-Site and Off-Site Bound States of the Discrete Nonlinear Schr\"odinger Equation and the Peierls-Nabarro Barrier}
\author{M. Jenkinson and M.I. Weinstein}

\begin{document}

\maketitle

\begin{abstract}
We construct multiple families of solitary standing waves of the discrete cubically nonlinear Schr\"{o}dinger equation (DNLS) in dimensions $d=1,2$ and $3$. These states are obtained via a bifurcation analysis about the continuum (NLS) limit. One family consists {\it on-site symmetric} (vertex-centered) states; these are spatially localized solitary standing waves which are symmetric about any fixed lattice site. The other spatially localized states are {\it off-site symmetric}.  Depending on the spatial dimension, these may be bond-centered, cell-centered, or face-centered. Finally, we show that the energy difference among distinct states of the same
frequency is exponentially small with respect to a natural parameter. This provides a rigorous bound for the so-called {\it Peierls-Nabarro} energy barrier.

\end{abstract}


\section{Introduction}

In this paper we study the  solitary standing waves of the  \underline{discrete} cubically nonlinear Schr\"odinger (DNLS) equation in dimensions $d=1,2$ and $3$. We construct, by bifurcation methods, families of on-site (vertex-centered)  symmetric and off-site (bond-, face-, and cell- centered) symmetric spatially localized discrete standing wave solutions.

We begin with a brief discussion of the continuum nonlinear Schr\"odinger (NLS) equation:
\begin{align}
& i \partial_t u(x,t) = - \Delta_x u(x,t) - |u(x,t)|^2 u(x,t),\ \ x \in \mathbb{R}^d, \ t \in \mathbb{R}\ .  \label{eqn:introNLS} 
\end{align}
NLS has, for any frequency $\omega<0$, a standing wave solution:
\begin{align}
u_\omega(x,t) = e^{- i\omega t} \psi_{| \omega |}(x), \label{eqn:cbreather}
\end{align}
where $\psi_{| \omega |}(x)$ is the unique solution to the nonlinear elliptic problem
\begin{align}
- \Delta_x \psi_{| \omega |} - \left| \psi_{| \omega |} \right|^2 \psi_{| \omega |}\ =\ \omega\ \psi_{| \omega |}, \quad u\in H^1(\mathbb{R}^d) \label{eqn:introtiNLS}
\end{align}
which is real-valued, positive, radial symmetric and decreasing to zero at spatial infinity
\cite{Bgn:99,K:89,S:77,SS:99,T:06,W:83}; see also Proposition \ref{prop:Psi}.  We refer to this solution as the {\it ground state} of NLS. Note that $\psi_{| \omega |}(x) = \sqrt{| \omega |}\ \psi_1 \left(\sqrt{| \omega |}x\right)$. 
Since NLS is Galilean invariant, these solutions can be ``boosted'' to generate solitary {\it traveling} waves. For any velocity, $v$, and frequency $\omega$,
\begin{equation}
e^{iv\cdot(x-vt)}\ u_\omega(x-2vt,t)\ =\ e^{-i\omega t} e^{iv\cdot(x-vt)}\ \psi_{ |\omega |}(x-2vt)\ ,
\label{boosted}\end{equation}
is a solution to NLS.

\medskip

We now formulate DNLS on the lattice $h \mathbb{Z}^d$, where $h > 0$ is the lattice spacing parameter. Define  $u = \{ u_n \}_{n \in \mathbb{Z}^d}$, $t \in \mathbb{R}$, and introduce
the difference operator, $\delta_j$:
\begin{equation}
(\delta_j u)_n  = u_{n+e^{(j)}}-u_n\ ,
\label{diffj}
\end{equation}
where $e^{(j)}$ is the unit vector in the $j$th coordinate direction. The
 the $d$-dimensional discrete (centered-difference) Laplacian, $\delta^2=\sum_{j=1}^d\delta_j^2$, is given by:
\begin{align}
\left(\delta^2 {u}\right)_n = \sum_{|j-n|=1}  u_j - 2 d\ u_n.
\end{align}
The summation is over lattice points in $\mathbb{Z}^d$ which are a unit distance from the point $n\in\mathbb{Z}^d$.
The discrete NLS equation (DNLS) is the system:
\begin{align}
i \partial_t u_n = - h^{-2} (\delta^2 {u})_n  - |u_n |^2 u_n\ ,\ n\in\mathbb{Z}^d.  \label{eqn:introDNLS}
\end{align}
DNLS is a Hamiltonian system, expressible in the form
\begin{align}
i\partial_t{ u} &= \frac{ \delta\mathcal{H}[u,\overline{u}] }{\delta \overline{u}},\ \ {\rm where}\label{dnls-hamsys}\\
\mathcal{H}_{\textrm{DNLS}}=\mathcal{H}[{u},\overline{u}] &=  \frac{1}{h^2}  \sum_{j = 1}^d \sum_{n\in\mathbb{Z}^d}|(\delta_j{u})_n|^2-\frac12 |u_n|^4\ .
\label{dnls-ham}\end{align}
The initial value problem
\begin{align}
& i \partial_t u_n(t) = - h^{-2} (\delta^2 u)_n(t)  - |u_n(t) |^2 u_n(t) \ ,\ n\in\mathbb{Z}^d,  \ \ t \geq 0 \nonumber \\
& u_n(0) = f_n \in l^2(\mathbb{Z}^d), \label{eqn:IVPdnls}
\end{align}
is globally well-posed, in the sense that for each ${f} = \{ f_n \}_{n \in \mathbb{Z}^d} \in l^2(\mathbb{Z}^d)$ there exists a unique global solution $u(t) = \{ u_n(t) \}_{n \in \mathbb{Z}^d} \in C^1([0,\infty), l^2(\mathbb{Z}^d))$ to \eqref{eqn:IVPdnls}. This result follows from a standard contraction mapping argument applied to the equivalent integral equation formulation of the initial value problem;
see, for example, \cite{KLS:12}. Their proof is formulated in one dimension but applies in arbitrary dimension, since $\|{f}\|_{l^\infty(\mathbb{Z}^d)}\lesssim\|{ f}\|_{l^2(\mathbb{Z}^d)}$.

Furthermore, the functionals $\mathcal{H}_{\rm DNLS}[{u},\overline{u}]$ and
\begin{equation}
\mathcal{N}_{\textrm{DNLS}}=\mathcal{N}[{u},\overline{u}]=\sum_{n\in\mathbb{Z}^d} |u_n|^2
\label{N-l2}\end{equation}
are conserved quantities (time - invariant) for solutions of DNLS.

\medskip
Discrete nonlinear dispersive systems such as DNLS arise in the context of nonlinear optics, {\it e.g.} \cite{ADRT:94,ADLR:94,ESMB:98, ESMB:99,NV:11}, dynamics of biological molecules, {\it e.g.} \cite{DK:75, ES:85}, and condensed matter physics. For example, they arise in the study of intrinsic localized modes in anharmonic crystal lattices, {\it e.g.} \cite{BS:75, TKS:88}. See also \cite{STKB:02, HDC:10,KMT:11}. In these fields, discrete systems arise either as phenomenological models or as {\it tight-binding} approximations; see also \cite{MPS:08, PS:10}. There is also a natural interest in such systems as discrete numerical approximations of continuum equations.
\medskip

In analogy with NLS \eqref{eqn:introDNLS}, DNLS is known to have discrete solitary standing waves
\cite{ES:85,K:09,W:99}
 \begin{align}
u_n(t) = e^{- i \omega t} g_n,\ \ n\in\mathbb{Z}^d \quad {\rm where}\ \omega < 0\ .
 \end{align}
Here, ${g}$ satisfies the discrete elliptic problem, an infinite system of algebraic equations:
 \begin{align}
 - h^{-2} (\delta^2{g})_n - |g_n|^2 g_n\ =\  \omega\ g_n
  \quad \quad \|{g}\|_{l^2(\mathbb{Z}^d)}^2=\sum_{n \in \mathbb{Z}^d} | g_n |^2 < \infty. \label{eqn:tiDNLS}
 \end{align}
\medskip

The goal of this paper is to present a detailed study of {\it on-site symmetric} (vertex-centered) and {\it off-site symmetric} (bond-centered, face-centered or cell-centered) discrete solitary standing waves in dimensions $1, 2$ and $3$. We are motivated by the general question of the effects of discretness and, in particular,  the following\medskip

\noindent{\it Question:}\ Are there discrete solitary \underline{traveling} waves of DNLS?
\medskip

\begin{figure}[H]
\centering
\resizebox{.9\textwidth}{!}{\includegraphics{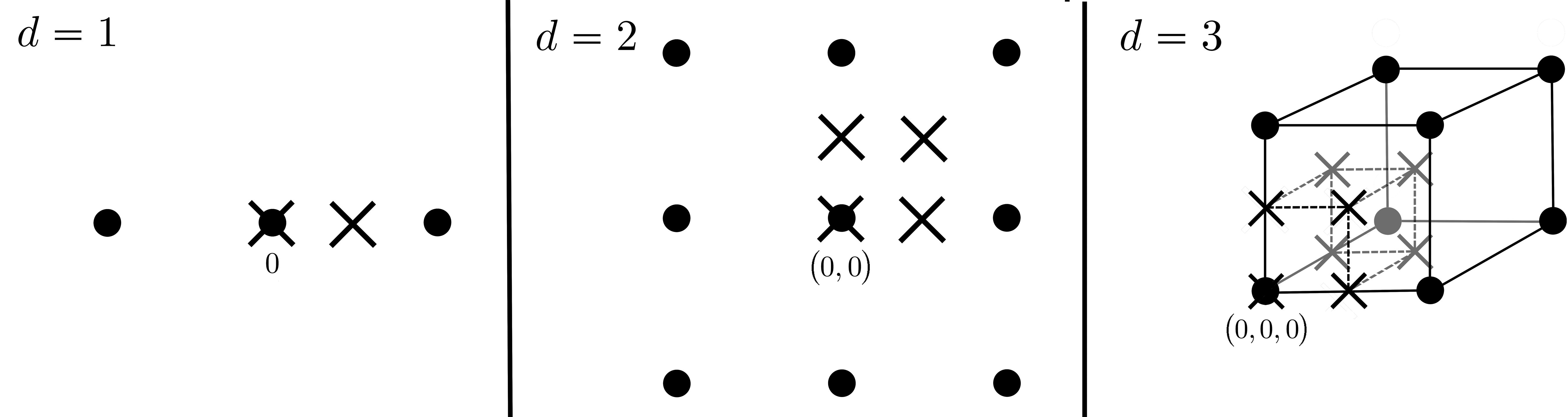} }
\caption{Centerings of solitary standing waves in dimensions $d=1,2,3$. }
\label{fig:sol-max-vs-t}
\end{figure}

This question has been studied at least since pioneering article of Peyrard and Kruskal \cite{PK:84} for the propagation of discrete kinks for discrete $\phi^4$ model.
Numerical evidence strongly supports the claim that there are \emph{no} discrete traveling waves of DNLS.  Figure \ref{fig:sol-max-vs-t} displays several simulations which shed light. For a range of values of $h$, we solve the initial value problem for DNLS \eqref{eqn:introDNLS} in spatial dimension $d = 1$ with initial condition
\begin{equation}
u_n(0)=  e^{iv\cdot x_n}\ \psi_{| \omega |}(x_n),\ \ x_n = nh ,\ n \in\mathbb{Z}\ ,
\label{discrete-ic}
\end{equation}
obtained by evaluating $e^{iv x}\psi_{| \omega |}(x)$ at the lattice sites $x_n =n h,\ n \in\mathbb{Z}$.
For the continuum limit, the solution is given by \eqref{boosted}. Plotted in figure \ref{fig:sol-max-vs-t} is the location of the ${\rm argmax}_{n \in\mathbb{Z}}|u_n (t)|$\ versus $t$ for the parameter choices $v= 0.1$ and $\omega = 1$.
The continuum limit ($h=0$) corresponds to the dashed straight line of slope $2 v = 0.2$, the (group) speed of the solitary wave envelope $\psi^{|\omega |}$ in \eqref{boosted}. For a range of $t\ge0$, the other curves start out linearly but then level off and approach a constant value 
$x_n =x_{n_\star(h)}$.
%
%
An examination of the time-evolving solution profiles shows a rightward moving localized structure, which continuously radiates small amplitude dispersive waves (``phonons," or lattice vibrational modes), slows down and eventually relaxes to a discrete solitary wave, which is ``pinned'' to a lattice site.

The phenomenology developed in the physics literature to explain this phenomenon is as follows. The discrete dynamical system, DNLS, supports solitary standing waves which are centered ``on-site'' and ``off-site". Note, for  the continuum (translation invariant NLS,\ \eqref{eqn:introNLS} ) limit, that the values of time-invariant quantities
\[\mathcal{H}_{\textrm{NLS}}[u]=\int_{\mathbb{R}^d}\ |\nabla u|^2-\frac12|u|^4\ dx,\quad
 \mathcal{N}_{\textrm{NLS}}[u]=\int_{\mathbb{R}^d}\ |u|^2\ dx
\]
for $u=u_\omega(x+x_0,t)$ are independent of $x_0$.
  For DNLS, in contrast,  the on-site waves are of lower energy and are therefore stable.  An amount of energy equal to this energy difference, the {\it Peierls-Nabarro (PN) barrier}, must be expended to move a discrete soliton from one lattice site to an adjacent one. This energy is dissipated through the radiation of small amplitude dispersive waves (phonons) to infinity. Since a quantum of energy is lost at each transition, the translating localized wave structure eventually no longer exceeds the PN barrier and converges asymptotically to a discrete on-site (stable) solitary wave. A mathematically rigorous study of these phenomena is an open problem. \medskip

\begin{figure}[H]
  \begin{minipage}[c]{0.55\textwidth}
    \includegraphics[width=\textwidth]{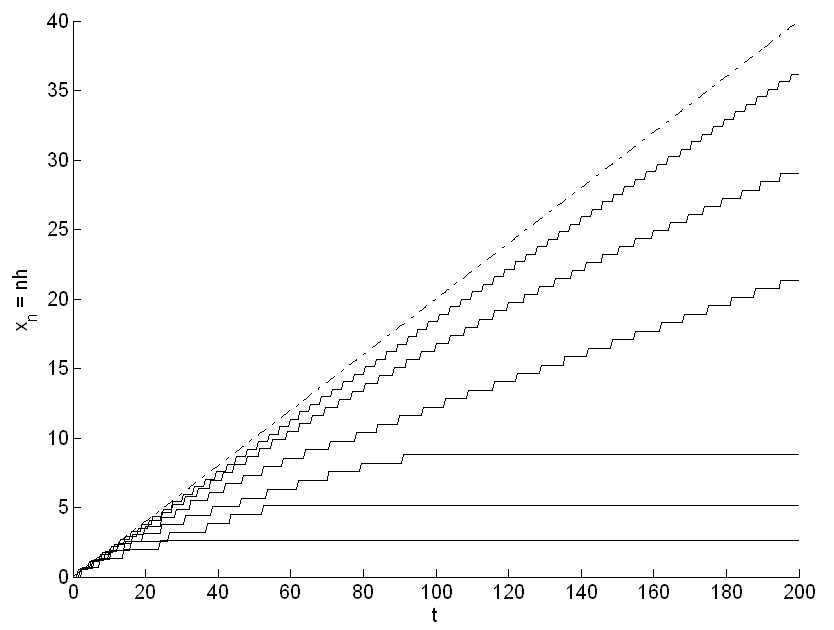}
  \end{minipage}
  \begin{minipage}[c]{0.22\textwidth}
    \caption{ Position of peak magnitude for traveling pulses of DNLS initial value problem in spatial dimension $d = 1$, with $h=.54, .58, .61, .6295, .64 $ and $.66$ and initial condition \eqref{discrete-ic} for $v = 0.1$ and $\omega = 1$. } \label{fig:onsiteoffsite}
  \end{minipage}
\end{figure}


\begin{remark}
Another class of discrete nonlinear systems which exhibits similar phenomena are  lattice-nonlinear Klein-Gordon models, {\it e.g.} the discrete sine-Gordon equation and the discrete $\phi^4$ equation \cite{PK:84,KW:00}.
 A study of the latter stages of asymptotic relaxation to a stable on-site kink, via radiative (non-dissipative) decay phenomena is pursued in \cite{KW:00}. The mechanism is resonant coupling of discrete and continuum modes and resulting radiation damping, studied  in the setting of continuum nonlinear wave equations in, for example,
 \cite{WY:96,SW:99,BS:03,SW:04}.
\end{remark}
\begin{remark}
We note that there are discrete systems which have translating, non-deforming solitary traveling waves: (a) The Ablowitz-Ladik (AL) lattice, a discrete integrable system resulting from a special discretization of the one-dimensional NLS \cite{AL:76} (see also discussion in section \ref{previous}),   (b) FPU lattices in the supersonic regime \cite{FP:99, FW:94, PR:11, H:10} and (c) the nonlinear Klein-Gordon lattice \cite{BZ:06} in the supersonic regime.
\end{remark}
 \bigskip

   The goal of this work is to obtain a precise understanding of the on-site symmetric and off-site symmetric discrete solitary standing waves.
 For simplicity, we first describe our results for $d=1$. Two and three dimensional lattices are discussed later in this section and in greater detail in section \ref{section:higherdim}.\medskip

 \begin{definition}[On-site symmetric and off-site symmetric states in one spatial dimension]
\label{defn:onoff}\\
Let ${g} = \{ g_n \}_{n \in \mathbb{Z}}$.
\begin{enumerate}
\item A solution to equation (\ref{eqn:tiDNLS}) is referred to as on-site symmetric if for all $ n \in \mathbb{Z}$, it satisfies
\begin{align}
g_{n} = g_{-n}.
\end{align}
 In this case, ${g}$ is symmetric about $n = 0$. \medskip

\item A solution to equation (\ref{eqn:tiDNLS}) is referred to as off-site symmetric or bond-centered symmetric if for all $n \in \mathbb{Z}$, it satisfies
\begin{align}
 g_{n} = g_{-n + 1}.
\end{align}
In this case, ${g}$ is symmetric about the point halfway between $ n = 0$ and $n = 1$.
\end{enumerate}
\end{definition}

\begin{figure}[H]
  \begin{minipage}[c]{0.75\textwidth}
    \includegraphics[width=\textwidth]{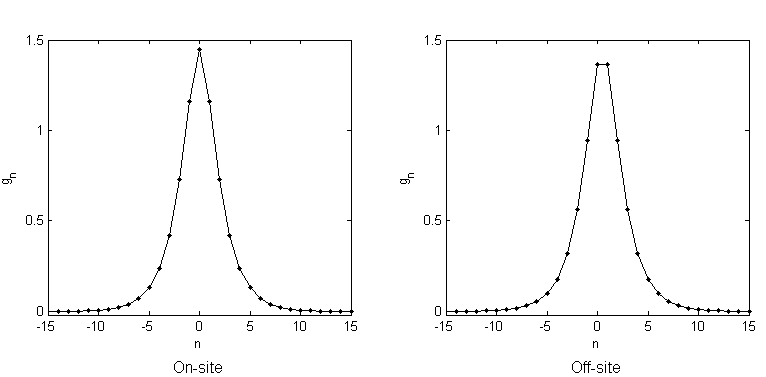}
  \end{minipage}\hfill
  \begin{minipage}[c]{0.22\textwidth}
    \caption{ On-site symmetric (left) and off-site symmetric or bond-centered (right) standing wave solutions of discrete NLS for $d = 1$. Discrete solitary waves profiles are defined by the values at discrete points, $nh$ (here $h= 0.6$ and $\omega = 1$). Plotted are linearly interpolated profiles.  } \label{fig:onsiteoffsite}
  \end{minipage}
\end{figure}

We seek spatially localized on-site and off-site solutions of \eqref{eqn:tiDNLS} in
 two related distinguished limits,  expressed in terms of the frequency $\omega$ and lattice spacing $h$:
 \medskip

 \begin{enumerate}
\item[(L1)] {\it Continuum limit; frequency $\omega < 0$ fixed and lattice (grid) spacing $h \to 0$.}  This is a limit of  interest in numerical computations. Solutions are expected to approach those of continuum NLS.

\item[(L2)] {\it Homogenized long wave limit; fixed lattice spacing $h$ and $\omega \to 0$.}
Here, there is large scale separation between the width of the discrete standing wave and the lattice spacing.
Solutions are expected to approach a (homogenized or averaged)  continuum NLS equation.
\end{enumerate}
\medskip

 The two limits (L1) and (L2) may be studied together through the introduction of a single parameter.
 Introduce
 \begin{align}
 & \omega = - \epsilon^2 \neq 0, \qquad \quad  g_n = h^{-1} G_n, \qquad \alpha \equiv \epsilon \ h \ . \label{alpha-def}
\end{align}
 Then,  $G=\{G_n\}_{n\in\mathbb{Z}^d}$ satisfies the nonlinear eigenvalue problem
\begin{align}
- \alpha^2 G_n = - (\delta^2 G )_n - |G_n|^2 G_n\ ,\qquad  G\in l^2(\mathbb{Z}^d)\ .\label{eqn:tiDNLSrescaled}
\end{align}

\noindent Thus, the limits (L1) and (L2) are reduced to the study of DNLS for lattice spacing $h=1$ and $\alpha \to 0$.

\begin{figure}[H]
  \begin{minipage}[c]{0.67\textwidth}
    \includegraphics[width=\textwidth]{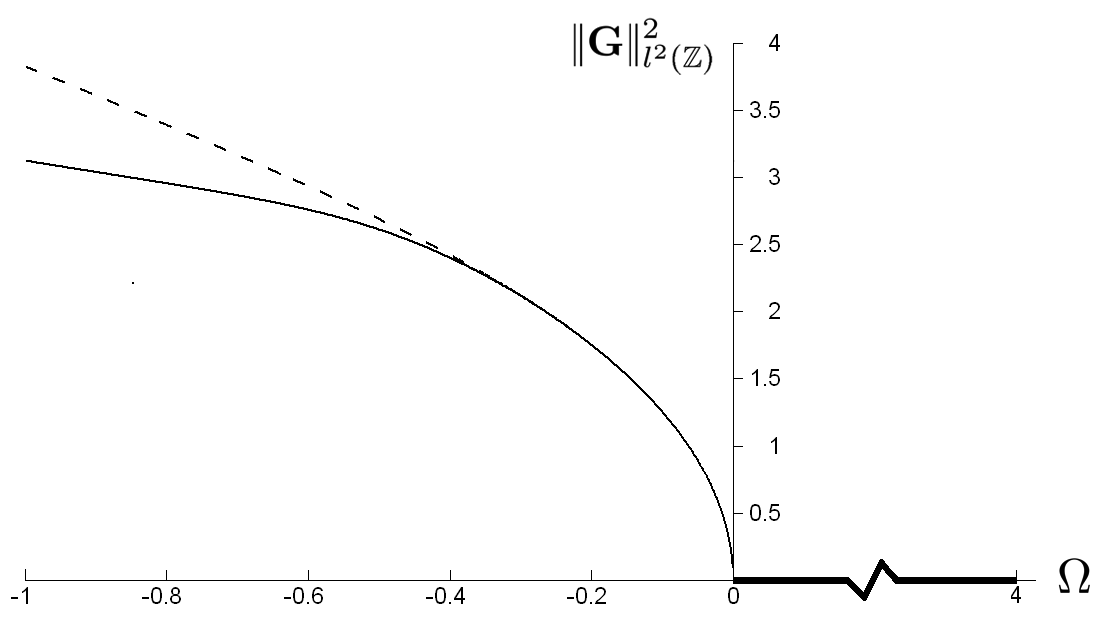}
  \end{minipage}\hfill
  \begin{minipage}[c]{0.3\textwidth}
    \caption{ Bifurcation of on-site (solid line) and off-site (dashed line) solutions of \eqref{eqn:tiDNLSrescaled} from the continuous spectrum with $l^2(\mathbb{Z})$ norm as a function of $\Omega = - \alpha^2$. The thick dark line marks the finite band of continuous spectrum, the interval $[0,4]$, of the discrete Laplacian, $-\delta^2$.
    } \label{fig:bifurcation}
  \end{minipage}
\end{figure}

Nonlinear bound states arise as bifurcations of non-trivial localized states from the zero state at frequency $\alpha^2=0$, the endpoint of the continuous spectrum; see Figure \ref{fig:bifurcation}. \\

\noindent The main results of this paper concern the localized solutions of \eqref{eqn:tiDNLSrescaled} for $\alpha^2$ small:
\begin{enumerate}
\item {\bf Theorems \ref{th:main} and \ref{th:maingeneral},\ Bifurcation of onsite and off-site states of DNLS :}\  In dimensions $d=1,2$ or $3$, there exist families of on-site (vertex-centered) symmetric and off-site (bond- , face- and cell- centered) symmetric solitary standing waves  of \eqref{eqn:tiDNLSrescaled}, which bifurcate from the continuum limit ($\alpha\downarrow0$)  ground state solitary wave of NLS. See figure \ref{fig:onsiteoffsite}.\smallskip

\item {\bf Theorem \ref{th:PN},\  Exponential smallness of the Peierls-Nabarro barrier:}\  Fix $s \in (0,1)$ and arbitrarily close to zero. Then, there exist positive constants $\alpha_0$ and $C>0$ such that for all $0 < \alpha < \alpha_0$, we have:
\begin{align}
& \Big|\ \mathcal{N} \left[G^{\alpha, \rm on} \right] - \mathcal{N} \left[ G^{\alpha, \rm off} \right] \ \Big| 
 \lesssim \alpha^{2-d} \ e^{ - C /\alpha}, \nonumber \\
&
\Big|\ \mathcal{H} \left[ G^{\alpha, \rm on} \right] - \mathcal{H} \left[ G^{\alpha, \rm off} \right] \ \Big| \lesssim \alpha^{2-d} \ e^{ - C /\alpha} \ ;
\end{align}
see figure \ref{fig:bifurcation}.
Here, 
\begin{align}
\mathcal{N}[G] \equiv \mathcal{N} \left[G,\overline{G} \right] = \sum_{n \in \mathbb{Z}^d} |G_n|^2,  \label{eqn:power2}
\end{align}
and
\begin{align}
\mathcal{H}[G] \equiv \mathcal{H} \left[G, \overline{G} \right] = \sum_{j = 1}^d \sum_{n \in \mathbb{Z}^d} | G_{n + e^{(j)}} - G_n |^2 - \frac{1}{2} | G_n |^4 \label{eqn:hamil2}
\end{align}
are the corresponding square $l^2(\mathbb{Z}^d)$ norm (Power) and Hamiltonian of \eqref{eqn:tiDNLSrescaled} (for effective lattice spacing $h = 1$). The quantities
\begin{align}
\mathcal{N} \left[G^{\alpha, \rm on} \right] - \mathcal{N} \left[G^{\alpha, \rm off}\right] \ \ {\rm and} \ \ \mathcal{H} \left[G^{\alpha, \rm on}\right] - \mathcal{H} \left[G^{\alpha, \rm off}\right]
\end{align}
are related to the PN barrier; see the earlier discussion. Theorem \ref{th:PN} applies for $d = 1, 2, 3$ where $G^{\alpha, {\rm on}}$ corresponds to a vertex-centered state and $G^{\alpha, {\rm off}}$ corresponds to a bond-, face-, or cell-centered state.

\end{enumerate}
\medskip

\begin{remark}
Plotted in Figure \ref{fig:bifurcation} are the bifurcating  curves of on- and off-site states for the one-dimensional DNLS
equation showing the lower value of $\mathcal{N}$ for the on-site state.  Physical heuristics and these numerical simulations suggest that the on-site state is the least energy (``ground'') state:\medskip

\noindent {\bf Conjecture 2:} Let $d=1,2$ or $3$. There exist constants $\alpha_0, C > 0$ such that for all $0 < \alpha < \alpha_0$ ,
\begin{align}
& 0 < \mathcal{N} [ G^{\alpha, \rm off} ] - \mathcal{N} [ G^{\alpha, \rm on} ]  \lesssim \alpha^{2 - d} \ e^{- C /\alpha}, \qquad 0 < \mathcal{H} [ G^{\alpha, \rm on} ] - \mathcal{H} [ G^{\alpha, \rm off}] \lesssim \alpha^{2 - d} \ e^{- C /\alpha}. \label{eqn:conjsign}
\end{align}
\end{remark}

\subsection{Strategy of proofs of Theorem \ref{th:main} - Theorem \ref{th:PN}} \label{section:strategy}

We outline the strategy for dimension $d=1$. As noted above in (L1) and (L2), the limit $\alpha \to 0$ in (\ref{eqn:tiDNLS2}) is related to the continuum NLS limit. In order to compare 
 the spatially discrete and spatially continuous problems,  it is  natural to work, respectively, with the discrete and continuous Fourier transforms. These are both functions of a continuum variable (momentum, respectively, quasi-momentum).

 Let $\widehat{ g}(q)=\mathcal{F}_D[ g](q)$ denote the discrete Fourier transform on $\mathbb{Z}$ of the sequence $g=\{g_n\}_{n\in\mathbb{Z}}$ and 
  let $\widetilde{f}(q)=\mathcal{F}_c[f](q)$ denote the continuous Fourier transform on $\mathbb{R}^d$ of $f:\mathbb{R}^d\to\mathbb{C}$; see section \ref{section:notation} for definitions and Appendix \ref{subsection:DFT}
    for a discussion of key properties.  The following proposition characterizes on-site symmetric and off-site symmetric states on $\mathbb{Z}$ in terms of  the form of their discrete Fourier transforms.  The higher dimensional analogue ($d\ge1$) is stated  in Proposition \ref{prop:sigmacentered}.
     The proof is given in appendix \ref{subsection:details}.  \medskip

\begin{proposition}
\label{prop:off-on}
\begin{itemize}
\item[(a)] Let $G = \{G_n \}_{n \in \mathbb{Z}}\in l^2(\mathbb{Z})$ be real and on-site symmetric in the sense of Definition \ref{defn:onoff}. Then, $\widehat{G}(q) = \mathcal{F}_{_D}[ G ],$ the discrete Fourier transform of $G$, is real-valued and symmetric. Conversely, if $\widehat{G}(q)$ is real and symmetric, then ${\mathcal{F}^{-1}_{_D}[ \widehat{G} ]}$, its inverse discrete Fourier transform, is real and on-site symmetric.

\item[(b)] If $G = \{G_n \}_{n \in \mathbb{Z}} \in l^2(\mathbb{Z})$ is real and off-site symmetric in the sense of Definition \ref{defn:onoff}, then
\begin{equation}
\widehat{G}(q) = e^{- i q/2} \widehat{K}(q),
\label{dft-off}
\end{equation}
 where $\widehat{K}(q)$ is real and symmetric. Conversely, if $\widehat{G}(q) = e^{- i q/2} \widehat{K}(q)$, where $\widehat{K}(q)$ is real and symmetric, then ${\mathcal{F}^{-1}_{_D}[\ \widehat{G}\ ]}$ is real and off-site symmetric.
\end{itemize}
\end{proposition}
\medskip

To prove our main results, we first rewrite the equation for a DNLS standing wave profile (onsite or offsite) $g=g^{\sigma,\alpha}\in l^2(\mathbb{Z})$,  \eqref{eqn:tiDNLSrescaled}, in discrete Fourier space
 for $G(q) = e^{-i\sigma q} \widehat{K}(q)$,
  where $K=\widehat{K}^{\sigma,\alpha}$ satisfies $K(q+2\pi)=\widehat{K}(q)$. 
 Here, $\sigma=0$ corresponds to the on-site case and $\sigma=1/2$ to the off-site case. Note that $\widehat{K}(q)$ is determined by its restriction, $\widehat{\phi}(q)$, to
  the fundamental cell $q\in\mathcal{B}=[-\pi,\pi]$  (Brillouin zone); we occassionally suppress the dependence on $\alpha$ or $\sigma$ for notational convenience.
  
Define the rescaled quasimomentum: $Q=q/\alpha$.  We obtain the following equation for  $\widehat{\Phi}(Q)=\widehat{\phi}(q/\alpha)$, defined for $Q\in\mathcal{B}_\alpha=[-\pi/\alpha,\pi/\alpha]$, the rescaled (stretched)  Brillouin zone:
\begin{align}
\mathcal{D}^{\sigma,\alpha}[\widehat{\Phi}]\equiv \left[ 1 + M_{\alpha}( Q) \right]\ \widehat{\Phi}( Q ) -   \   \frac{\chi_{_{\mathcal{B}_{_{\alpha}}}}( Q)}{4 \pi^2}\
\left(\ \widehat{\Phi}  * \widehat{\Phi} * \widehat{\Phi}\ \right)(Q) + R_1^{\sigma} [ \widehat{\Phi} ](Q)  = 0, \label{eqn:Phieqn3-intro}
\end{align}
 Here, $\chi_{_{\mathcal{B}_{_{\alpha}}}}(Q)$ is characteristic function of $\mathcal{B}_\alpha$, $M_{\alpha}( Q)$ is the scaled  Fourier symbol of the discrete Laplacian:
 \begin{align} 
 M_{\alpha}(Q) &\equiv \frac{1}{\alpha^2} M(Q \alpha) =  \frac{4}{\alpha^2}  \sin^2 \left( \frac{ Q \alpha}{2} \right) \label{eqn:M2def-intro}
 \end{align}
 and 
 \begin{align}
R_1^{\sigma} [ \widehat{\Phi} ](Q) &= -\frac{\chi_{_{\mathcal{B}_{_{\alpha}}}}( Q)}{4 \pi^2}\ \sum_{m = \pm 1} \ e^{2 m \pi i \sigma} \ \left(\ \widehat{\Phi} * \widehat{\Phi} * \widehat{\Phi}\ \right)  ( Q - 2 m \pi / \alpha).  \label{eqn:R1rewrite-intro}
\end{align}
It is important to note that the nonlinear operator, $\widehat{\Phi}\mapsto\mathcal{D}^{\sigma,\alpha}[\widehat{\Phi}]$, depends on $\sigma$, which designates the case of on-site or off-site states, only through the ``$\pm1$ side-band'' term $R_1^{\sigma}$. 

Reasoning formally,  we see that  \eqref{eqn:Phieqn3-intro} converges to: 
\begin{equation}
(1+|Q|^2)\widetilde{\psi}_1(Q)\ -\ 
 \frac{1}{4 \pi^2}\ \left(\widetilde{\psi_1}*\widetilde{\psi_1}*\widetilde{\psi_1}\right)(Q)=0,
\label{FT-NLS}\end{equation}
  the equation for the Fourier transform on $\mathbb{R}$ of the continuum solitary wave, $\psi_1(x)$. In observing this we have used: (a)  $M_\alpha(Q)\to|Q|^2$, for bounded $Q$,  as $\alpha\to0$, (b) $\mathcal{B}_\alpha\to\mathbb{R}$  as $\alpha\to0$, and
(c)   $\chi_{_{\mathcal{B}_{{\alpha}}}}(Q) R_1^{\sigma}[\widehat\Phi](Q)\to0$ as $\alpha\to0$ for $\widehat{\Phi}(Q)$ localized; see Lemma \ref{lemma:expconvo}.
Therefore, at leading order in $\alpha$ the behavior of  $\widehat{\Phi}^{\sigma,\alpha}$ appears to be $\widetilde{\psi}_1(Q)$.

In fact, we show in Theorems \ref{th:main} and \ref{th:maingeneral} that for  any  non-negative integer $J$, there  solutions of (\ref{eqn:tiDNLS2}), $\widehat{\Phi}=\widehat{\Phi^{\alpha, \sigma}}(Q)$ for $\sigma=0$ (on-site), and $\sigma=1/2$ (offsite), of the form:
\begin{align}
&  \widehat{\Phi^{\alpha,\sigma}}(Q) = e^{- i \alpha Q \sigma} \left[ \widetilde{\psi_1} \left(Q\right) + \alpha^2 \ F_1 [ \widetilde{\psi_1} ] \left( Q \right) + \dots + \alpha^{2J} \ F_J [ \widetilde{\psi_1} ] \left( Q \right) + \widehat{E_J^{\alpha, \sigma}}\left( Q\right) \right],\label{eqn:ansatz}
\end{align}
The mappings $\widetilde{\psi_1}\mapsto F_j[\widetilde{\psi_1}], j\ge1,$ are nonlocal nonlocal mappings defined below. 
For any $J\ge1$, the sum: $\widetilde{\psi_1} \left(Q\right) + \alpha^2 \ F_1 [ \widetilde{\psi_1} ] \left( Q \right) + \dots + \alpha^{2J} \ F_J [ \widetilde{\psi_1}]$  is  independent of $\sigma$ and $\|\widehat{E_J^{\alpha, \sigma}}\|_{L^2(-\pi/\alpha,\pi/\alpha)}$ is of order $\alpha^{2J+1}$ as $\alpha\downarrow0$ . Hence, up to the phase factor related to the centering of the wave relative to the spatial lattice, the difference between on-site and off-site states is  {\it beyond all polynomial orders in $\alpha$} for $\alpha\downarrow0$.

From the previous discussion, the on-site and off-site discrete standing solitary standing waves, $G^{\alpha,{\rm on}}$ and $G^{\alpha,{\rm off}} $ can  be constructed from \eqref{eqn:ansatz}, by rescaling $Q=q/\alpha$ and   inverting the discrete Fourier transform,

Theorem \ref{th:PN}, which bounds the  differences 
$\mathcal{N}[G^{\alpha,{\rm on}}]-\mathcal{N}[G^{\alpha,{\rm off}}]$ and $\mathcal{H}[G^{\alpha,{\rm on}}]-\mathcal{H}[G^{\alpha,{\rm off}}]$,
related to the Peireles-Nabbaro barrier,   are exponentially small, is derived using the continuity of $\mathcal{N}[\cdot]$ and $\mathcal{H}[\cdot]$, and the Plancherel identity as follows: 
\begin{align}
 \bigg| \mathcal{N} [G^{\alpha,{\rm off}}] - \mathcal{N} [G^{\alpha,{\rm on}}] \bigg|^2 \lesssim  \alpha \  \int_{\mathbb{R}} 
 (1+|Q|^2)\ |\ \widehat{\Phi^{\alpha,1/2}}(Q) - \widehat{\Phi^{\alpha,0}}(Q)\ |^2\ dQ \label{eqn:Nest-introd}
\end{align} 
 Proposition \ref{prop:expdiffEonoff} bounds $\widehat{\Phi^{\rm diff}}(Q) \equiv \widehat{\Phi^{\alpha,1/2}}(Q) - \widehat{\Phi^{\alpha,0}}(Q)$, whose equation 
 may be derived from the difference of the equations: $\mathcal{D}^{\alpha,\sigma}[\widehat{\Phi^{\alpha,\sigma}}]Q) =0$, $\sigma=0,1/2$; see \eqref{eqn:Phieqn3-intro}. The norm of $\widehat{\Phi^{\rm diff}}$ is shown to be  controlled by that of the $\pm1$ side-band terms, $R_1^\sigma$. The latter is shown to be exponentially small, using the exponential decay, uniformly in $\alpha$ small, of $\widehat{\Phi^{\alpha,\sigma}}(Q), \ \sigma=0,1/2$; see    Appendix \ref{section:expogeneral} . 
   We obtain:
 \begin{align}
\int_{\mathbb{R}} 
 (1+|Q|^2)\ |\ \widehat{\Phi^{\alpha,1/2}}(Q) - \widehat{\Phi^{\alpha,0}}(Q)\ |^2\ dQ \lesssim e^{- C / \alpha} . \label{eqn:diffexpo-intro}
\end{align}

Finally we remark that the relation of \eqref{eqn:Phieqn3-intro} to the continuum limit NLS equation \eqref{FT-NLS} was based on formal convergence argument, as $\alpha\downarrow0$, for fixed scaled quasimomentum, $Q\in [-\pi/\alpha,\pi/\alpha]$. To make the arguments rigorous we use a Lypapunov-Schmidt reduction strategy. We first solve the quasi-momentum components of $\widehat{E_J^{\alpha, \sigma}}\left( Q\right)$ for $\alpha^{r-1}\le |Q|\le \pi/\alpha, \ (0<r<1)$ (high frequency
 components of $\widehat{E_J^{\alpha, \sigma}}$) in terms 
of those for $0\le |Q|\le \alpha^{r-1}$ (low frequency components of $\widehat{E_J^{\alpha, \sigma}}$ ). The solutions of the low-frequency equation can be studied  perturbatively about the continuum NLS limit using the implicit function theorem.

\medskip

\subsection{Previous work on solitary standing waves of DNLS and relation to the present work}\label{previous}

Campbell and Kivshar  \cite{CK:93} provided an early formal calculation of the energy difference between off-site and on-site discrete solitary waves (PN barrier) in a small perturbation of the  Ablowitz-Ladik (AL-DNLS) system. AL-DNLS,
a completely integrable Hamiltonian system, is the particular  discretization of 1D NLS in which $|u_n(t)|^2$ is replaced by $\frac12(|u_{n-1}(t)|^2+|u_{n+1}(t)|^2)$ in \eqref{eqn:introDNLS}.
  AL-DNLS has localized discrete {\it breather} solutions which can be centered about {\it any} point in the continuum,  $\mathbb{R}$. The results in \cite{CK:93} suggest that the off-site state of DNLS has larger Hamiltonian energy than the on-site state for fixed $l^2$ norm, an {\it effective PN barrier} which is exponentially small in the perturbative parameter. Also in the perturbed AL-DNLS setting, Kapitula and Kevrekidis \cite{KK:01} perform linear spectral analysis about on- and off-site states and reach conclusions consistent with those in \cite{CK:93}.

Weinstein  \cite{W:99} studied, by variational methods, the existence of discrete solitary standing waves of DNLS in $\mathbb{Z}^d$, with general homogeneous polynomial nonlinearity; see also \cite{W:83,MW:96}. Here, such waves are realized as  minimizers (nonlinear ground states) of the Hamiltonian, $\mathcal{H}_{\textrm{DNLS}}$, subject to fixed $l^2$ norm.
Such ground states, when they exist, are nonlinearly orbitally stable.
 Particular attention is given to conditions on the nonlinearity and spatial dimension for which there is an {\it excitation threshold}, a positive $l^2$ threshold below which there does not exist any bound state \cite{MW:96,W:99}.  This variational method of construction does not give information on whether nonlinear ground states are on-site or off-site.

Variational methods can also be used to construct on-site and off-site solitary standing waves solutions of \eqref{eqn:tiDNLSrescaled} by appropriately constraining the function classes in which critical points are sought. Bambusi and Penati \cite{BP:10} construct  on-site and off-site solitary standing wave solutions near the continuum limit ($0<\alpha^2\ll1$) of the one-dimensional DNLS by combining a finite element approach with the above variational formulation. Herrmann \cite{H:11} constructs on-site and off-site solitary standing waves for any $\alpha^2>0$ as the infinite period limit of variationally obtained periodic standing waves; see also \cite{P:06}. Chong, Pelinovsky, and Schneider  \cite{CPS:12} construct, in an asymptotic limit, on-site and off-site states which lie near formal collective coordinate (variational) approximations.

\medskip

From the perspective of the general dynamics of DNLS \eqref{eqn:introDNLS}, our results complement
earlier work and extend it to higher dimensions.
Our bifurcation results  provide a step in the direction of a better dynamical understanding in that the construction can be used, for
$\alpha^2>0$ and small,  in a perturbative spectral analysis of the linearized dynamics separately about on-site and off-site states of general DNLS-type equations in dimensions $d=1,2,3$. A challenging aspect of this path is that  the energy difference between on- and off-site states (PN barrier), and other properties, is exponentially small in $\alpha$;
 see Figure \ref{fig:bifurcation}.

\subsection{Outline of the paper}
{\ \ }

\noindent In Section \ref{section:notation}, we first provide a summary of basic facts about the discrete Fourier transform and some of its properties, and also provide a list of notations and conventions used throughout the paper.
\\ In Section \ref{section:NLS}, we summarize the properties of the continuum NLS solitary standing wave which we will use in the following sections. 
\\ In Section \ref{section:mainresults}, we state our result on the bifurcation of onsite and offsite solitary waves  for spatial dimension $d = 1$ (Theorem \ref{th:main}; see figure \ref{fig:onsiteoffsite}),  and general dimension $d = 1, 2, 3$  (Theorem \ref{th:maingeneral}). The  exponentially small bound on the PN-barrier,  the $l^2$ energy difference between onsite and offsite solutions is stated in Theorem \ref{th:PN}.\\
 Sections \ref{section:1dproof} through \ref{section:rescalinglow0} contain the proof of Theorem \ref{th:main}. In particular, Section \ref{section:formalexpansion} contains a formal asymptotic analysis of DNLS which we use in the proof.
 \\ Sections \ref{section:rigorousexpansionproof} and \ref{section:rescalinglow0} construct the error in the rigorous asymptotic expansion of the solution to DNLS.\\
  In Section \ref{section:rigorousexpansionproof}, we obtain equations for the error and and construct the high frequency component of the error as a function of the low frequency components and the small parameter $\alpha$.\\
   In Section \ref{section:rescalinglow0} we construct the low frequency component of the error and map our asymptotic expansion back to the solution to DNLS, completing the proof of Theorem \ref{th:main}. \\
   In Section \ref{section:PNbarrier}, we prove Theorem \ref{th:PN}. \\
 In Section \ref{section:higherdim}, we provide details which generalize the proofs found in Sections \ref{section:1dproof} through \ref{section:PNbarrier} to general spatial dimension $d = 1,2,3$, completing the proofs of Theorems \ref{th:maingeneral} and \ref{th:PN}. \\ 
   There are several appendices. In Appendix \ref{subsection:DFT} we discuss properties of the discrete Fourier transform. In Appendix \ref{IFT} we provided a formulation of the implicit function theorem, which we can apply directly to our setting. Appendix \ref{section:expogeneral} addresses generally the exponential decay of solitary waves in the Fourier variable, a property which we use often. Appendices \ref{subsection:details} through \ref{subsection:details2} contain various technical tools and details used.

\section*{Acknowledgments} This work was supported in part by National Science Foundation grants DMS-10-08855,  DMS-1412560   and DGE-1069420 (the Columbia Optics and Quantum Electronics IGERT), and a grant from the Simons Foundation (\#376319, MIW).  The authors thank O. Costin, R. Frank, P. Kevrekides and J. Marzuola for very stimulating discussions.
\bigskip

\subsection{Preliminaries, Notation and Conventions}
\label{section:notation}

 Define the discrete Fourier transform (DFT) of the sequence
  ${f} = \{ f_n \}_{n \in \mathbb{Z}^d} \in l^1(\mathbb{Z}^d) \cap l^2(\mathbb{Z}^d)$
    by
\begin{align}
\widehat{f}(q) = \mathcal{F}_{_D} [{f} ](q) \equiv \sum_{n \in \mathbb{Z}^d}  f_n e^{- i q \cdot n} \ ,\ q \in \mathbb{R}^d . \label{eqn:DFT}
 \end{align}
 Since $\widehat{f}(q + 2 \pi e^{(j)}) = \widehat{f}(q)$, where  $e^{(j)}$ denotes the $j^{th}$ standard basis element in $\mathbb{R}^d$, we shall view $\widehat{f}(q)$ as being defined on the torus $\mathbb{T}^d \simeq \mathcal{B}/ \pi \mathbb{Z}^d$, where $\mathcal{B}$ is fundamental period cell  ({\it Brillouin zone}),
 \begin{equation}
\mathcal{B} \equiv [-\pi, \pi]^d \ .
 \label{brillouin}\end{equation}
 Thus, $\widehat{f}$ is completely determined by its values on $\mathcal{B}$.
 We shall also make use of the scaled {\it Brillouin zone},
 \begin{equation}
 \mathcal{B}_\alpha\ \equiv \Big[-\frac{\pi}{\alpha}, \frac{\pi}{\alpha}\Big]^d \quad \quad {\rm with} \quad \quad \mathbb{T}^d \simeq \mathcal{B}_{\alpha} \Big/ \frac{\pi}{\alpha} \mathbb{Z}^d.
 \label{scaled-brillouin}\end{equation}
The inverse discrete Fourier transform is defined by
 \begin{align}
 f_n = \left( \mathcal{F}^{-1}_{_D} [\widehat{f}] \right)_n = \frac{1}{(2 \pi)^d} \int_{\mathcal{B}} \widehat{f}(q) e^{i q \cdot n} dq\ . \label{eqn:invDFT}
 \end{align}
A summary of key properties of the discrete Fourier transform is included in Appendix \ref{subsection:DFT}.\medskip

\medskip
A function $F(q), \ q \in \mathbb{R}^d$ is said to be $2 \pi-$ {\it periodic } if for all $ j \in \{1, \dots, d \} : F(q) = F(q + 2 \pi e^{(j)})$. A function $F(q), \ q \in \mathbb{R}^d$ is said to be $2 \pi \sigma-$ {\it pseudo-periodic } if there exists a $\sigma \in \mathbb{R}$ such that $F(q) = e^{2 \pi i \sigma} F(q + 2 \pi e^{(j)}),\ j=1,\dots,d$. Throughout the paper, we shall follow the convention that $C > 0$ and $C_j$, $j \in \mathbb{N}$  refer to constants, which may not necessarily be the same constant between any two inequalities.
\medskip

For $a$ and $b$ in $L^1_{_{\rm loc}}(\mathbb{R}^d)$, we define the \emph{convolution on $\mathcal{B}_\alpha$} by
\begin{align}
    \left(a *_{_{\alpha}} b\right)(q) = \int_{\mathcal{B}_{\alpha}} a(\xi) b(q- \xi) d\xi, \label{eqn:periodicconvolution}
\end{align}

\medskip
\begin{proposition}
\label{prop:pseudoperiodicity}
Let $a(q)$,$b(q)$, and $c(q)$ $\ 2 \pi \sigma / \alpha -$ denote pseudo-periodic functions with fundamental cell $\mathcal{B}_{\alpha}$. Then,  their convolution satisfies the usual commutative and associative properties:
\begin{align}
a *_{\alpha} b = b *_{\alpha} a \quad \quad {\rm and } \quad \quad (a *_{\alpha} b) *_{\alpha} c = a *_{\alpha} ( b *_{\alpha} c).
\end{align}
\end{proposition}
\medskip

\noindent The standard convolution on $\mathbb{R}^d$ is defined by
\begin{align}
(f * g) (q) = \int_{\mathbb{R}^d} f(\xi) g (q - \xi) d \xi\ .
\end{align}

\noindent For $f \in L^2(\mathbb{R}^d)$, the continuous Fourier transform and its inverse are given by
\begin{align}
\widetilde{f}(q) = \mathcal{F}_{_C} [ f ](q) \equiv \int_{\mathbb{R}^d} f(x) e^{- i q \cdot x} dx, \qquad {\rm and} \qquad
 f(x) = \mathcal{F}_{_C}^{-1} [\widetilde{f}](x) = \frac{1}{(2 \pi)^d} \int_{\mathbb{R}^d} \widetilde{f}(q) e^{i q \cdot x} dq\ .
 \end{align}
 \medskip

\noindent The following are notations used throughout:
\begin{enumerate}
 \item  The inner product,
 \begin{align}
 \left\langle f, g \right\rangle = \left\langle f, g \right\rangle_{L^2(\mathcal{B})} = \int_{\mathcal{B}} \overline{f(q)} \ g(q) \ dq,
 \end{align}
 where $\overline{f}$ is the complex conjugate of $f$. \\

\item $ \| {f} \|^2_{l^2(\mathbb{Z}^d)}  = \sum_{n \in \mathbb{Z}^d} | f_n |^2 . $ \\

\item $L^2(A)$, the space of functions satisfying $\| f \|_{L^2(A)} = \left( \int_A | f(x) |^2 dx \right)^{1/2} < \infty $. \\

\item $H^1(A)$, the space of functions satisfying $\| f \|_{L^2(A)} = \left( \| f \|^2_{L^2(A)} + \| \nabla f \|^2_{L^2(A)}  \right)^{1/2} < \infty $.

\item $L^{2,a}(A)$, the space of functions satisfying $\| f \|_{L^{2,a}(A)} = \left(\int_A (1 + |q|^2)^a |f(q)|^2 dq \right)^{1/2} < \infty$, where $L^{2,0}(A) = L^2(A)$. \\

\item $L^{2,a}_{_{\rm even}}(A)$, the space of functions $f \in L^{2,a}(A)$ which are even. That is, for any $\tau_j = \pm 1,$ $ j = 1,\ \dots \,d$, $f(\tau_1 x_1, \ \dots \, \tau_d x_d) = f(x_1,\ \dots \,x_d)$. We also refer to $L^2_{_{\rm even}}(A) = L^{2,0}_{_{\rm even}}(A)$. We also refer to these as symmetric functions. \\

\item $H^a(\mathbb{R}^d)$, the space of functions such that $\widetilde{f} = \mathcal{F}_{_C}[f] \in L^{2,a}(\mathbb{R}^d)$, with $\left\| f \right\|_{H^a(\mathbb{R}^d)} \equiv \| \widetilde{f} \|_{L^{2,a}(\mathbb{R}^d)}$. \\

\item For $f:\mathbb{Z}^d\to\mathbb{C}$, $f=\{f_n\}_{n\in\mathbb{Z}^d}$, define the forward  difference operators:\ 
$
\left( \delta_j f \right)_n = f_{n + e^{(j)}} - f_n ,\ \ j=1, \dots, d\ .
$
Here, $e^{(j)}$ is the standard unit vector in the $j$th coordinate direction.\\ When $d = 1$, we also use the short-hand
$
\left( \delta f\right)_n = \left( \delta_1 f \right)_n = f_{n + 1} - f_n.
$

\item For $d = 1$ and $f= \{ f_n \}_{n \in \mathbb{Z}}$, the one-dimensional discrete Laplacian is given by:
 \begin{align*}
\left( \delta^2 f \right)_n = f_{n + 1} + f_{n - 1} - 2 f_n\ .
 \end{align*}
 For general dimension, $d$:
 \begin{align*}
\left( \delta^2 f\right)_n =  \sum_{|j-n|=1} f_j - 2d\ f_n\ .
 \end{align*} 
 The summation is over nearest neighbor lattice points.\\

\item $\chi_{_A}(x) = \left\{
     \begin{array}{lr}
       1 & : x \in A \\
       0 & : x \notin A
     \end{array}
   \right.$, the indicator function for a set $A$, and
    $ \overline{\chi}_{_A} = 1 - \chi_{_A} $. \\

\item $f(\alpha) = \mathcal{O}(\alpha^{\infty})$ if for all $n \geq 1$, $ f(\alpha) = \mathcal{O}(\alpha^n)$. \\

\end{enumerate}

Throughout this article, we shall make frequent use of the following inequalities.

\begin{enumerate}
\item Fix $a > d/2$. Then, if  $f_1,\ f_2\in H^a(\mathbb{R}^d)$ then, the product $f_1 f_2\in H^a(\mathbb{R}^d)$.  Moreover, we have
\begin{align*}
\left\| f_1 f_2 \right\|_{H^a(\mathbb{R}^d)} \lesssim \left\| f_1 \right\|_{H^a(\mathbb{R}^d)} \ \left\| f_2 \right\|_{H^a(\mathbb{R}^d)} .
\end{align*}
\item Equivalently by the Plancherel identity, we have that 
if  $\widetilde{f_1},\ \widetilde{f_2}\in L^{2,a}(\mathbb{R}^d)$ then, their convolution $\widetilde{f_1} * \widetilde{f_2}\in L^{2,a}(\mathbb{R}^d)$
 and we have:
\begin{align}
\left\| \widetilde{f_1} * \widetilde{f_2} \right\|_{L^{2,a}(\mathbb{R}^d)} \lesssim \left\| \widetilde{f_1} \right\|_{L^{2,a}(\mathbb{R}^d)} \ \left\| \widetilde{f_2} \right\|_{L^{2,a}(\mathbb{R}^d)}. \label{eqn:algebra0}
\end{align}

\end{enumerate}

\section{Properties of the NLS  ground state standing waves} \label{section:NLS}

The following results summarize properties of the NLS solitary standing wave (``soliton'') and its Fourier transform on $\mathbb{R}^d$. See, for example,  references \cite{SS:99,Bgn:99,T:06}. \medskip

\begin{proposition}[NLS Ground State]
\label{prop:psi}
There exists a unique positive $H^1(\mathbb{R}^d)$ solution $\psi_{| \omega |}(x)$ to \eqref{eqn:introtiNLS} which is real-valued, symmetric about $x = 0$ and decaying to zero at infinity. Moreover,

\begin{enumerate}
\item $\psi_{| \omega |}(x)$ is in Schwartz class, $\mathcal{S}(\mathbb{R}^d)$ and is  exponentially decaying:
    \begin{align}
    | \psi^{|\omega|}(x) | \lesssim \frac{e^{- \sqrt{ | \omega | }  \ |x|}}{ |x|^{\frac{d-1}{2}}}.
    \end{align}
\item $ \psi_{| \omega |}(x) = \sqrt{| \omega |} \ \psi_1(\sqrt{| \omega |}x)$.

\item Uniqueness up to Phase and Spatial-Translation:\
Any solution of \eqref{eqn:introtiNLS} is of the form
$e^{i\theta}\ \psi_{| \omega |}(x - x_0) $
for some $\theta\in\mathbb{R}$ and $x_0 \in \mathbb{R}^d$.

\end{enumerate}
\end{proposition}
\bigskip

We shall also require the detailed properties of  the Fourier transform of $\psi_{| \omega |}$:\medskip

\begin{proposition}[Fourier transform of NLS Ground State]
\label{prop:Psi}
{\ }

\noindent The Fourier transform, $\widetilde{\psi_{| \omega |}}(q) = \mathcal{F}_c[\psi_{| \omega |}](q)$, satisfies the equation
\begin{align}
    \omega\ \widetilde{\psi_{| \omega |}}(q) = |q|^2 \widetilde{\psi_{| \omega |}}(x)\ -\ \frac{1}{4 \pi^2}\ \widetilde{\psi_{| \omega |}} * \widetilde{\psi_{| \omega |}} * \widetilde{\psi_{| \omega |}}(q),\ \qquad q\in\mathbb{R}^d.
\label{Psi-eqn}    \end{align}
and has the following properties:
\begin{enumerate}
    
\item  Scaling: 
\begin{align}
 \widetilde{\psi_{| \omega |}}(q) = \left(\frac{1}{| \omega |}\right)^{(d-1)/2} \widetilde{\psi_1} \left(\frac{q}{\sqrt{| \omega |}}\right)  \label{eqn:Psi-omega}.
 \end{align}

\item Exponential decay bound: Let $a > d/2$. Then, there exists a positive constant $C_0$ such that
\begin{equation}
\left\| e^{C_0 |q| } \ \widetilde{\psi_1}(q) \right\|_{L^{2,a}(\mathbb{R}_q^d)} \lesssim 1. 
\label{Psi-bound}
\end{equation}

 \end{enumerate}
\end{proposition}
The   exponentially weighted $L^{2,a}$ bound, \eqref{Psi-bound}, follows from Lemma \ref{lemma:expogeneral} in Appendix \ref{section:expogeneral}.

\medskip

In our bifurcation analysis, an important role is played by the the operator $f\mapsto L_+f$, the linearization of the stationary NLS equation with $\omega = 1$, \eqref{eqn:introtiNLS}, about $\psi_1$. Here,
\begin{equation}
L_+ \equiv
1 - \Delta_x -  3 (\psi_1(x))^2\ ,
\label{L+def}
\end{equation}
In Fourier space, $\widetilde{f}(q)\mapsto \widetilde{L_+} \widetilde{f}(q)$, where
\begin{equation}
\widetilde{L_+} = 1 + |q|^2 - \frac{3}{4 \pi^2}\ \widetilde{\psi_1} * \widetilde{\psi_1} *
\label{L+Fourier}
\end{equation}
In particular, we require a characterization  of the $L^2(\mathbb{R}^d)-$ kernel of $L_+$;  see \cite{W:85,K:89,Bgn:99,T:06}.
\begin{proposition}
\label{prop:Lplus}
Assume  $1\le d\le3$. 
\begin{enumerate}
\item The continuous spectrum of $L_+$ is given by the half-line $[1,\infty)$.
\item Zero is an isolated eigenvalue with corresponding eigenspace, 
\[ {\rm kernel}(L_+)={\rm span}\{\partial_{x_j}\psi_1(x),\ j=1,\dots,d\ \}.\]
\item Equivalently, the kernel of $\widetilde{L_+} $  is spanned by the functions $q_j \widetilde{\psi_1}(q),\ j=1,\dots, d$.
\item $L_+: H^{a}_{_{\rm even}}(\mathbb{R}^d)\to H^{a-2}_{\rm even}(\mathbb{R}^d)$ is an isomorphism.
\item $\widetilde{L_+}: L^{2,a}_{_{\rm even}}(\mathbb{R}^d)\to L^{2,a-2}_{_{\rm even}}(\mathbb{R}^d)$ is an isomorphism.
\end{enumerate}
\end{proposition}
 
\section{Main results}
\label{section:mainresults}
\medskip

We begin with precise statements of our results on the existence of discrete solitary standing waves in spatial dimensions $d = 1, 2, 3$. We first give the simpler statement in dimension $d=1$,  referring to the two types of solutions of Definition \ref{defn:onoff}; see figure \ref{fig:onsiteoffsite}.\medskip

\begin{theorem}({\it Discrete DNLS solitary waves on $\mathbb{Z}$})
\label{th:main}
 Let  $\psi_1(x)$ denote the ground state of NLS; see Proposition \ref{prop:psi} with $\omega = - 1$ and  $\widetilde{\psi_1}$ its  Fourier transform; see Proposition \ref{prop:Psi}. Consider the nonlinear eigenvalue problem governing real-valued one-dimensional DNLS standing waves:
\begin{align}
& - \alpha^2 G^{\alpha}_n = - (\delta^2 G^{\alpha})_n - (G^{\alpha}_n)^3,
\qquad G^{\alpha} \in l^2(\mathbb{Z}). \label{eqn:tiDNLS2}
\end{align}
\begin{enumerate}
 \item Fix an integer $J\ge0$. There exist mappings $\mathcal{G}_j: L^2(\mathbb{R}) \rightarrow L^2(\mathbb{R})$, for $ j=0,1,\dots, J$ and a positive  constant $\alpha_0 = \alpha_0[J] > 0$ such that for all $0 < \alpha < \alpha_0$, the following holds:\\
  There exist two families of real-valued symmetric solutions to (\ref{eqn:tiDNLS2}). These are on-site (vertex-centered) and off-site (bond-centered) solutions of DNLS \eqref{eqn:tiDNLS2}.  To leading order in $\alpha^2$ satisfy: $ G^{\alpha, {\rm on} }_n,\ 
 G^{\alpha, {\rm off} }_n\ \approx\ \mathcal{F}_D^{-1}\left[\widetilde{\psi_1}\right](n)$. More precisely, 
\begin{align}
\begin{array}{lll}
& \textrm{\rm {\bf On-site symmetric \ (vertex-centered):}} \\
& \hspace{3cm}  G^{\alpha, {\rm on} }_n = \sum_{j = 0}^J \ \alpha^{2j} \  \mathcal{G}_j \left[ \widetilde{\psi_1} \right](n) + \mathcal{E}^{\alpha, J, {\rm on } }_n, \quad \quad \quad  n \in \mathbb{Z},\\
 & \hspace{3cm}  {\rm where}\ \  \left\| \mathcal{G}_j [ \widetilde{\psi_1} ] (\cdot) \right\|_{l^2(\mathbb{Z})} \sim \alpha^{1/2}, \quad \quad  \quad  \left\|  \mathcal{E}^{\alpha, J, {\rm on}} \right\|_{l^2(\mathbb{Z})} \lesssim \alpha^{2J + 5/2}, \\
& \textrm{\rm {\bf Off-site symmetric \ (bond-centered):}}  \\
& \hspace{3cm}  G^{\alpha, {\rm off} }_n =
G^{\alpha, {\rm on} }_n = \sum_{j = 0}^J \ \alpha^{2j} \  \mathcal{G}_j \left[   \widetilde{\psi_1} \right](n- 1/2)  + \mathcal{E}^{\alpha, {\rm off} }_n, \quad \quad \quad  n \in \mathbb{Z}, \\
   &  \hspace{3cm} {\rm where}\ \  \left\|  \mathcal{G}_j [ \widetilde{\psi_1}]  \left( \cdot - \frac{1}{2} \right) \right\|_{l^2(\mathbb{Z})} \sim  \alpha^{1/2}, \quad \quad \quad \left\| \mathcal{E}^{ \alpha, J, {\rm off}}  \right\|_{l^2(\mathbb{Z})} \lesssim \alpha^{2J + 5/2}. 
\end{array} \label{eqn:solutions}
\end{align}
%
%
%
     \end{enumerate}
\end{theorem}

In Definition \ref{defn:sigmacentered}, we generalize the notion of on-site symmetric (vertex-centered) and off-site symmetric (bond-centered) waves to the notion of $\sigma-$ centered waves.
%
\medskip

\begin{theorem}
\label{th:maingeneral}[Discrete solitary waves in $\mathbb{Z}^d$]
 Let  $\psi_1(x)$ denote the ground state of NLS; see Proposition \ref{prop:psi} with $\omega = - 1$. Denote by $\widetilde{\psi_1}$ the Fourier transform of $\psi_1$ (see Proposition \ref{prop:Psi}). Consider the nonlinear eigenvalue problem governing real-valued discrete solitary waves of DNLS:
\begin{equation}
 - \alpha^2 G^{\alpha}_n = - (\delta^2 G^{\alpha})_n - (G^{\alpha}_n)^3, \quad \quad n \in \mathbb{Z}^d \label{eqn:tiDNLS2general},\qquad G^{\alpha} \in l^2(\mathbb{Z}^d).
\end{equation}
Fix $0 \leq J < \infty, \ \ J \in \mathbb{N}$. There exists a constant $\alpha_0 = \alpha_0[J] > 0$ and $J > 0$ functionals $\mathcal{G}_j : L^2(\mathbb{R}^d) \rightarrow L^2(\mathbb{R}^d)$ such that for all $0 < \alpha < \alpha_0$, the following holds:
\begin{enumerate}
\item  For every $d-$ tuple $\sigma \in \{ 0, 1/2 \}^d$ (see Table \ref{table:centering}), there exists a real-valued $\sigma-$ centered  solution of \eqref{eqn:tiDNLS2}; see Definition \ref{defn:sigmacentered}:
\begin{align}
& G^{\alpha, \sigma}_n =  \sum_{j = 0}^J \ \alpha^{2j} \ \mathcal{G}_j \left[ \widetilde{\psi_1} \right](n - \sigma) +\ \mathcal{E}^{\alpha, J, \sigma}_n, \ \ {\rm where} \label{eqn:solutionsgeneral} \\
& \left\| \mathcal{G}_j [\widetilde{\psi_1}] \left( \cdot - \sigma \right) \right\|_{l^2(\mathbb{Z}^d)} \sim \alpha^{1 - d/2},\qquad \left\| {\mathcal{E}}^{\alpha, J \sigma} \right\|_{l^2(\mathbb{Z}^d)} \lesssim \alpha^{2J + 3 - d/2 }. 
\end{align} \\
%

\item There are $2^d$ solutions, one for each $\sigma \in \{ 0 , 1/2 \}^d$. Modulo reflections, there are $d + 1$ distinct centerings which are labeled in Table \ref{table:centering}. 
\end{enumerate}
\end{theorem}

\begin{table}[h]
\begin{centering}
\small{
\begin{tabular}{ |c|c|c| }
\hline
$ d $ & $ \sigma \in \mathbb{Z}^d/2 $ & {\bf Centering} \\
\hline
\hline
\multirow{2}{*}{1} & 0 & Vertex-centered (On-site) \\
 & 1/2 & Bond-centered (Off-site) \\
\hline
\hline
\multirow{4}{*}{2} & (0, 0) & Vertex-centered (On-site) \\
 & (0, 1/2) & Bond-centered (Off-site) \\
 & (1/2, 0) & Bond-centered (Off-site) \\
 & (1/2, 1/2) & Cell-centered (Off-site) \\
\hline
\hline
\multirow{8}{*}{3} & (0, 0, 0) & Vertex-centered (On-site) \\
 & (0, 0, 1/2) & Bond-centered (Off-site) \\
 & (0, 1/2, 0) & Bond-centered (Off-site) \\
 & (1/2, 0, 0) & Bond-centered (Off-site) \\
 & (0, 1/2, 1/2) & Face-centered (Off-site) \\
 & (1/2, 0, 1/2) & Face-centered (Off-site) \\
 & (1/2, 1/2, 0) & Face-centered (Off-site) \\
 & (1/2, 1/2, 1/2) & Cell-centered (Off-site) \\
\hline
\end{tabular}
\label{table:centering}
\caption{Centering of discrete solitions in dimensions $d = 1, \ 2, \ 3$.} }
\end{centering}
\end{table}

\noindent {\bf Conjecture 3:} The $2^d$ solutions in Theorem \ref{th:maingeneral} are locally unique for each $\sigma$ up to lattice translations by some $M \in \mathbb{Z}^d$. That is, if $\{ J^{\alpha, \sigma}_n \}_{n \in \mathbb{Z}^d}$ is a solution to \eqref{eqn:tiDNLS2general} with $\alpha < \alpha_0$, it must be centered about $ \sigma + M $ for $\sigma \in \{ 0, 1/2 \}^d$ and $M \in \mathbb{Z}^d$, and we must have $J^{\alpha, \sigma}_n = \pm G^{\alpha, \sigma}_{n + M}$. \\

 \begin{remark} \label{remark:main2}
In Theorem \ref{th:maingeneral}, we assume $1\le d\le3$.
 We do not expect there to be a bifurcation from at zero frequency for dimensions $d\ge4$. Indeed, the $\alpha^2\to0$ rescaled-limit of such bifurcating states is the solitary standing wave solution of continuum NLS, satisfying  $-\Delta u+u-u^3=0$. A well-known  argument based on ``Pohozaev'' / virial identities shows that the equation $-\Delta u+u-u^p=0$ has  $H^1(\mathbb{R}^d)$ solutions  only if $p<(d+2)/(d-2)$. For the cubic case, $p=3$, this implies $ d\le3$; see, for example,  \cite{S:77,SS:99}. Therefore, there can be no bifurcation for $d\ge4$. \\
\end{remark}


\begin{theorem}
\label{th:PN} [Exponential smallness of Peierls-Nabarro barrier]
Let $\sigma_1, \sigma_2 \in \{ 0, 1/2 \}^d$. There exist constants $\alpha_0 > 0, C$ and $D>0$ such that for all $0 < \alpha < \alpha_0$,
\begin{align}
& \bigg| \mathcal{N} [ G^{\alpha,  \sigma_1} ] - \mathcal{N} [ G^{\alpha,  \sigma_2} ] \bigg| =  \left|  \| G^{\alpha,  \sigma_1} \|^2_{l^2(\mathbb{Z}^d)} - \|G^{\alpha, \sigma_2} \|^2_{l^2(\mathbb{Z}^d)} \right|\ \le\ D \alpha^{2 - d} \ e^{- C / \alpha},\quad \textrm{and} \nonumber \\
  &  \bigg| \mathcal{H} [ G^{\alpha,  \sigma_1} ] - \mathcal{H} [ G^{\alpha,  \sigma_2} ] \bigg| \ \le \ D \alpha^{2 - d} \ e^{- C / \alpha}. \label{eqn:pntheorem}
\end{align}
\end{theorem}


\begin{remark}[Connection to results on stability]

Results of  \cite{GSS:86,W:86} imply 
that  conditions for the stability of a positive and decaying solitary standing wave solution, $G^{\alpha, \sigma}$, is orbitally Lyapunov stable provided
\begin{itemize}
\item[(S1)] $L_+^{\rm disc} $ has one negative eigenvalue.

\item[(S2)] The following ``slope'' condition holds:
\begin{align}
\frac{d}{d \alpha} \mathcal{N} \left[ G^{\alpha,\sigma} \right]  = - \frac{1}{\alpha^2} \frac{d}{d \alpha} \mathcal{H} \left[ G^{\alpha,\sigma} \right]  > 0. \end{align}
\end{itemize}

Conjecture 2, \eqref{eqn:conjsign}, implies that the onsite DNLS standing wave satisfies (S1) and therefore that the curve $\alpha\mapsto \mathcal{N} \left[ G^{\alpha,\sigma=0} \right]$ determines the stability or instability of the on-site wave. 

\end{remark}

\section{Beginning of the proof of Theorem \ref{th:main}; formulation of DNLS in Fourier space for $d=1$}
\label{section:1dproof}

For ease of presentation, we focus initially on the one-dimensional case ($d=1$). We adapt the current discussion to dimensions $d=2,3$ (Theorem \ref{th:maingeneral}) in section \ref{section:higherdim}.\medskip

Applying the discrete Fourier transform
 \eqref{eqn:DFT} to equation \eqref{eqn:tiDNLS2}, governing $G=G^\alpha$, we obtain an equivalent equation for the discrete Fourier transform, $\widehat{G}(q)=\widehat{G^{\alpha}}(q)$:
\begin{align}
     \widehat{DNLS}[\widehat{G}](q)\ \equiv\ &\big[ \alpha^2 +  M(q) \big] \widehat{G}(q) - \left(\frac{1}{2 \pi} \right)^2 \widehat{G} *_{_1} \widehat{G} *_{_1} \widehat{G}(q) = 0, \nonumber\\
 &    \widehat{G}(q + 2\pi) = \widehat{G}(q),\ \label{eqn:tiDNLS2realfourier}
\end{align}
where we recall the definition of the convolution $f*_{_1}g$ on $\mathcal{B} = \mathcal{B}_1$ in \eqref{eqn:periodicconvolution}
and its properties (Appendix \ref{subsection:DFT}).
Here, $M(q)$ denotes the (discrete) Fourier symbol of the 1-dimensional discrete Laplacian with unit lattice-spacing:
\begin{align}
M(q) \ \widehat{G}(q)\ =\ \widehat{(\delta^2 G)}(q) =  4 \sin^2(q/2) \ \widehat{G}(q) \ . \label{eqn:Mdef}
\end{align}
 By Proposition \ref{prop:off-on}  we have that
\begin{enumerate}
\item if $G$ is onsite symmetric, then $\widehat{G}(q)= \widehat{K}(q)$, where $ \widehat{K}(q)$ is real and symmetric, and
\item if $G$ is offsite symmetric, then $\widehat{G}(q)=e^{-iq/2} \widehat{K}(q)$, where $\widehat{K}(q)$ is real and symmetric.
\end{enumerate}
We therefore seek $\widehat{G}(q)$ in the form
\begin{align}
\widehat{G^{\sigma}}(q)\ &=\ e^{-i\sigma q}\ \widehat{K^{\sigma}}(q),\ \ \sigma=0,1/2\label{Gaq}
\\ \widehat{K^{\sigma}}(q)&= \widehat{K^{\sigma}}(-q),\ \ \overline{\widehat{K^{\sigma}}(q)}=\widehat{K^{\sigma}}(q)
 \label{Hrealsym}
\end{align}
Substitution of \eqref{Gaq} into \eqref{eqn:tiDNLS2realfourier} yields
\begin{align}
      &\big[ \alpha^2 +  M(q) \big] \widehat{K^\sigma}(q) - \left(\frac{1}{2 \pi} \right)^2 \widehat{K^\sigma} *_{_1} \widehat{K^\sigma} *_{_1} \widehat{K^\sigma}(q) = 0, \quad q \in \mathbb{R},  \label{Heqn}\\
& \widehat{K^{\sigma}}(q+2\pi)=e^{2\pi i\sigma}\ \widehat{K^{\sigma}}(q), \qquad \sigma=0,1/2\ .
\label{Hpseudo}\end{align}
The ``Bloch'' phase factor, $e^{2\pi i\sigma}$ (equal to $\pm1$) encodes the on-site and off-site cases.  
\bigskip

\begin{lemma}\label{pisigma-pseudo}
Let $A(q)$, defined on $\mathbb{R}$, be $2\pi\sigma-$ pseudo-periodic, {\it i.e.} $A(q+2\pi)=e^{2\pi i\sigma}\ A(q)$. Then,
 $A(q)$ is completely determined by its values on $\mathcal{B}=[-\pi,\pi]$ and has the representation:
 \begin{align}
A(q) =  \sum_{m \in \mathbb{Z}} \chi_{_{\mathcal{B}}}(q - 2 m \pi) A(q - 2  m \pi) e^{2 \pi i m \sigma}.  \label{eqn:Atiling}
\end{align}
\end{lemma}
\noindent{\bf Proof of Lemma \ref{pisigma-pseudo}:} Since $A(q)=e^{2\pi i \sigma} A(q-2\pi )$, we have $A(q)=e^{2\pi i m\sigma} A(q-2\pi m)$ for all $m\in\mathbb{Z}$. Hence, 
\begin{align*}
A(q) = A(q) \sum_{m \in \mathbb{Z}} \chi_{_{\mathcal{B}}}(q - 2 m \pi) = \sum_{m \in \mathbb{Z}} \chi_{_{\mathcal{B}}}(q - 2 m \pi) A(q - 2  m \pi) e^{2 \pi i m \sigma}. \qquad \qquad \Box
\end{align*}

By Lemma \ref{pisigma-pseudo} we may express $\widehat{K^{\sigma}}(q)$, for any $q \in \mathbb{R}$, explicitly in terms of its values on $q \in \mathcal{B}$.  In particular,  we set
\begin{align}
\widehat{K^{\sigma}}(q) = \widehat{\phi^{\sigma}}(q), \qquad q \in \mathcal{B}.  \label{H-ansatz}
\end{align}
Extending \eqref{H-ansatz} to $q \in \mathbb{R}$, we have:
\begin{align}
& \widehat{K^\sigma}(q) \equiv \sum_{m \in \mathbb{Z}} \chi_{_{\mathcal{B}}}(q - 2 m \pi) \  \widehat{\phi^{ \sigma}} (q - 2 m \pi) \ e^{2 m \pi i \sigma},  \qquad  {\rm and} \qquad \widehat{G^\sigma}(q) = e^{-i\sigma q} \ \widehat{K^\sigma}(q)  \label{eqn:ansatz3}.
\end{align}
Note that $\widehat{\phi^{\sigma}}(q - 2 m \pi) = \chi_{_{\mathcal{B}}}(q - 2 m \pi) \ \widehat{\phi^{\sigma}}(q - 2 m \pi)$ is supported on $ \{ q: q \in 2 m \pi + \mathcal{B} = [(2 m - 1) \pi, (2 m + 1) \pi] \}$. Therefore,
\begin{align}
 \chi_{_{\mathcal{B}}}(q)  \ \widehat{K^{\sigma}}(q) = \widehat{\phi^{\sigma}}(q), \qquad {\rm and} \qquad \chi_{_{\mathcal{B}}}(q) \ \widehat{G^{\sigma}}(q) = e^{- i q \sigma} \ \widehat{\phi^{\sigma}}(q). \label{eqn:decomp22}
\end{align}
Equations \eqref{eqn:ansatz3} encode the required $2\pi-$ periodicity of
 $\widehat{G^\sigma}(q)$ and the $2\pi\sigma-$ pseudo-periodicity of $\widehat{K^\sigma}(q)$ on all $\mathbb{R}$,. Furthermore,
 \begin{equation}
 \textrm{$\widehat{G^\sigma}(q)$ and $\widehat{K^\sigma}(q)$ are completely specified by $\widehat{\phi^\sigma}(q)$ for $q\in\mathcal{B}$.}
 \label{eqn:phitoG}
 \end{equation}

 With a view toward deriving an equation from \eqref{Heqn} determining $\widehat{\phi^\sigma}(q)$ for $ q\in\mathcal{B}$, we require a lemma which  facilitates simplification of the convolution terms in \eqref{Heqn}.

\begin{lemma}
\label{lemma:decomp}
Let  $\widehat{A}(q), \widehat{B}(q), \widehat{C}(q)$ be bounded $2\pi\sigma-$ pseudo-periodic functions of $q\in\mathbb{R}$.
Then
\begin{align}
\chi_{_{\mathcal{B}}}(q) \widehat{A} *_{_1} \widehat{B} *_{_1} \widehat{C} (q) = \chi_{_{\mathcal{B}}}(q) \sum_{m = -1}^1 e^{2 \pi i m\sigma} \widehat{A} *_{_1} \left[\ \widehat{B} *_{_1}  \left( \chi_{_{\mathcal{B}}} \widehat{C} \right)\right]  ( q - 2 m \pi). \label{eqn:decomp2}
\end{align}
Note that since $\chi_{_{\mathcal{B}}}(q)$ is not pseudo-periodic, the convolution in \eqref{eqn:decomp2} is not associative.
\end{lemma}

\noindent {\bf Proof of Lemma \ref{lemma:decomp}:} Observe that $q, \xi, \zeta \in \mathcal{B} = [-\pi,\pi]$ implies that $q - \xi - \zeta \in [- 3 \pi,  3 \pi]$. Therefore, $q, \xi, \zeta \in \mathcal{B}$ implies $q - \xi - \zeta - 2 m \pi \in \mathcal{B} = [- \pi, \pi]$ if and only if
$m\in \{ -1, 0, 1 \}$, and that  $\chi_{_{\mathcal{B}}}(q - \xi - \zeta - 2 m \pi ) = 0$ for $ m \not\in \{ -1, 0, 1 \}$. Applying Lemma \ref{pisigma-pseudo} to $\widehat{C}(q)$ we have
\begin{align}
& \chi_{_{\mathcal{B}}}(q)  \widehat{A} *_{_1} \left[ \widehat{B} *_{_1} \widehat{C}\right](q) \nonumber \\
& = \chi_{_{\mathcal{B}}}(q) \int_{\mathcal{B}} d\xi \int_{\mathcal{B}}  d\zeta\ \widehat{A}(\xi)  \widehat{B}(\zeta) \sum_{m  \in \mathbb{Z}} \chi_{_{\mathcal{B}}}(q - \xi - \zeta - 2 m \pi)\ \widehat{C}(q - \xi - \zeta - 2 m \pi) e^{2 m \pi i \sigma} \nonumber \\
& = \chi_{_{\mathcal{B}}}(q) \int_{\mathcal{B}} d\xi \int_{\mathcal{B}} d\zeta\ \widehat{A}(\xi)  \widehat{B}(\zeta) \sum_{m = -1}^{1} \chi_{_{\mathcal{B}}}(q - \xi - \zeta - 2 m \pi)\ \widehat{C}(q - \xi - \zeta - 2 m \pi) e^{2 m \pi i \sigma} \nonumber \\
& = \chi_{_{\mathcal{B}}}(q) \sum_{m = -1}^1 e^{2 m \pi i \sigma} \widehat{A} *_{_1} \left[\widehat{B} *_{_1}   \left( \chi_{_{\mathcal{B}}} \widehat{C} \right)\right] (q - 2m \pi). \hspace{2cm} \Box
\end{align} 

\noindent Applying Lemma \ref{lemma:decomp} and \eqref{eqn:ansatz3}, we have:

 \begin{proposition} \label{prop:phieqn}
Equation \eqref{Heqn} for $\widehat{K^{\sigma}}(q)$ on $q \in \mathbb{R}$ is equivalent to the following equation for the compactly supported function  $\widehat{\phi^{\sigma}}(q)=\chi_{_{\mathcal{B}}}(q)\ \widehat{\phi^{\sigma}}(q)$: 
\begin{align}
& [\alpha^2 + M(q)] \ \widehat{\phi^{\sigma}}(q) - \frac{\chi_{_{\mathcal{B}}}(q)}{4\pi^2}  
\  \left(\ \widehat{\phi^{\sigma}} *  \widehat{\phi^{\sigma}} *  \widehat{\phi^{\sigma}}\ \right) (q) 
\nonumber\\
&\qquad\qquad - \ 
\frac{\chi_{_{\mathcal{B}}}(q)}{4\pi^2}  
\sum_{m = \pm1} \ e^{2 m \pi i \sigma} \ \left(\ \widehat{\phi^{\sigma}} *  \widehat{\phi^{\sigma}} *  \widehat{\phi^{\sigma}}\ \right) (q - 2 m \pi) = 0
\label{eqn:phieqn}
\end{align}
where
$M(q) \widehat{G}(q)\ =\ 4 \sin^2(q/2)\ \widehat{G}(q) = \widehat{\delta^2 G}(q) $; see \eqref{eqn:Mdef}.
\end{proposition}

\medskip 

\noindent {\bf Proof of Proposition \ref{prop:phieqn}:} We project \eqref{Heqn} onto $\mathcal{B}$, decompose $\widehat{K^{\sigma}}(q)$ with \eqref{eqn:ansatz3}, and apply Lemma \ref{lemma:decomp} to \eqref{Heqn} get
\begin{align}
 [\alpha^2 + M(q)] \ \widehat{\phi^{\sigma}}(q) - \frac{\chi_{_{\mathcal{B}}} }{4 \pi^2} \sum_{m = -1}^1 e^{2 \pi i m\sigma} \widehat{\phi^{\sigma}} *_{_1} \left[ \widehat{\phi^{\sigma}} *_{_1}  \left( \chi_{_{\mathcal{B}}} \widehat{\phi^{\sigma}} \right) \right]  ( q - 2 m \pi) = 0. \label{eqn:phieqn2}
\end{align}
We observe that the multiplication of \eqref{eqn:phieqn2} by $\overline{\chi}_{_{\mathcal{B}}}(q)$ implies \emph{a-priori} that
\begin{align}
\left[ \alpha^2 + M(q) \right] \ \overline{\chi}_{_{\mathcal{B}}}(Q) \ \widehat{\phi^{\sigma}}(q) = 0 \qquad \Longrightarrow \qquad \overline{\chi}_{_{\mathcal{B}}}(Q) \ \widehat{\phi^{\sigma}}(q) = 0,  \label{eqn:Phispt}
\end{align}
where we have used that $M(q) \geq 0$. Thus,  we may write for $ m = -1, 0, 1$,
\begin{align}
\widehat{\phi^{\sigma}} *_{_{1}} \left[ \widehat{\phi^{\sigma}} *_{_{1}}  \left( \chi_{_{\mathcal{B}}} \ \widehat{\phi^{\sigma}} \right) \right] (q - 2 m \pi)  & = \left( \chi_{_{\mathcal{B}}} \ \widehat{\phi^{\sigma}} \right) *  \left( \chi_{_{\mathcal{B}}} \  \widehat{\phi^{\sigma}} \right) *  \left( \chi_{_{\mathcal{B}}} \  \widehat{\phi^{\sigma}} \right)  (q - 2m \pi)  \nonumber \\
&  = \left(    \widehat{\phi^{\sigma}}   *   \widehat{\phi^{\sigma}}  *  \widehat{\phi^{\sigma}}   \right) (q - 2 m \pi ) ,
\end{align}
This completes the proof of Proposition \ref{prop:phieqn}. $\Box$ \\

\subsection{Rescaled equation for $\widehat{\phi^{\sigma}}$} \label{section:rescaledphi}
As discussed in Section \ref{section:strategy}, we expect that for $\alpha \ll 1$:
\begin{align}
\widehat{\phi^{\sigma}}(q) \sim \widetilde{\psi_{\alpha^2}}(q) =   \ \widetilde{\psi_1} \left( \frac{q}{\alpha} \right), \quad \quad 
\end{align}
where $\widetilde{\psi_1}$ denotes the Fourier transform of the continuum NLS solitary wave.
We therefore study \eqref{eqn:phieqn} using rescaling which makes explicit the relation between
DNLS and the continuum (NLS) limit for $\alpha$ small:
 We introduce:
\begin{align}
& \textrm{\rm Rescaled momentum:} \quad \quad & Q \equiv q/\alpha,\ \ 
Q\in \mathcal{B}_{\alpha} = [-\pi/\alpha,\pi/\alpha],  \nonumber \\
& \textrm{\rm Rescaled projection:} \quad \quad  &  \chi_{_{\mathcal{B}_{_{\alpha}}}}(Q) \equiv \chi_{_{\mathcal{B}}}(Q \alpha) = \chi_{_{ \left[ - \frac{\pi}{\alpha}, \frac{\pi}{\alpha} \right] }}(Q), 
 \nonumber \\
& \textrm{\rm Rescaled wave:} \quad \quad  &  \widehat{\Phi^{\sigma}}(Q) \equiv \widehat{\phi^{\sigma}}(Q \alpha) = \widehat{\phi^{\sigma}}( q), \nonumber \\
& \textrm{\rm Rescaled discrete Fourier symbol:} \quad \quad  & M_{\alpha}(Q) \equiv \frac{1}{\alpha^2} M(Q \alpha) =  \frac{4}{\alpha^2}  \sin^2 \left( \frac{ Q \alpha}{2} \right). \label{eqn:M2def}
\end{align}

The following proposition is a formulation of Proposition \ref{prop:phieqn}  in terms of functions of the rescaled quasi-momentum, $Q$:

\begin{proposition}
\label{prop:Phieqn}
Equation \eqref{Heqn} for $\widehat{K^{\sigma}}(q)$ on $q \in \mathbb{R}$ is equivalent to the following equation for  $\widehat{\Phi^{\sigma}}(Q)=\chi_{_{\mathcal{B}_\alpha}}(q)\ \widehat{\Phi^{\sigma}}(Q)$, compactly supported on $\mathcal{B}_\alpha=[-\pi/\alpha,\pi/\alpha]$: 
\begin{align}
&\mathcal{D}^{\sigma,\alpha}[\widehat{\Phi^{\sigma}}](Q)\ \equiv\ [1 + M_\alpha(Q)] \ \widehat{\Phi^{\sigma}}(Q) - \frac{\chi_{_{\mathcal{B}_\alpha}}(Q)}{4\pi^2}  
\  \left(\ \widehat{\Phi^{\sigma}} *  \widehat{\Phi^{\sigma}} *  \widehat{\Phi^{\sigma}}\ \right) (Q) 
\ +\ R_1^{\sigma} [ \widehat{\Phi^{\sigma}} ] (Q) = 0\ ,
\label{eqn:Phieqn}
\end{align}
where $R_1^{\sigma} [ \widehat{\Phi^{\sigma}} ]$ contains the $\pm1$-sideband contributions:
\begin{align}
R_1^{\sigma} [ \widehat{\Phi^{\sigma}} ] (Q) &\equiv  -  \frac{\chi_{_{\mathcal{B}_{_{\alpha}}}}( Q)}{4 \pi^2} \ \sum_{m = \pm 1} \ e^{2 m \pi i \sigma} \  \left(\ \widehat{\Phi^{\sigma}} *  \widehat{\Phi^{\sigma}} *  \widehat{\Phi^{\sigma}}\ \right)  ( Q - 2 m \pi / \alpha), \label{eqn:R1def}
\end{align}
and 
$M_\alpha(Q)=\ \frac{4}{\alpha^2} \sin^2(\frac{\alpha Q}{2})$; see \eqref{eqn:M2def}. 
\end{proposition}

\noindent To prove  Proposition \ref{prop:Phieqn} we need to re-express the convolutions in \eqref{eqn:phieqn} in terms of $\widehat{\Phi^{\sigma}}(Q)$. 
For this we use the following lemma, proved by change of variables. 
\begin{lemma}
\label{lemma:convresc}
Suppose that $\widehat{a}(q) = \widehat{A}(Q)$, $\widehat{b}(q) = \widehat{B}(Q)$, and $\widehat{c}(q) = \widehat{C}(Q)$, where $Q=q/\alpha$.  Then
\begin{align}
\left( \widehat{a} * \widehat{b} * \widehat{c} \right)  (q)  = \alpha^2 \ \left( \widehat{A} * \widehat{B} * \widehat{C} \right) (Q) .
\end{align}
\end{lemma}

Applying the rescalings \eqref{eqn:M2def} and Lemma \ref{lemma:convresc} to \eqref{eqn:phieqn} and  then dividing by $\alpha^2$, we obtain \eqref{eqn:Phieqn}.\   \\

Note that for small $|\alpha|$, the scaled symbol, $M_\alpha(Q)=\frac{4}{\alpha^2} \sin^2 \left( \frac{Q \alpha}{2} \right)$, has the expansion in powers of $\alpha^2$ for  fixed  $Q \in \mathbb{R}$:
\begin{align}
M_{\alpha}(Q)  = 2 \sum_{j = 0}^{\infty} \frac{  \alpha^{2j}  \ (-1)^{j} \  |Q |^{2j + 2} }{ (2j + 2)!}= \ |Q|^2 - \frac{\alpha^2 \ | Q |^4}{12} + \frac{ \alpha^4 \ |Q|^6}{360} + \mathcal{O} \left( \alpha^6 \ |Q|^8 \right).
\label{Malpha-expand}
\end{align}

Using truncations of the expansion of $M_\alpha(Q)$ we shall, for any $J=0,1,2,\dots$,  construct $\widehat{\Phi}^\sigma(Q)$ in the form of a finite expansion in power of $\alpha^{2j},\  j=0,\dots, J$, with an error term of 
which is of order $\alpha^{2J+2}$
 plus a corrector of higher order. For each $J$, the polynomial expansion in $\alpha^2$ is {\sl independent of $\sigma$}. The construction is summarized in the following:\\
%
%

\begin{proposition} \label{prop:rigorousexpansion}
Fix $J \geq 0$, $a > 1/2$, and $\sigma \in \{ 0, 1/2 \}$. Then there exist a constant $\alpha_0 = \alpha_0[a, J, \sigma] > 0$, and  $J$ mappings $F_j: L_{_{\rm even}}^{2,a}(\mathbb{R}) \rightarrow L_{_{\rm even}}^{2,a}(\mathbb{R}), \ j = 0, \dots, J$, and a unique, real-valued function $\widehat{E_J^{\alpha, \sigma}} \in L_{_{\rm even}}^{2,a}(\mathbb{R})$ such that for all $0 < \alpha < \alpha_0$, 
\begin{align}
\widehat{\Phi^{\sigma}}( Q)\ =\ \chi_{_{\mathcal{B}_{_{\alpha}}}}(Q) \ \widetilde{\psi_1}(Q)\ +\ \sum_{j = 1}^J \ \alpha^{2j} \  \chi_{_{\mathcal{B}_{_{\alpha}}}}(Q) \ F_j \left[ \widetilde{\psi_1} \right] (Q)  + \widehat{E_J^{\alpha, \sigma}}(Q), \label{eqn:rigorousexpansion}
\end{align}
solves equation \eqref{eqn:Phieqn}
with the error bound:
\begin{align}
\left\| \widehat{E_J^{\alpha, \sigma}} \right\|_{L^{2,a}(\mathbb{R})} \lesssim \alpha^{2J + 2}
\end{align}
$F_j,\ j\ge1$, defined in Proposition \ref{prop:orderj} below,  is independent of $\sigma$ and $\alpha$, and 
$
 \widehat{E_J^{\alpha, \sigma}}( Q) = \chi_{_{\mathcal{B}_{_{\alpha}}}}(Q) \ \widehat{E_J^{\alpha, \sigma}}( Q).
$
\end{proposition}
 \medskip

\begin{remark}
By Proposition \ref{prop:rigorousexpansion}, since the polynomial expansion in $\alpha^2$ is completely determined by $\widetilde{\psi_1}(Q)$, 
$\widehat{\Phi^{\sigma}}(Q)$ is completely specified once we have contructed  $\widehat{E_J^{\alpha, \sigma}}(Q)$ for $Q \in \mathbb{R}$.
And, in turn by the rescalings $Q = q/\alpha$ and $\widehat{\Phi^{\sigma}}(Q) = \widehat{\phi^{\sigma}}(Q \alpha) = \widehat{\phi^{\sigma}}(q)$, \eqref{eqn:phitoG} implies that
$\widehat{G^\sigma}(q)$ and $\widehat{K^\sigma}(q)$ are completely specified by $\widehat{E_J^{\alpha, \sigma}}(q/\alpha)$ for $q \in \mathbb{R}$.
Therefore, Proposition \ref{prop:rigorousexpansion} completely characterizes $\widehat{G^{\sigma}}(q)$. \end{remark}
\medskip

The proof of Proposition \ref{prop:rigorousexpansion} extends over sections \ref{section:formalexpansion} through \ref{section:completion}. In Section \ref{section:formalexpansion}, we formally derive and construct the functionals $F_j \left[ \cdot \right]$, appearing in the expansion \eqref{eqn:rigorousexpansion}. We then contruct and bound the corrector $\widehat{E_J^{\alpha, \sigma}}(Q)$ in Sections \ref{section:rigorousexpansionproof} and \ref{section:rescalinglow0} by using  a Lyapunov-Schmidt reduction strategy.

\subsection{Formal asymptotic expansion for $\widehat{\Phi^{\sigma}}$, the solution of \eqref{eqn:Phieqn}} \label{section:formalexpansion}

A solution to \eqref{eqn:Phieqn}-\eqref{eqn:R1def}, $\widehat{\Phi}$, is compactly supported on $\mathcal{B}_\alpha$. Our approach to solving 
\eqref{eqn:Phieqn}-\eqref{eqn:R1def} is to first construct 
formal power series solution, $F^\alpha(Q)$,  of the related equation:
\begin{align}
 \left[ 1 + M_{\alpha}( Q) \right]\ F( Q ) - \ \frac{1}{4 \pi^2}
\left(F *F *F\right)( Q) = 0\  \label{eqn:continuumF}
\end{align}
in powers of $\alpha^2$.
Each term in this power series, $F_j^\alpha(Q)$ , will have support on all $\mathbb{R}$ and can be shown to decay exponentially as  $|Q|\to\infty$.
The deviation of \eqref{eqn:continuumF} from \eqref{eqn:Phieqn} are terms of the form:
\begin{equation}
- \left(1-{\chi}_{_{\mathcal{B}_{_{\alpha}}}}(Q) \right) \left[ 1 + M_{\alpha}(Q) \right] F(Q) + \frac{1}{4 \pi^2} \ \left(1-{\chi}_{_{\mathcal{B}_{_{\alpha}}}}(Q) \right) (F^\alpha*F^\alpha*F^\alpha)(Q) + R^{\alpha,\sigma}_1[F^\alpha](Q) ,
\label{F-alpha}
\end{equation}
whose norms can be shown to be beyond all polynomial orders in $\alpha$
 as $\alpha\to0$, {\it i.e.} $\mathcal{O}(\alpha^m)$, for all $m\ge1$,  in $L^{2,a}(\mathbb{R};dQ)$ with $ a>1/2$. Therefore, we expect that if 
 $ \widehat{\Phi^{\alpha,\sigma}}(Q)$ is a solution of \eqref{eqn:Phieqn}, then
the function  $ {\chi}_{_{\mathcal{B}_{_{\alpha}}}}(Q) \ F^\alpha(Q)$, where 
 $F^\alpha$ solves \eqref{eqn:continuumF} formally solves  \eqref{eqn:Phieqn} with an error which is beyond all polynomial orders in $\alpha^2$, {\it i.e.}
 \begin{equation}
\left\| \mathcal{D}^{\alpha,\sigma}\left[\ {\chi}_{_{\mathcal{B}_{_{\alpha}}}} \ F^\alpha\ \right]\ \right\|_{L^{2,a}(\mathbb{R};dQ)}
\ =\  \mathcal{O}(\alpha^\infty)
\label{sol-beyond-all}
\end{equation}

 Using the power series expansion of $M_\alpha(Q)$ in \eqref{Malpha-expand}, we now construct  $F^\alpha(Q)$ in the form:
\begin{align}
& F^\alpha(Q)= \sum_{ j = 0}^{\infty} \alpha^{2j} \ F_j (Q), \label{eqn:2series}
\end{align}
The sequence of truncated sums, 
$S_J^{\sigma}=\sum_{ j = 0}^{\infty} \alpha^{2j} \ F_j$, is a sequence of approximate solutions with decreasing residuals:
 $\left\| \mathcal{D}^{\alpha,\sigma}\left[S_J^{\sigma}\right]\ \right\|_{L^{2,a}}=\mathcal{O}(\alpha^{2J+2})$. In the coming sections, construct a solution
 $\Phi^{\sigma,\alpha}=S_J^{\sigma}+E^{\sigma,\alpha}_J$ by a Lyapunov-Schmidt procedure.

We  now turn to the construction of the terms in  series \eqref{eqn:2series}.  Substitution of \eqref{eqn:2series} into \eqref{F-alpha}, we obtain a hierarchy of equations for $F_j$.
\medskip

\begin{flalign}
 & \mathcal{O}(\alpha^{0}) {\bf \ equation:} \hspace{1.5cm} \left[ 1 + | Q |^2 \right] \ F_0(Q) -  \left( \frac{1}{2 \pi} \right)^2 \ F_0 * F_0 * F_0  (Q) = 0. & \label{eqn:order0}
\end{flalign}
Equation \eqref{eqn:order0} is the Fourier transform of continuum NLS \eqref{Psi-eqn}. Denote by
\begin{align}
F_0(Q) = \widetilde{\psi_1}(Q) = \mathcal{F}_{_C} \left[ \psi_1 \right](Q),
\end{align}
where $\psi_1(x)$ is the unique (up to translation)
 \emph{positive} and decaying solution of NLS. $\psi_1(x)$ is real-valued and radially symmetric about some point, which we take to be $x=0$. By Proposition \ref{prop:Psi}  there exists $C_0>0$ such that 
 \begin{equation}
 e^{C_0 | Q| } \ \widetilde{\psi_1}(Q) \in L^{2,a}(\mathbb{R})\ .\label{C0-def}\end{equation}
 
 At each order in $\alpha^2$, we shall derive an equation for $F_j(Q)$ of the general form:
 \begin{align}
\widetilde{L_+}  \ F^{\sharp}(Q) = F^{\flat}(Q).\label{Fsharp}
\end{align}
It is important for us to understand how decay properties $F^{\flat}(Q)$ propagate to the solution 
$F^{\sharp}(Q)$.
\begin{proposition}  \label{prop:lplusinvert}
Fix $a > 1/2$. Suppose that $F^{\flat}(Q) \in L_{_{\rm even}}^{2,a}(\mathbb{R};dQ)$ and that there exists a constant $C_\flat > 0$ such that $e^{C_\flat |Q|} F^{\flat}(Q) \in L^{2,a}(\mathbb{R};dQ)$. Then there exists a solution of \eqref{Fsharp},  $F^{ \sharp}(Q) \in L_{_{\rm even}}^{2,a}(\mathbb{R};dQ)$.
Furthermore, we have $e^{C_\flat |Q|} F^{\sharp}(Q) \in L^{2,a}(\mathbb{R};dQ)$.
\end{proposition}
\medskip

 Since $F^{\flat}(Q)$ is even it is $L^2(\mathbb{R};dQ)$ orthogonal to the kernel of $\widetilde{L_+} ={\rm span}\{Q\widetilde{\psi_1}(Q)\}$. Therefore, $F^{\sharp}=\left( \widetilde{L_+} \right)^{-1} F^\flat\in
L_{_{\rm even}}^{2,a+2}(\mathbb{R};dQ)$;
  see  Proposition \ref{prop:Lplus}.  A detailed proof that the exponential decay rate is preserved is given in  \cite{J:15}. The idea is to break $F^{\sharp}$ into its  low ($|Q|\le\epsilon^{-1}$) and high  ($|Q|\ge\epsilon^{-1}$) frequency components, $F_{{\rm lo},\epsilon}$ and  $F_{{\rm hi},\epsilon}$. The norm $\|e^{C_\flat|Q|}F_{{\rm lo},\epsilon}\|_{L^{2,a}}$ is bounded in terms of $\|F^\sharp\|_{L^{2,a}}$. While the norm $\|e^{C_\flat|Q|}F_{{\rm hi},\epsilon}\|_{L^{2,a}}$ is controlled  by a boot-strap argument using that $\chi(|Q|\ge\epsilon^{-1})(1+|Q|^2)^{-1}\ F_0*F_0*F_{{\rm hi},\epsilon}$ has  $L^{2,a}-$ norm which is bounded by $\sim \epsilon^2\ \|e^{C_\flat|Q|}F_{{\rm hi},\epsilon}\|_{L^{2,a}}$.
\medskip

We now turn to the hierarchy of equations  at order $\alpha^{2j}$, beginning with $j=1$.
 We find
\begin{align}
& \left[ 1 + | Q |^2 \right] \ F_1(Q) -  3\ \frac{1}{4 \pi^2}\ \widetilde{\psi_1} * \widetilde{\psi_1} * F_1  (Q) = \frac{ | Q |^4}{12} \ \widetilde{\psi_1}(Q),
\end{align}
or
\begin{flalign}
 & \mathcal{O}(\alpha^{2}) {\bf \ equation:} \hspace{3cm} \widetilde{L_+} \ F_1(Q) = \frac{ | Q |^4}{12} \ \widetilde{\psi_1}(Q). &  \label{eqn:order1}
\end{flalign}
Here, $L_+=-\partial_x^2+1-3\psi_1^2(x)$, is the linearization of the continuum NLS operator about $\psi_1(x)$. 
By Proposition \ref{prop:Lplus}, 
$
F_1(Q) = \left( \widetilde{L_+} \right)^{-1} \ \left(  \frac{ | Q |^4}{12} \ \widetilde{\psi_1}(Q) \right) \in L_{_{\rm even}}^{2,a}(\mathbb{R}).$ Since \eqref{eqn:order1} has real-valued forcing, $F_1$ is real-valued. 
Let $C_\flat=3C_0/4$ and note $\|e^{C_\flat|Q|}|Q|^4\widetilde{\psi_1}\|_{L^{2,a}}\lesssim  
\|e^{C_0 |Q|}\widetilde{\psi_1}\|_{L^{2,a}}$
Therefore,  $e^{\frac{3C_0}{4}|Q|} F_1(Q)\in L^{2,a}(\mathbb{R};dQ)$ for $a>1/2$.

\medskip
We now proceed to  inductively construct and bound the  sequence $F_j(Q),\ j\ge1$ using Proposition \ref{prop:lplusinvert} and the following two lemmata, proved in detail in \cite{J:15}:

\begin{lemma} \label{lemma:expoconvogen}
Fix $a > 1/2$. Suppose that  $\widetilde{f_1}, \widetilde{f_2} \in L_{_{\rm even}}^{2,a}(\mathbb{R})$. Then $\widetilde{f_1} * \widetilde{f_2} \in L_{_{\rm even}}^{2,a}(\mathbb{R} )$. Suppose further that there exist $c_1 , c_2 > 0$ such that $e^{c_1 |Q|} \widetilde{f_1}(Q) \in L^{2,a}(\mathbb{R})$, $e^{c_2  |Q|} \widetilde{f_2}(Q) \in L^{2,a}(\mathbb{R}_Q)$. Then for $c_3 = \min(c_1, c_2)$, we have $e^{c_3 |Q|} \ \widetilde{f_1} * \widetilde{f_2}(Q) \in L_{_{\rm even}}^{2,a}(\mathbb{R}_Q)$ and 
\begin{align}
\left\| \ e^{c_3 |Q|} \ (\widetilde{f_1} * \widetilde{f_2}) (Q) \ \right\|_{L^{2,a}(\mathbb{R}_Q)} \lesssim \left\| \ e^{c_3 |Q|} \ \widetilde{f_1}(Q)  \ \right\|_{L^{2,a}(\mathbb{R}_Q)} \left\| \ e^{c_3 |Q|}  \   \widetilde{f_2} (Q) \ \right\|_{L^{2,a}(\mathbb{R}_Q)}.
\end{align}  
\end{lemma} 

Lemma \ref{lemma:expoconvogen} is a direct consequence of \eqref{eqn:algebra0}, appropriately distributing the exponential weights, and $c_3 |Q| - c_1 |Q - \xi | - c_2 |\xi| \leq 0$. 
\medskip

\begin{lemma} \label{lemma:polyexpo}
Fix $a > 1/2$ and $k \in \mathbb{N}$. Suppose that  $\widetilde{f} \in L_{_{\rm even}}^{2,a}(\mathbb{R})$ and that there exists $c_1>0$ such that $e^{c_1 |Q|} \widetilde{f}(Q) \in L^{2,a}(\mathbb{R})$. Then
$ | Q |^{2k} \ \widetilde{f} \in L_{_{\rm even}}^{2,a}(\mathbb{R})$ and for any $0 < c_2 < c_1$, we have
\begin{align}
\left\| e^{c_2 |Q| } |Q|^{2k} \widetilde{f} \right\|_{L^{2,a}(\mathbb{R})} \leq \left( \frac{2k}{c_1 - c_2} \right)^{2k} \ e^{- 2k}\ \left\| \ e^{c_1 |Q|}  \   \widetilde{f} (Q) \ \right\|_{L^{2,a}(\mathbb{R}_Q)} . 
\end{align}
\end{lemma} 
Lemma \ref{lemma:polyexpo} follows from $ e^{c_2 |Q| } |Q|^{2k} | \widetilde{f}(Q) | \leq \left( \frac{2k}{c_1 - c_2} \right)^{2k} \ e^{- 2k} \cdot e^{c_1 |Q| } | \widetilde{f}(Q) |$ and then taking the $L^{2,a}$ norm. 
 
\medskip

\begin{proposition} \label{prop:orderj}
Let $ j \geq 1$. The equation for $F_j$ at order $\mathcal{O}(\alpha^{2j})$, independent of $\alpha$ and $\sigma$, is given by
\begin{flalign}
 & \mathcal{O}(\alpha^{2 j}) {\bf \ equation:} &  \widetilde{L_+} \ F_j(Q) & = 2 \sum_{k = 0}^{j-1} \frac{  (-1)^{k -  j + 1} \  | Q |^{2j - 2k + 2} \ F_k (Q) }{ (2j - 2k + 2)!}, & \nonumber \\
& & & + \frac{1}{( 2 \pi)^2} \sum_{ \substack{ k + l + z = j \\ 0 \leq k, l, z < j  }} \ F_k * F_l * F_z (Q) \equiv H_j \left[F_0, \dots, F_{j-1} \right](Q) , & \label{eqn:orderj}
 \end{flalign}
 and has the unique solution 
 \begin{align}
 F_j & = \left( \widetilde{L_+} \right)^{-1} \bigg( H_j \left[ F_0 , \dots, F_{j-1} \right] \bigg) \in L_{_{\rm even}}^{2,a}(\mathbb{R};dQ). 
 \end{align}
 Furthermore, $F_j$ is real-valued and $e^{C_j |Q| } \ F_j(Q) \in L^{2,a}(\mathbb{R};dQ)$, where
 $C_j \equiv C_0 \left( \frac{1}{2} + \frac{1}{2^{j+1}} \right) \geq \frac{C_0}{2}$ and $C_0>0$ is as in \eqref{C0-def}.   \\
\end{proposition}

\noindent {\bf Proof of Proposition \ref{prop:orderj}:} We induct to solve at each order in $\alpha^{2j}$. Let $F_0(Q) \equiv \widetilde{\psi_1}(Q)$, which solves \eqref{eqn:order0}, is real-valued, and satisfies $e^{C_0 |Q|} F_0(Q) \in L^{2,a}(\mathbb{R};dQ)$. Fix $m \geq2$ and assume that for $1 \leq j \leq m - 1$, $F_j(Q) \in L_{_{\rm even}}^{2,a}(\mathbb{R})$ satisfies \eqref{eqn:orderj} and is real-valued. Furthermore, assume that
\begin{align}
e^{C_j |Q|} F_j(Q) \in L^{2,a}(\mathbb{R};dQ), \qquad C_j \equiv C_0 \left( \frac{1}{2} + \frac{1}{2^{j+1}} \right) \geq \frac{C_0}{2}. \label{j-bounds}
\end{align}
We have already proven above that these inductive hypotheses hold for $ j = 1$. We expand
\begin{align}
M_{\alpha}(Q)  = 2 \sum_{j = 0}^{\infty} \frac{  \alpha^{2j}  \ (-1)^{j} \  |Q |^{2j + 2} }{ (2j + 2)!}, \hspace{2cm}  F^\alpha(Q)= \sum_{ j = 0}^{\infty} \alpha^{2j} \ F_j (Q),
\end{align}
and substitute into \eqref{eqn:continuumF}. Using \eqref{eqn:order0} for $F_0(Q) \equiv \widetilde{\psi_1}(Q)$ 
 we obtain 
%
 %
%
\begin{align}
& \sum_{j = 1}^{\infty} \ \alpha^{2j} \ \widetilde{L_+} \ F_j(Q) 
 = 2 \sum_{j = 1}^{\infty} \sum_{k = 0}^{\infty} \frac{ \alpha^{2j + 2k} (-1)^{j + 1} |Q|^{2j + 2} }{(2j + 2)!} F_k(Q) + \sum_{j = 1}^{\infty} \frac{\alpha^{2j}}{( 2 \pi)^2}  \ \sum_{ \substack{ k + l + z = j \\ 0 \leq k, l, z < j  }} \ F_k * F_l * F_m (Q)  \nonumber \\
& = \sum_{j = 1}^{\infty} \alpha^{2j} \bigg( 2 \sum_{k = 0}^{j-1} \frac{  (-1)^{k -  j + 1} \  | Q |^{2j - 2k + 2} }{ (2j - 2k + 2)!}  \ F_k (Q)  + \frac{1}{ 4 \pi^2 }  \ \sum_{ \substack{ k + l + z = j \\ 0 \leq k, l, z < j  }} \ F_k * F_l * F_z (Q) \bigg). \label{eqn:derive2}
\end{align}
Applying the inductive hypothesis \eqref{eqn:orderj}  for $1 \leq j \leq m - 1$ and dividing by $\alpha^{2m}$, \eqref{eqn:derive2} becomes
\begin{align}
&  \widetilde{L_+} \ F_m(Q)  + \alpha^2 \ \bigg[  \sum_{j = m + 1}^{\infty} \ \alpha^{2j - 2 (m + 1)} \ \widetilde{L_+} \ F_j(Q) \bigg]
\nonumber \\
& = 2 \sum_{k = 0}^{m - 1} \frac{  (-1)^{k -  m + 1} \  | Q |^{2m - 2k + 2} \ F_k (Q) }{ (2m - 2k + 2)!} + \frac{1}{( 2 \pi)^2} \sum_{ \substack{ k + l + z = m \\ 0 \leq k, l, z < m  }} \ F_k * F_l * F_z (Q)    \nonumber \\
&  \ \ +  \alpha^2 \ \bigg[  \sum_{j = m + 1}^{\infty} \alpha^{2j - 2(m+1)} \bigg( 2 \sum_{k = 0}^{j-1} \frac{  (-1)^{k -  j + 1} \  | Q |^{2j - 2k + 2} }{ (2j - 2k + 2)!}  \ F_k (Q)  + \frac{1}{ 4 \pi^2 }  \ \sum_{ \substack{ k + l + z = j \\ 0 \leq k, l, z < j  }} \ F_k * F_l * F_z (Q) \bigg) \bigg]. \label{eqn:derive3}
\end{align}
Since $2j - 2(m+1) \geq 0$ for $j \geq m + 1$, the bracketed terms with coefficient $\alpha^2$ are $\mathcal{O}(\alpha^2)$. Therefore the terms of order precisely $\alpha^{2m}$ are given by\eqref{eqn:orderj}. This establishes the case: $ j = m$. 

%

\medskip

We now prove that \eqref{eqn:orderj} has a solution, $F_m$ satisfying \eqref{j-bounds} with $j=m$.  
 First, applying Lemmata \ref{lemma:expoconvogen} and \ref{lemma:polyexpo} to the right hand side  of \eqref{eqn:orderj} for $j=m$, $H_m$,  we have that $H_m\in L_{_{\rm even}}^{2,a}$ with the bound:

\begin{align}
\left\| \ e^{C_m |Q| } \ H_m \left[  F_0 , \dots, F_{m-1} \right] (Q)  \ \right\|_{L^{2,a}(\mathbb{R}; dQ)} \lesssim  \lambda_m,\ \ 
\end{align}
where {\small{ $C_m=C_0 \left( \frac{1}{2} + \frac{1}{2^{m+1}} \right)$ and 
\begin{align}
\lambda_m  \equiv \ & 2 \ \sum_{k = 0}^{m - 1} \frac{   e^{- (2m - 2k + 2)} }{ (2m - 2k + 2)!} \ \left[ \frac{2m - 2k + 2}{C_k - C_m} \right]^{2m - 2k + 2}  \ \left\| \ e^{C_k |\cdot|} \ {F_k} \ \right\|_{L^{2,a}(\mathbb{R})} \nonumber \\
& + \frac{1}{( 2 \pi)^2} \sum_{ \substack{ k + l + z = m \\ 0 \leq k, l, z < m  }} \left\| \ e^{ C_k |\cdot | } \  {F_k}   \ \right\|_{L^{2,a}(\mathbb{R})} \ \left\| \ e^{ C_l |\cdot | } \  {F_l}    \ \right\|_{L^{2,a}(\mathbb{R})}  \ \left\| \ e^{ C_z |\cdot | } \  {F_z}  \ \right\|_{L^{2,a}(\mathbb{R})}. 
\end{align}
}}
Proposition \ref{prop:lplusinvert} implies that there exists a unique solution $F_m(Q) \in L_{_{\rm even}}^{2,a}(\mathbb{R})$ to equation \eqref{eqn:orderj} with $e^{C_m |Q|} F_m(Q) \in L^{2,a}(\mathbb{R})$. Finally, to see that $F_m$ is real-valued, note that equation \eqref{eqn:orderj} for $F_m$ is a linear with inhomogeneous forcing on the right-hand-side given by $H_m \left[F_0, \dots, F_{m -1} \right]$, which is necessarily real-valued for $F_j$ real-valued, $ j = 0, \dots, m-1$. This completes the proof of Proposition \ref{prop:orderj}. 
$\Box$

\section{Rigorous justification of asymptotic series \ref{eqn:rigorousexpansion} and proof of Proposition \ref{prop:rigorousexpansion}} \label{section:rigorousexpansionproof} 
Fix $J \geq 0$ and define the truncated asymptotic expansion
\begin{align}
S_J^{\alpha}(Q) \equiv \sum_{j = 0}^J  \ \alpha^{2j} \ \chi_{_{\mathcal{B}_{_{\alpha}}}}(Q) \  F_j \left[ \widetilde{\psi_1} \right] (Q). \label{eqn:Sdef}
\end{align}
where
\begin{align}
F_0 [ \widetilde{\psi_1} ] (Q) \equiv \widetilde{\psi_1}(Q), \qquad {\rm and} \qquad F_j  [ \widetilde{\psi_1} ] (Q) \equiv F_j(Q), \quad \quad j \geq 1,
\end{align}
with $F_j \in L^{2,a}(\mathbb{R})$ prescribed in Proposition \ref{prop:orderj}. Note that \eqref{eqn:Sdef} is the projection of the first $J+1$ terms of the formal asymptotic expansion $F^{\alpha}$ in \eqref{eqn:2series}. Note that
\begin{align}
(I - \chi_{_{\mathcal{B}_{_{\alpha}}}}) | F_j(Q) | = (I - \chi_{_{\mathcal{B}_{_{\alpha}}}})  e^{- C_j |Q|} \cdot  e^{C_j |Q|} | F_j(Q) | \leq e^{- \pi C_j / \alpha} \cdot e^{C_j |Q|} | F_j(Q) | , 
\end{align}
from which we see that the Fourier tail neglected in \eqref{eqn:Sdef} is exponentially small in $\alpha$. In the subsequent sections, we use the following consequence of Proposition \ref{prop:orderj}. 

\begin{proposition} \label{prop:S}
For $S_J^{\alpha}$ defined in \eqref{eqn:Sdef}, we have $ S_J^{\alpha} \in L^{2,a}$ and, for $\alpha > 0$ sufficiently small, the bounds
\begin{align}
& \left\| S_J^{\alpha} \right\|_{L^{2,a}(\mathbb{R})} \lesssim 1, \qquad {\rm and} \qquad \left\| e^{C_S  |\cdot|} \ S_J^{\alpha} \right\|_{L^{2,a}(\mathbb{R})} \lesssim 1, 
\end{align}
where $C_S = \min\{ C_0, ..., C_J \} $, and where $C_j$ are the constants prescribed in Proposition \ref{prop:orderj}. 
\end{proposition}

\subsection{Equation for the remainder, $\widehat{E_J^{\alpha, \sigma}}$}

In order to prove Proposition \ref{prop:rigorousexpansion}, we seek an equation for $\widehat{E_J^{\alpha, \sigma}}(Q) = \widehat{\Phi^{\sigma}}( Q) -  S_J^{\alpha}(Q)$. 
The equation for $\widehat{\Phi^{\sigma}}$ is \eqref{eqn:Phieqn} given in Proposition \ref{prop:Phieqn}. 
Substituting into \eqref{eqn:Phieqn}, we obtain the following equation for $\widehat{E_J^{\alpha, \sigma}}$. \\

\begin{proposition} \label{prop:Eeqn}
Equation \eqref{eqn:Phieqn} is equivalent to the following closed equation for $\widehat{E_J^{\alpha, \sigma}}(Q)$ on $Q \in \mathbb{R}$:
\begin{align}
\left[ 1 + M_{\alpha}( Q) \right] \ \widehat{E_J^{\sigma, \alpha}}( Q ) - 3 \ \chi_{_{\mathcal{B}_{_{\alpha}}}}( Q)  \  \left( \frac{1}{2 \pi} \right)^2
\widetilde{\psi_1} * \widetilde{\psi_1} * \widehat{E_J^{\alpha, \sigma}} ( Q) = \mathcal{R}_{J,1}^{\sigma} \left[ \alpha, \widehat{E_J^{\alpha, \sigma}} \right] (Q), \label{eqn:Eeqn}
\end{align}
where $ \mathcal{R}_{J,1}^{\sigma} \left[ \alpha, \widehat{E_J^{\alpha, \sigma}} \right] $ is defined in \eqref{eqn:HJdef}.  \\
\end{proposition}

\begin{remark}
Note that the operator on the left-hand side has formal limit $\widetilde{L_+}$, where $L_+$ is the linearized continuum NLS operator displayed in Proposition \ref{prop:Lplus}.
\end{remark}
\medskip

\noindent {\bf Proof of Proposition \ref{prop:Eeqn}:}
Substitution of $ \widehat{\Phi^{\sigma}} = S_J^{\alpha} + \widehat{E_J^{\alpha, \sigma}}$ into \eqref{eqn:Phieqn} yields, after some manipulation, \eqref{eqn:Eeqn}, 
where
\begin{align}
\mathcal{R}_{J,1}^{\sigma} \left[ \alpha, \widehat{E_J^{\alpha, \sigma}} \right]  \equiv \mathcal{D}^{\sigma,\alpha}[S_J^{\alpha}]  + R_{_{\rm L}}^{\sigma} \left[ \alpha, \widehat{E_J^{\alpha, \sigma}} \right] + R_{_{\rm NL}}^{\sigma} \left[ \alpha, \widehat{E_J^{\alpha, \sigma}} \right] . \label{eqn:HJdef}
\end{align}
$R_{_{\rm L}}^{\sigma} $ contains terms which are linear in $\widehat{E_J^{\alpha, \sigma}}$ but which are higher order in $\alpha$, and $R_{_{\rm NL}}^{\sigma} $ contains terms which are nonlinear in $\widehat{E_J^{\alpha, \sigma}}$; they are respectively given by
\begin{align}
& R_{_{\rm L}}^{\sigma} \left[ \alpha, \widehat{E_J^{\alpha, \sigma}} \right] (Q) \equiv   \chi_{_{\mathcal{B}_{_{\alpha}}}}( Q)  \ \frac{3}{4 \pi^2} \ \bigg[ \ \sum_{m = -1}^1 \ e^{2 m \pi i \sigma} \   S_J^{\alpha} * S_J^{\alpha} * \widehat{E_J^{\alpha, \sigma}} (Q - 2 m \pi / \alpha)  - \ \widetilde{\psi_1} * \widetilde{\psi_1}  * \widehat{E_J^{\alpha, \sigma}}(Q) \bigg], \nonumber \\
& R_{_{\rm NL}}^{\sigma} \left[ \alpha, \widehat{E_J^{\alpha, \sigma}} \right] (Q) \equiv \chi_{_{\mathcal{B}_{_{\alpha}}}}( Q)  \  \left( \frac{1}{2 \pi} \right)^2 \ \sum_{m = -1}^1 \ e^{2 m \pi i \sigma} \ \bigg[ 3  \ S_J^{\alpha} * \widehat{E_J^{\alpha, \sigma}} * \widehat{E_J^{\alpha, \sigma}} (Q - 2 m \pi / \alpha)  \nonumber \\
& \hspace{7cm} + \widehat{E_J^{\alpha, \sigma}} * \widehat{E_J^{\alpha, \sigma}} * \widehat{E_J^{\alpha, \sigma}} (Q - 2 m \pi / \alpha) \bigg]. \label{eqn:RLNLdef}
\end{align}
Note that $\mathcal{R}_{J,1}^{\sigma} \left[ \alpha, \widehat{E_J^{\alpha, \sigma}} \right] = \chi_{_{\mathcal{B}_{_{\alpha}}}} \ \mathcal{R}_{J,1}^{\sigma} \left[ \alpha, \widehat{E_J^{\alpha, \sigma}} \right]$. This completes the proof of Proposition \ref{prop:Eeqn}. $\Box$

\medskip

\subsection{Coupled system for high and low frequency components of $\widehat{E_J^{\alpha, \sigma}}$} \label{section:decomp}
We now embark on the construction of a solution  $\widehat{E_J^{\alpha, \sigma}} \in L^{2,a}(\mathbb{R})$ to \eqref{eqn:Eeqn} for $\alpha > 0$ sufficiently small. Our strategy is to formulate the equation for $\widehat{E_J^{\alpha, \sigma}}$ as an equivalent coupled system for its high and low frequency components. Let $r$ be such that $0 < r < 1$. Define the sharp spectral cutoff functions
\begin{align}
& \chi_{_{\rm lo}}(Q) =  \chi \left( |Q| \leq \alpha^{r-1} \right) \ \ {\rm and}\ \  \chi_{_{\rm hi}}(Q) =  \chi \left( |Q| > \alpha^{r-1} \right), \label{lo-hi-cut}\\
& \  {\rm where}\ \  1 = \chi_{_{\rm lo}}(Q) + \chi_{_{\rm hi}}(Q),
\nonumber\end{align}
Note that $\chi_{_{\rm lo}}(Q) \ \chi_{_{\mathcal{B}_{_{\alpha}}}}(Q) = \chi_{_{\rm lo}}(Q) $, while $\chi_{_{\rm hi}}(Q) \ \chi_{_{\mathcal{B}_{_{\alpha}}}}(Q) =  \chi_{_{\mathcal{B}_{_{\alpha}}}}(Q)  - \chi_{_{\rm lo}}(Q)$. 
For general $\widehat{A}( Q )$, defined for $ Q \in \mathbb{R} $, we introduce its localizations near and away from $ Q =0$:
\begin{align}
&\widehat{ A}{_{_{\rm lo}}}( Q )\ =\ \left(\chi_{_{\rm lo}} \widehat{A}\right)( Q ) \equiv \chi_{_{\rm lo}}( Q ) \widehat{A}(q), \qquad {\rm and} \qquad \widehat{ A}{_{_{\rm hi}}}( Q ) = \left(\chi_{_{\rm hi}} \widehat{A}\right)( Q ) \equiv \chi_{_{\rm hi}}( Q ) \widehat{A}(q),
\end{align}
In particular, we use   $\chi_{_{\rm lo}}$ and $\chi_{_{\rm hi}}$ to localize $\widehat{E_J^{\alpha, \sigma}}$ on $| Q | \leq \alpha^{r-1}$ and $ |Q| > \alpha^{r-1} $:
\begin{align}
& \widehat{E^{\alpha, \sigma}_{_{\rm lo}}}( Q ) = \chi_{_{\rm lo}}( Q ) \widehat{E_J^{\alpha, \sigma}}( Q )\ \ {\rm and}\ \  \widehat{E^{\alpha, \sigma}_{_{\rm hi}}}( Q ) = \chi_{_{\rm hi}}(Q ) \widehat{E_J^{\alpha, \sigma}}( Q ),
\end{align}
Thus, we can express $\widehat{E_J^{\alpha, \sigma}}$ as follows:
\begin{align}
& \widehat{E_J^{\alpha, \sigma}}( Q ) = \widehat{E^{\alpha,\sigma}_{_{\rm lo}}}( Q ) + \widehat{E^{\alpha, \sigma}_{_{\rm hi}}}( Q ) \label{eqn:decomp}.
\end{align}

\noindent {\bf NOTE:} {\it Since the analysis for the cases $\sigma = 0$ (onsite) and $\sigma = 1/2$ (offsite) in sections \ref{section:decomp} - \ref{section:solutionlow} are very similar, in order to keep the notation less cumbersome we omit the superscripts $\alpha$ and $\sigma$, when the context is clear, and shall instead write:} {\small{
\begin{align}
\widehat{E}( Q )=\widehat{E_J^{\alpha, \sigma}}( Q ), \quad \quad
\widehat{E_{_{\rm lo}}}( Q )=\widehat{E^{\alpha, \sigma}_{_{\rm lo}}}( Q ), \quad \quad
\widehat{E_{_{\rm hi}}}( Q )=\widehat{E^{\alpha, \sigma}_{_{\rm hi}}}( Q ). \quad \quad
\label{eqn:drop-super} \end{align} }}

The following Proposition is obtained by applying the spectral projections $\chi_{_{\rm lo}}$ and $\chi_{_{\rm hi}}$ to \eqref{eqn:Eeqn}.

\bigskip

\begin{proposition}
\label{prop:highlow}
Equation (\ref{eqn:Eeqn}) is equivalent to the following coupled system of equations for the low and high frequency components, $\widehat{E_{_{\rm lo}}}$ and $\widehat{E_{_{\rm hi}}}$,  of $\widehat{E}$ on $Q \in \mathbb{R}$:
\begin{flalign}
& \textbf{Low Frequency Equation:} \nonumber \\
&  \hspace{3cm} \left[ 1 + M_{\alpha}(Q) \right] \  \widehat{E_{_{\rm lo}}} ( Q ) - \chi_{_{\rm lo}}( Q )   \frac{3}{4 \pi^2}   \left( \widetilde{\psi_1} * \widetilde{\psi_1} * \widehat{E_{_{\rm lo}}}  (Q) + \widetilde{\psi_1} * \widetilde{\psi_1} * \widehat{E_{_{\rm hi}}}  (Q) \right) & \nonumber \\ 
& \hspace{7cm} = \chi_{_{\rm lo}}( Q ) \ \mathcal{R}_{J,1}^{\sigma} \left[ \alpha, \widehat{E_{_{\rm lo}}} + \widehat{E_{_{\rm hi}}} \right] ( Q), \label{eqn:low} \\
& \textbf{High Frequency Equation:} \nonumber \\
&  \hspace{3cm} \left[ 1 + M_{\alpha}(Q) \right] \  \widehat{E_{_{\rm hi}}} ( Q ) - \chi_{_{\rm hi}}( Q )  \ \chi_{_{\mathcal{B}_{_{\alpha}}}}(Q) \  \frac{3}{4 \pi^2}   \left( \widetilde{\psi_1} * \widetilde{\psi_1} * \widehat{E_{_{\rm lo}}}  (Q) + \widetilde{\psi_1} * \widetilde{\psi_1} * \widehat{E_{_{\rm hi}}}  (Q) \right) & \nonumber \\
& \hspace{7cm} = \chi_{_{\rm hi}}( Q ) \ \mathcal{R}_{J,1}^{\sigma} \left[ \alpha, \widehat{E_{_{\rm lo}}} + \widehat{E_{_{\rm hi}}} \right] ( Q).   \label{eqn:high}
\end{flalign}
Here, $\mathcal{R}_{J,1}^{\sigma}$ is defined in \eqref{eqn:HJdef}. 
\end{proposition} \\

\subsection{Solving for $\widehat{E_{_{\rm hi}}} \left[\alpha, \widehat{E_{_{\rm lo}}} \right] $ and reduction to a closed equation for the low-frequency components, $\widehat{E_{_{\rm lo}}}$}
\label{subsect:lyapschm}

We shall solve (\ref{eqn:low}) and (\ref{eqn:high}) via a Lyapunov-Schmidt reduction strategy \cite{N:01}; see the discussion of section \ref{section:strategy}. We first solve for $\widehat{E_{_{\rm hi}}} =
\widehat{E_{_{\rm hi}}} \left[ \alpha , \widehat{E_{_{\rm lo}}} \right]$ as a functional of $\widehat{E_{_{\rm lo}}}$ and $\alpha$, with  $\alpha$ sufficiently small. 
We view the equation of $\widehat{E_{_{\rm hi}}}$ as depending on parameters $\alpha \in \mathbb{R}, |\alpha| \ll 1 $ and a function $\Gamma( Q ) = \widehat{E_{_{\rm lo}}}( Q ) \in L^{2,a}(\mathbb{R})$: 
\begin{align}
\left[ 1 + M_{\alpha}(Q) \right] \ \widehat{E_{_{\rm hi}}}( Q)  - \chi_{_{\rm hi}}( Q) \ \chi_{_{\mathcal{B}_{_{\alpha}}}}(Q)  \  \frac{3}{4 \pi^2}   \bigg( \widetilde{\psi_1} * \widetilde{\psi_1} *  \Gamma (Q) + \widetilde{\psi_1} * \widetilde{\psi_1} *  \widehat{E_{_{\rm hi}}}(Q)  \bigg) \nonumber \\
- \chi_{_{\rm hi}}(q) \ \mathcal{R}_{J,1}^{\sigma} [ \alpha, \Gamma + \widehat{E_{_{\rm hi}}} ] ( Q ) = 0\ . \label{eqn:EhatGam}
\end{align}
In the following proposition, we construct $\widehat{E_{_{\rm lo}}} \mapsto \widehat{E_{_{\rm hi}}} \left[ \alpha, \widehat{E_{_{\rm lo}}} \right]$. Note that since $0 < r < 1$, 
$  \underset{ \alpha \to 0}{\lim} \ \alpha^{2 - 2r} = 0.  $

\begin{proposition}
\label{prop:ifthigh} Set $0 < r < 1$. 
\begin{enumerate}
\item There exist constants $\alpha_0, \beta_0 >  0$,  such that for all $\alpha \in (0, \alpha_0)$, equation \eqref{eqn:EhatGam} defines a unique mapping
\begin{align}
&(\alpha, \Gamma) \mapsto \widehat{E_{_{\rm hi}}}[\alpha, \Gamma],
\nonumber\\
&\widehat{E_{_{\rm hi}}}: [0,1] \times B_{_{   \beta_0 }}(0)  \rightarrow L^{2,a}(\mathbb{R}) \nonumber
\end{align} 
where $B_{_{  \beta_0 }}(0) \subset L^{2,a}(\mathbb{R})$ and $\widehat{E_{_{\rm hi}}}[\alpha, \Gamma]$
 is the unique solution to the high frequency equation \eqref{eqn:EhatGam} (see also \eqref{eqn:high}). In particular, if $\Gamma \in L^{2,a}_{_{\rm even}}$, then $\widehat{E_{_{\rm hi}}}[\alpha, \Gamma] \in L^{2,a}_{_{\rm even}}$ .\
 \item Moreover, this mapping is $C^1$ with respect to $\Gamma$ and for all $(\alpha, \Gamma) \in [0, \alpha_0) \times B_{_{\beta_0 }}(0)$, and there exists some constant $C>0$ such that
\begin{align}
& \left\| \widehat{E_{_{\rm hi}}}[\alpha, \Gamma] \right\|_{L^{2,a}(\mathbb{R})} \lesssim \ \alpha^{2 -2r}  \ \left\| \Gamma \right\|_{L^{2,a}(\mathbb{R})} + e^{- C/ \alpha^{1-r}}.\label{eqn:Ehighestlyap}\\
& \left\| \ D_{\Gamma} \widehat{E_{_{\rm hi}}}[\alpha, \Gamma] \ \right\|_{L^{2,a}(\mathbb{R}) \longrightarrow L^{2,a}(\mathbb{R})} \lesssim \ \alpha^{2 - 2r} , \label{eqn:DEhighest}
\end{align}
where the implicit constants are dependent on $\alpha_0$ and $\beta_0$. 

\item $\widehat{E_{_{\rm hi}}}[\alpha, \Gamma]$ is supported on $ Q \in  \left[ - \frac{\pi}{\alpha}, - \alpha^{r-1} \right) \cap \left( \alpha^{r-1}, \frac{\pi}{\alpha} \right]$ such that $\widehat{E_{_{\rm hi}}}[\alpha, \Gamma] = \chi_{_{\rm hi}} \ \chi_{_{\mathcal{B}_{_{\alpha}}}} \ \widehat{E_{_{\rm hi}}}[\alpha, \Gamma] $. \\

\end{enumerate}
\end{proposition}

\noindent {\bf Proof of Proposition \ref{prop:ifthigh}:} Since $0< 1 \leq 1 + M_{\alpha}(Q)  $ for all  $\alpha$ positive and small, we may rewrite (\ref{eqn:EhatGam}) (see also (\ref{eqn:high})) as
\begin{align}
\widehat{E_{_{\rm hi}}}(Q ) -   \chi_{_{\rm hi}}( Q )  \ [ 1 + M_{\alpha}(Q) ]^{-1} \bigg( \chi_{_{\mathcal{B}_{_{\alpha}}}}(Q) \  \frac{3}{4 \pi^2}  \ \left[ \widetilde{\psi_1} * \widetilde{\psi_1} *  \Gamma (Q) + \widetilde{\psi_1} * \widetilde{\psi_1} * \widehat{E_{_{\rm hi}}}(Q) \right]  \nonumber \\
- \mathcal{R}_{J,1}^{\sigma} [ \alpha, \Gamma + \widehat{E_{_{\rm hi}}}]( Q ) \bigg) = 0. \label{eqn:highrewrite}
\end{align}

We apply the implicit function theorem to obtain $\widehat{E_{_{\rm hi}}}$ as a functional of $\alpha$ and $\Gamma$. The key steps in the proof are given in Propositions \ref{prop:R1estimate} through \ref{prop:higheven}. The proofs of these propositions are deferred until Appendix \ref{section:technicalprops}. We first give bounds on the latter two terms in equation \eqref{eqn:highrewrite}. 
\medskip

\begin{proposition}
\label{prop:R1estimate}
 There exists a constant $\alpha_0 > 0$, such that for all $(\Gamma,\alpha)$
  with $0<\alpha<\alpha_0$,  and $\Gamma\in L^{2,a}(\mathbb{R})$, we have the bounds
  {\small{
\begin{align}
& \bigg\| \ \chi_{_{\rm hi}} \ \chi_{_{\mathcal{B}_{_{\alpha}}}} \  [ 1 + M_{\alpha}(\cdot)  ]^{-1} \ \widetilde{\psi_1} * \widetilde{\psi_1} *   \Gamma \  \bigg\|_{L^{2,a}(\mathbb{R})} \lesssim \ \alpha^{2 - 2r} \left\| \Gamma \right\|_{L^{2,a}(\mathbb{R})} , \label{eqn:invbound0} \\
& \bigg\| \ \chi_{_{\rm hi}} \ \chi_{_{\mathcal{B}_{_{\alpha}}}} \ [ 1 + M_{\alpha}(\cdot)  ]^{-1} \ \widetilde{\psi_1} * \widetilde{\psi_1} *   \widehat{E_{_{\rm hi}}} \ \  \bigg\|_{L^{2,a}(\mathbb{R})} \lesssim \ \alpha^{2 - 2r} \left\| \widehat{E_{_{\rm hi}}} \right\|_{L^{2,a}(\mathbb{R})} , \label{eqn:invbound0pt5} \\
& \bigg\| \ \chi_{_{\rm hi}} \ [ 1 + M_{\alpha}(\cdot) ]^{-1} \ \mathcal{R}_{J,1}^{\sigma} [\alpha, \Gamma +  \widehat{E_{_{\rm hi}}}] \ \bigg\|_{L^{2,a}(\mathbb{R})} \lesssim \ e^{- C/ \alpha^{1-r}} \ \nonumber \\
& +  \alpha^{2 -2r} \ \bigg[   \left\| \Gamma \right\|_{L^{2,a}(\mathbb{R})} + \left\| \widehat{E_{_{\rm hi}}} \right\|_{L^{2,a}(\mathbb{R})} +
  \left\| \Gamma \right\|^2_{L^{2,a}(\mathbb{R})} + \left\| \Gamma \right\|_{L^{2,a}(\mathbb{R})} \left\| \widehat{E_{_{\rm hi}}} \right\|_{L^{2,a}(\mathbb{R})} + \left\| \widehat{E_{_{\rm hi}}} \right\|^2_{L^{2,a}(\mathbb{R})}  \nonumber \\
& \quad \quad + \left\| \Gamma \right\|^3_{L^{2,a}(\mathbb{R})} + \left\| \Gamma \right\|^2_{L^{2,a}(\mathbb{R})} \left\| \widehat{E_{_{\rm hi}}} \right\|_{L^{2,a}(\mathbb{R})}
\ +\ \left\| \Gamma \right\|_{L^{2,a}(\mathbb{R})} \left\| \widehat{E_{_{\rm hi}}} \right\|^2_{L^{2,a}(\mathbb{R})} + \left\| \widehat{E_{_{\rm hi}}} \right\|^3_{L^{2,a}(\mathbb{R})}  \bigg] .
\label{inv-bound}
\end{align}
}}
\end{proposition}

We may write equation \eqref{eqn:highrewrite} as $
\mathcal{A}[ \alpha, \Gamma, \widehat{E_{_{\rm hi}}} ]( Q)  = 0$, 
where $\mathcal{A}: (0, \infty) \times L^{2,a}(\mathbb{R}) \times L^{2,a}(\mathbb{R}) \longrightarrow L^{2,a}(\mathbb{R})$ is defined by 
\begin{align}
& \mathcal{A} \left[ \alpha, \Gamma , \widehat{E_{_{\rm hi}}} \right]( Q ) \equiv \widehat{E_{_{\rm hi}}}( Q)  -   \chi_{_{\rm hi}}(Q) \ [ 1 + M_{\alpha}( Q)  ]^{-1} \ \nonumber \\
& \hspace{2cm} \cdot  \bigg( \frac{3}{4 \pi^2}  \ \chi_{_{\mathcal{B}_{_{\alpha}}}}(Q) \ \bigg[   \widetilde{\psi_1} * \widetilde{\psi_1} *  \Gamma ( Q) + \widetilde{\psi_1} * \widetilde{\psi_1} * \widehat{E_{_{\rm hi}}}(Q) \bigg]  + \mathcal{R}_{J,1}^{\sigma}[ \alpha,  \Gamma +  \widehat{E_{_{\rm hi}}}](Q ) \bigg). \label{eqn:Ahigh}
\end{align}
By Proposition \ref{prop:R1estimate}, 
we can extend $\mathcal{A}$ as a continuous $L^{2,a}(\mathbb{R})$ operator valued function of $\alpha \in [0, \infty)$ 
as follows: {\small{
\begin{align}
\label{Adef2}
\mathcal{A}[ \alpha, \Gamma, \widehat{E_{_{\rm hi}}} ]( Q)  \equiv\left\{
\begin{array}{ll}
\widehat{E_{_{\rm hi}}}(q)  -   \chi_{_{\rm hi}}(Q) \ [ 1 + M_{\alpha}( Q)  ]^{-1} \   \bigg( \frac{3}{4 \pi^2} \  \chi_{_{\mathcal{B}_{_{\alpha}}}}(Q) \ \bigg[   \widetilde{\psi_1} * \widetilde{\psi_1} *  \Gamma ( Q) & \\
\hspace{1.5cm} + \widetilde{\psi_1} * \widetilde{\psi_1} * \widehat{E_{_{\rm hi}}}(Q) \bigg]  + \mathcal{R}_{J,1}^{\sigma}[ \alpha,  \Gamma +  \widehat{E_{_{\rm hi}}}](Q ) \bigg), & {\rm for}\  \alpha > 0, \\
\widehat{E_{_{\rm hi}}}( Q ) & {\rm for}\  \alpha = 0.
\end{array}
\right. 
\end{align}
}}
We now summarize  the properties of the mapping $\mathcal{A}$ in the following proposition. 

\begin{proposition} \label{prop:Aprops}
\begin{enumerate}
\item The mapping  $\mathcal{A}: [0, \infty) \times L^{2,a}(\mathbb{R}) \times L^{2,a}(\mathbb{R}) \longrightarrow L^{2,a}(\mathbb{R})$
\[ (\alpha, \Gamma, \widehat{E_{_{\rm hi}}} )  \longmapsto \mathcal{A}[ \alpha,  \Gamma, \widehat{E_{_{\rm hi}}} ] ,\]
  is continuous at $(0,0,0)$.

\item $\mathcal{A} [0, 0, 0] = 0$ in $L^{2,a}(\mathbb{R})$.

\item \emph{Differential with respect to $\widehat{E_{_{\rm hi}}}$:}
For $\widehat{f}\in L^{2,a}(\mathbb{R})$, introduce the operator on $L^{2,a}(\mathbb{R})$ 
{\small{
\begin{align}
&  D_{Z}  \bigg( \mathcal{R}_{J,1}^{\sigma}[ \alpha, Z] \bigg) \ \widehat{f} ( Q)  
  =   \ R_{_{\rm L}}^{\sigma} \left[ \alpha,  \widehat{f} \right] ( Q)\nonumber\\
  &\qquad  + \chi_{_{\mathcal{B}_{_{\alpha}}}}( Q)  \  \left( \frac{1}{2 \pi} \right)^2 \ \sum_{m = -1}^1 \ e^{2 m \pi i \sigma} \ \bigg[ 6  \ S_J^{\alpha} * Z * \widehat{f} (Q - 2 m \pi / \alpha) 
 + 3  \ Z * Z * \widehat{f} (Q - 2 m \pi / \alpha)     \bigg]. \label{eqn:DShi}
\end{align}
}}
     For any $(\alpha, \Gamma, \widehat{E_{_{\rm hi}}} ) \in [0, \infty) \times L^{2,a}(\mathbb{R})  \times L^{2,a}(\mathbb{R})  $, $\mathcal{A}$ is Fr\'{e}chet differentiable with respect to $\widehat{E_{_{\rm hi}}}$ with 
 $D_{\widehat{E_{_{\rm hi}}}} \mathcal{A}[\alpha,\Gamma,\widehat{E_{_{\rm hi}}}] $, continuous at 
 $(0,0,0)$, 
{\small{
\begin{align}
D_{\widehat{E_{_{\rm hi}}}} \mathcal{A}[ \alpha, \Gamma, \widehat{E_{_{\rm hi}}} ] =
\left\{
\begin{array}{ll}
I - \chi_{_{\rm hi}} \ \left[ 1 + M_{\alpha}(Q) \right]^{-1} \ \bigg[ \frac{3}{4 \pi^2} \  \chi_{_{\mathcal{B}_{_{\alpha}}}}  \    \widetilde{\psi_1} * \widetilde{\psi_1} *   \\
\hspace{2cm} + D_Z \bigg( \mathcal{R}_{J,1}^{\sigma}[ \alpha, \Gamma + \widehat{E_{_{\rm hi}}}] \bigg) \bigg],\quad & \textrm{for $\alpha > 0$ }\\
I \  & \textrm{for $\alpha=0$}\ \ .
\end{array}
\right.
 \label{eqn:DfAhighproof}
\end{align}
}}


\item $D_{\widehat{E_{_{\rm hi}}}} \mathcal{A} [0, 0, 0] = I$ is an isomorphism of $L^{2,a}(\mathbb{R})$ onto $L^{2,a}(\mathbb{R})$.

\item \emph{Differential with respect to $\Gamma$:} For any $(\alpha, \Gamma, \widehat{E_{_{\rm hi}}} ) \in [0, \infty) \times L^{2,a}(\mathbb{R}) \times L^{2,a}(\mathbb{R})$,   the mapping  $\Gamma \mapsto \mathcal{A}[\alpha, \Gamma, \widehat{E_{_{\rm hi}}} ]$ is Fr\'{e}chet differentiable. Here, 
{\small{
\begin{align}
D_{\Gamma} \mathcal{A}[ \alpha, \Gamma, \widehat{E_{_{\rm hi}}} ] =
\left\{
\begin{array}{ll}
 - \chi_{_{\rm hi}} \ \left[ 1 + M_{\alpha}(Q) \right]^{-1} \ \bigg[ \frac{3}{4 \pi^2} \  \chi_{_{\mathcal{B}_{_{\alpha}}}}  \    \widetilde{\psi_1} * \widetilde{\psi_1} *  \\
\hspace{2cm} + D_Z \bigg( \mathcal{R}_{J,1}^{\sigma}[ \alpha, \Gamma +  \widehat{E_{_{\rm hi}}}] \bigg) \bigg],\quad & \textrm{for $\alpha > 0$ }\\
0 \  & \textrm{for $\alpha=0$}\ \ .
\end{array}
\right. \label{eqn:DFAlodef}
\end{align}
}}

\end{enumerate}
\end{proposition}
\medskip

 By Proposition \ref{prop:Aprops}, the mapping $\mathcal{A}$ satisfies the hypotheses of a variation of the implicit function theorem stated in Theorem \ref{th:IFT}. Therefore, there exist  $\alpha_0 , \beta_0 , \kappa > 0$ such that for all
 \begin{align}
 \alpha \in [0,\alpha_0) \ \ \textrm{and}\ \  \left\| \Gamma \right\|_{L^{2,a}(\mathbb{R})} < \beta_0 \ ,  \label{eqn:alpha0}
 \end{align}
 there exists a unique map,  differentiable with respect to $\Gamma$,
 \begin{align}
 &(\alpha, \Gamma ) \longmapsto \widehat{E_{_{\rm hi}}} \left[ \alpha, \Gamma \right],\nonumber\\
& \widehat{E_{_{\rm hi}}}: \left\{ \left(\alpha, \Gamma \right) \in [0,\alpha_0) \times L^{2,a}(\mathbb{R}) \Big|\ \
  \alpha\in[0,\alpha_0)\ \ \textrm{and}\ \  \left\| \Gamma  \right\|_{L^{2,a}(\mathbb{R})} < \beta_0 \right\} \longmapsto L^{2,a}(\mathbb{R}),  \nonumber \\
 & \left\| \widehat{E_{_{\rm hi}}} \left[\alpha, \Gamma \right] \right\|_{L^{2,a}(\mathbb{R})} \leq \kappa, \qquad {\rm with} \qquad \underset{(\alpha, \Gamma) \to (0,0) }{\lim} \ \widehat{E_{_{\rm hi}}} \left[ \alpha, \Gamma \right] = 0, \label{eqn:fhiprops}
 \end{align}
 which solves
  \begin{align}
 \mathcal{A} \left[\alpha, \Gamma, \widehat{E_{_{\rm hi}}} \left[ \alpha, \Gamma \right] \right](Q) = 0 \ \ {\rm for} \ \  \alpha\in[0,\alpha_0) \quad  \textrm{and} \quad \left\| \Gamma \right\|_{L^{2,a}(\mathbb{R})} < \beta_0   . \label{eqn:Asoln}
 \end{align}
Furthermore, the mapping $ \Gamma \longmapsto \widehat{E_{_{\rm hi}}} \left[ \alpha, \Gamma \right]$ is Fr\'echet differentiable with derivative $ D_{\Gamma} \widehat{E_{_{\rm hi}}} \left[ \alpha, \Gamma \right]$. \\

\begin{proposition}
 \label{prop:highiftests}
 The mapping $(\alpha, \Gamma ) \longmapsto \widehat{E_{_{\rm hi}}} \left[ \alpha, \Gamma \right] $, $\widehat{E_{_{\rm hi}}} :   [0,\alpha_0) \times B_{  \beta_0}(0)  \longmapsto L^{2,a}(\mathbb{R})$ satisfies the bounds
 \begin{align}
&  \left\| \widehat{E_{_{\rm hi}}} \left[ \alpha, \Gamma \right] \right\|_{L^{2,a}(\mathbb{R})} \lesssim \ \alpha^{2- 2r} \  \left\| \Gamma \right\|_{L^{2,a}(\mathbb{R})} + \ e^{- C / \alpha^{1-r}} \label{eqn:highiftests1} \\
&  \left\| D_{\Gamma} \widehat{E_{_{\rm hi}}} \left[ \alpha, \Gamma \right] \right\|_{L^{2,a}(\mathbb{R}) \longmapsto L^{2,a}(\mathbb{R})} \lesssim \ \alpha^{2 - 2r}, \ \label{eqn:highiftests2}
\end{align}
for some constant $C > 0$. \\
\end{proposition}
The mapping $ (\alpha, \widehat{E_{_{\rm lo}}}) \mapsto \widehat{E_{_{\rm hi}}}[\alpha, \widehat{E_{_{\rm lo}}} ]$ is supported on $ \left[ - \frac{\pi}{\alpha}, - \alpha^{r-1} \right) \cap \left( \alpha^{r-1}, \frac{\pi}{\alpha} \right]$. 

Finally we note that  the above proof can be adapted to show that the mapping $ \widehat{E_{_{\rm lo}}} \mapsto \widehat{E_{_{\rm hi}}}[\alpha, \widehat{E_{_{\rm lo}}} ] $ maps the space of even functions into itself.   \\

\begin{proposition}
\label{prop:higheven}
For $(\alpha, \Gamma) \in [0,\alpha_0) \times B_{   \beta_0}(0)$, the mapping $\Gamma \mapsto \widehat{E_{_{\rm hi}}} \left[ \alpha,\Gamma \right] $ from Proposition \ref{prop:ifthigh} maps  $\Gamma \in L_{_{\rm even}}^{2,a}(\mathbb{R})$ into $L_{_{\rm even}}^{2,a}(\mathbb{R})$. \\
\end{proposition}

\section{Analysis of the low frequency equation, governing $\widehat{E_{_{\rm lo}}}$} \label{section:rescalinglow0}

\subsection{Equation for $\widehat{E_{_{\rm lo}}}$ as a perturbation of the continuum NLS limit via the operator $\widetilde{L_+}$}
\label{section:rescalinglow}
Having constructed  $\widehat{E_{_{\rm lo}}}\mapsto\widehat{E_{_{\rm hi}}}[\alpha, \widehat{E_{_{\rm lo}}} ]$, we insert this map into equation (\ref{eqn:low}) and obtain the following closed equation for $\widehat{E_{_{\rm lo}}} (Q)$ on $Q \in \mathbb{R}$:
\begin{align}
&\left[ 1 + M_{\alpha}(Q) \right] \  \widehat{E_{_{\rm lo}}} ( Q ) - \chi_{_{\rm lo}}( Q )   \frac{3}{4 \pi^2}   \widetilde{\psi_1} * \widetilde{\psi_1} * \widehat{E_{_{\rm lo}}}  (Q)  & \nonumber \\ 
&  \quad  = \chi_{_{\rm lo}}( Q ) \ \mathcal{R}_{J,1}^{\sigma} \left[ \alpha, \widehat{E_{_{\rm lo}}} + \widehat{E_{_{\rm hi}}} [ \alpha, \widehat{E_{_{\rm lo}}} ] \right] ( Q) +   \chi_{_{\rm lo}}( Q )   \frac{3}{4 \pi^2}   \widetilde{\psi_1} * \widetilde{\psi_1} * \left( \widehat{E_{_{\rm hi}}}  [ \alpha, \widehat{E_{_{\rm lo}}} ] \right) (Q)  . \label{eqn:closed}
\end{align}
Here, $\mathcal{R}_{J,1}^{\sigma}$ is given in \eqref{eqn:HJdef}.

   Noting that the operator on the left-hand-side  of \eqref{eqn:closed} has a formal $\alpha\downarrow0$ limit equal to the linearized continuum NLS operator, $\widetilde{L_+}$,  we now rewrite \eqref{eqn:closed} as a small $\alpha$ perturbation of this limit:

\medskip

\begin{proposition}
\label{prop:rescale}
There exists $0\le\alpha_1\le\alpha_0$ and $0 < r < 1$ such that the following holds. For $\alpha\le\alpha_1$, equation  \eqref{eqn:closed} may be written as
\begin{align}
\widetilde{L_+} \widehat{E_{_{\rm lo}}} (Q) = \mathcal{R}_{J,2}^{\sigma}[\alpha, \widehat{E_{_{\rm lo}}}](Q). \label{eqn:lowrescaled}
 \end{align}
$\mathcal{R}_{J,2}^{\sigma} \left[\alpha, \widehat{E_{_{\rm lo}}} \right] (Q) $, displayed in \eqref{eqn:Htildedef}, is continuous at $(0,0) \in [0,\alpha_1) \times L^{2,a}(\mathbb{R}) $. Furthermore, the mapping $\widehat{E_{_{\rm lo}}} \mapsto \mathcal{R}_{J,2}^{\sigma} \left[ \alpha, \widehat{E_{_{\rm lo}}} \right]$ is Fr\'{e}chet differentiable with respect to $\widehat{E_{_{\rm lo}}}$, with  $D_{\widehat{E_{_{\rm lo}}}} \mathcal{R}_{J,2}^{\sigma}[ \alpha, \widehat{E_{_{\rm lo}}} ]$ displayed  in \eqref{eqn:DHtilde}. Finally, we have the bounds 
 \begin{align}
& \left\| \mathcal{R}_{J,2}^{\sigma}[\alpha, \widehat{E_{_{\rm lo}}}] \right\|_{L^{2,a-2 }(\mathbb{R})} \lesssim \ \alpha^{2J + 2} +\ \left\| \widehat{E_{_{\rm lo}}} \right\|_{L^{2,a}(\mathbb{R})}\ +\
  \left\| \widehat{E_{_{\rm lo}}} \right\|^2_{L^{2,a}(\mathbb{R})}\ +\   \left\| \widehat{E_{_{\rm lo}}} \right\|^3_{L^{2,a}(\mathbb{R})} , \label{eqn:Hestimate}\\
 & \left\| D_{\widehat{E_{_{\rm lo}}}} \mathcal{R}_{J,2}^{\sigma} \left[  \alpha, \widehat{E_{_{\rm lo}}} \right] \right\|_{_{L^{2,a}(\mathbb{R}) \to L^{2,a-2}(\mathbb{R})}} \lesssim  \ \alpha^{2r} + \alpha^{2 - 2r} \  + \ \left\| \widehat{E_{_{\rm lo}}} \right\|_{L^{2,a}(\mathbb{R})}\ +\  \left\| \widehat{E_{_{\rm lo}}} \right\|^2_{L^{2,a}(\mathbb{R})}  . \label{eqn:Hdiffestimate}
 \end{align}
 \end{proposition}

 \subsection{Proof of Proposition \ref{prop:rescale}; detailed derivation of equation \eqref{eqn:lowrescaled} for $\widehat{E_{_{\rm lo}}}$} \label{section:derivationphilo}
  We proceed to rewrite equation \eqref{eqn:closed} in the form \eqref{eqn:lowrescaled}. First, we use $
 \ \chi_{_{\rm hi}} = 1 - \chi_{_{\rm lo}}  \ $
to get
\begin{align}
\chi_{_{\rm lo}}( Q ) \ \widetilde{\psi_1} * \widetilde{\psi_1} * \widehat{E_{_{\rm lo}}}  (Q)  =   \widetilde{\psi_1} * \widetilde{\psi_1} * \widehat{E_{_{\rm lo}}}  (Q) - \chi_{_{\rm hi}}(Q) \   \widetilde{\psi_1} * \widetilde{\psi_1} * \widehat{E_{_{\rm lo}}}  (Q). \label{eqn:loderive1}
\end{align}
Next  note as a consequence of \eqref{eqn:closed} we have $\overline{\chi}_{_{\rm lo}}  \widehat{E_{_{\rm lo}}} = 0$. Therefore,  we may write:
\begin{align}
M_{\alpha}(Q) \ \widehat{E_{_{\rm lo}}} (Q)  
& = |Q|^2 \  \widehat{E_{_{\rm lo}}} (Q) + \chi_{_{\rm lo}} \left[ M_{\alpha}(Q) - | Q |^2 \right] \ \widehat{E_{_{\rm lo}}}(Q) .  \label{eqn:loderive2}
\end{align}
 We may now express the left-hand side of (\ref{eqn:closed}) via \eqref{eqn:loderive1} and \eqref{eqn:loderive2} as
\begin{align}
&\left[ 1 + M_{\alpha}(Q) \right] \  \widehat{E_{_{\rm lo}}} ( Q ) - \chi_{_{\rm lo}}( Q )   \frac{3}{4 \pi^2}   \widetilde{\psi_1} * \widetilde{\psi_1} * \widehat{E_{_{\rm lo}}}  (Q)\ 
 =\ \widetilde{L_+} \widehat{E_{_{\rm lo}}}(Q) - R_{_{\rm pert}} \left[ \alpha, \widehat{E_{_{\rm lo}}} \right] (Q). \label{eqn:LHSrescaled}
\end{align}
Here,  we have defined the (linear in $\widehat{E_{_{\rm lo}}}$) operator
\begin{align}
R_{_{\rm pert}} \left[ \alpha, \widehat{E_{_{\rm lo}}} \right] (Q) & \equiv   \left[ \  | Q |^2 - M_{\alpha}(Q)  \ \right] \ \widehat{E_{_{\rm lo}}}(Q) -  \chi_{_{\rm hi}}(Q) \  \frac{3}{4 \pi^2}  \   \widetilde{\psi_1} * \widetilde{\psi_1} * \widehat{E_{_{\rm lo}}}  (Q) \nonumber \\
& =   \chi_{_{\rm lo}} (Q) \ \left[ \  | Q |^2 - M_{\alpha}(Q)  \ \right] \ \widehat{E_{_{\rm lo}}}(Q) -  \chi_{_{\rm hi}}(Q) \  \frac{3}{4 \pi^2}  \   \widetilde{\psi_1} * \widetilde{\psi_1} * \widehat{E_{_{\rm lo}}}  (Q). \label{eqn:Rpertdef}
\end{align}
We may now rewrite \eqref{eqn:closed} as
\begin{align}
  \widehat{L _+} \widehat{E_{_{\rm lo}}}(Q)  & =  \chi_{_{\rm lo}}( Q ) \ \mathcal{R}_{J,1}^{\sigma} \left[ \alpha, \widehat{E_{_{\rm lo}}} + \widehat{E_{_{\rm hi}}} [ \alpha, \widehat{E_{_{\rm lo}}} ] \right] ( Q) + R_{_{\rm pert}} \left[ \alpha, \widehat{E_{_{\rm lo}}} \right] (Q). 
\end{align}
Now define
\begin{align}
\mathcal{R}_{J,2}^{\sigma} \left[ \alpha, \widehat{E_{_{\rm lo}}} \right](Q) & \equiv \chi_{_{\rm lo}}( Q ) \ \mathcal{R}_{J,1}^{\sigma} \left[ \alpha, \widehat{E_{_{\rm lo}}} + \widehat{E_{_{\rm hi}}} [ \alpha, \widehat{E_{_{\rm lo}}} ] \right] ( Q)  + R_{_{\rm pert}} \left[ \alpha, \widehat{E_{_{\rm lo}}} \right] (Q) , \label{eqn:Htildedef}
\end{align}
where $\mathcal{R}_{J,1}^{\sigma} $ is defined in \eqref{eqn:HJdef}, to get equation \eqref{eqn:lowrescaled}. Finally, the expression for $D_{\widehat{E_{_{\rm lo}}}} \mathcal{R}_{J,2}^{\sigma}[ \alpha, \widehat{E_{_{\rm lo}}} ] $ is displayed and bounded, together with  $\mathcal{R}_{J,2}^{\sigma}[ \alpha, \widehat{E_{_{\rm lo}}} ]$ in 
 Appendix \ref{section:lowests}.

\subsection{Solution of the low frequency equation}\label{section:solutionlow}
In order to solve equation (\ref{eqn:lowrescaled}), we make use of the following general lemma, a consequence of the implicit function theorem; see  Appendix \ref{IFT}.\\
\begin{lemma}
\label{lemma:generallinearop}
Let $\mathcal{L}: L^{2,a}_{_{\rm even}}(\mathbb{R}) \longmapsto L^{2,a-2}_{_{\rm even}}(\mathbb{R})$ be an isomorphism.  Assume that for some constant $\alpha_1$,  $\mathcal{R}: [0, \alpha_1)_{\alpha} \times L_{_{\rm even}}^{2,a}(\mathbb{R}) \longrightarrow L^{2,a-2}_{_{\rm even}}(\mathbb{R})$ is an operator which is continuous at $(0,0)$, is Fr\'{e}chet differentiable on $L_{_{\rm even}}^{2,a}(\mathbb{R}) $, and satisfies $\mathcal{R}[0,0] = 0$. Suppose also that $\mathcal{R}$ satisfies
\begin{align}
& \left\|  \mathcal{R}[\alpha, f]  \right\|_{L^{2,a-2}(\mathbb{R})} \lesssim   \gamma_1(\alpha) + \| f \|_{L^{2,a}(\mathbb{R})} + \| f \|^2_{L^{2,a}(\mathbb{R})} + \| f \|^3_{L^{2,a}(\mathbb{R})}  , \label{eqn:HIFTest} \\
& \left\| D_f \mathcal{R}[\alpha, f]  \right\|_{L^{2,a}(\mathbb{R}) \to L^{2,a-2}(\mathbb{R})} \lesssim   \gamma_2(\alpha)  + \|f \|_{L^{2,a}(\mathbb{R})} + \| f \|^2_{L^{2,a}(\mathbb{R})}  \label{eqn:HIFTestdiff},
\end{align}
for some continuous functions $\gamma_1(\alpha) \geq 0$ and $\gamma_2(\alpha) \geq 0$ satisfying $
\gamma_1(0) = \gamma_2(0 ) = 0. $

 Then there exists a constant $\alpha_2 \le \alpha_1$ such that for all $0<\alpha < \alpha_2$, the equation
\begin{align}
\mathcal{L} f(Q) = \mathcal{R}[\alpha, f](Q),
\end{align}
has a unique, even solution $f = f[\alpha] \in L^{2,a}_{_{\rm even}}(\mathbb{R})$ satisfying the estimate
\begin{align}
\| f[\alpha] \|_{L^{2,a}(\mathbb{R})} \lesssim \gamma_1(\alpha). \label{eqn:sigmaest}
\end{align}
\end{lemma} 

\noindent {\bf Proof of Lemma \ref{lemma:generallinearop}:} We shall apply the implicit function theorem. Define $\mathcal{J}: L^{2,a}_{_{\rm even}}(\mathbb{R}) \times \mathbb{R} \longrightarrow L^{2,a-2}_{_{\rm even}}(\mathbb{R})$:
\begin{align}
\mathcal{J}[\alpha, f](Q) \equiv \mathcal{L} f(Q) -  \mathcal{R}[\alpha, f](Q).
\end{align}
Note that $\mathcal{J}[0, 0](Q) \equiv0$. 
 We check the following properties of $\mathcal{J}$:
%
\begin{enumerate}
\item The mapping  $(\alpha, f)  \longmapsto \mathcal{J} [\alpha, f]$, $\mathcal{J}: L^{2,a}_{_{\rm even}}(\mathbb{R}) \times \mathbb{R} \longmapsto L^{2,a-2}_{_{\rm even}}(\mathbb{R})$ is continuous at $(0,0)$.

\item For $\alpha \neq 0$, $\mathcal{J}$ is Fr\'{e}chet differentiable with respect to $f$ with
\begin{align}
D_f \mathcal{J} [\alpha, f] = \mathcal{L} - D_f \mathcal{R}[\alpha, f],\label{dJ}
\end{align}
which  is continuous at $(0,0)$ in $L^{2,a}_{_{\rm even}}(\mathbb{R}) \times \mathbb{R}$.

\item $D_f \mathcal{J} [0, 0] = \mathcal{L}$ is an isomorphism of $L^{2,a}_{_{\rm even}}(\mathbb{R})$ onto $L^{2,a-2} _{_{\rm even}}(\mathbb{R})$.
\end{enumerate}

We now verify these properties. Since $\mathcal{L}$ and $\mathcal{R}$ are continuous at $(0,0)$, so is $\mathcal{J}$. It is simple to check for the map $f\mapsto\mathcal{J}[\alpha, f]$ that \eqref{dJ} holds.  
The boundedness of $\mathcal{L}$ and estimate $\eqref{eqn:HIFTest}$ give
\begin{align}
& \| \mathcal{J}[\alpha, f] \|_{L^{2,a-2}(\mathbb{R})} \leq  \| \mathcal{L} f \|_{L^{2,a-2}(\mathbb{R})} + \|  \mathcal{R}[\alpha, f] \|_{L^{2,a-2}(\mathbb{R})} \nonumber \\
& \leq \| \mathcal{L} \|_{L^{2,a}(\mathbb{R}) \to L^{2,a-2}(\mathbb{R})} \| f \|_{L^{2,a}(\mathbb{R})} + \|  \mathcal{R}[\alpha, f] \|_{L^{2,a-2}(\mathbb{R})} \nonumber \\
&
\leq C \left( \gamma_1(\alpha) + \| f \|_{L^{2,a}(\mathbb{R})} + \| f \|^2_{L^{2,a}(\mathbb{R})} + \| f \|^3_{L^{2,a}(\mathbb{R})} \right). 
\end{align}
This  implies that $\mathcal{J}[\alpha,f]$ is continuous at $(\alpha,f)=(0,0)$. Similarly, estimate $\eqref{eqn:HIFTestdiff}$ implies 
\begin{align}
& \| D_f \mathcal{J}[\alpha, f] - \mathcal{L} \|_{L^{2,a}(\mathbb{R}) \to L^{2,a-2}(\mathbb{R})} =
\| D_f \mathcal{R}[\alpha, f] \|_{L^{2,a}(\mathbb{R}) \to L^{2,a-2}(\mathbb{R})} \nonumber \\
 & \lesssim \ \gamma_2(\alpha) + \|f \|_{L^{2,a}(\mathbb{R})} + \| f \|^2_{L^{2,a}(\mathbb{R})} ,
 \end{align}
and thus  $D_f \mathcal{J}[0,0] = \mathcal{L}$. 

  By the above discussion, we can apply implicit function theorem as given in Appendix \ref{IFT}. Thus, there exists some $\alpha_2 > 1$ such that for all $ \alpha< \alpha_2$, there exists a mapping $\alpha \longmapsto f[\alpha]$, $f: \mathbb{R}_+ \longmapsto L^{2,a}_{_{\rm even}}(\mathbb{R})$ which solves
 \begin{align}
\mathcal{J} \big[\alpha, f[\alpha] \big](Q) = 0 \qquad \Longleftrightarrow \qquad \mathcal{L} f[\alpha](q) = \mathcal{R} \big[ \alpha, f[\alpha] \big](Q).
 \end{align}
Estimate \ref{eqn:sigmaest} follows from a more detailed proof of the above lemma as given in
 Appendix  \ref{subsection:details2}. This completes the proof of Lemma \ref{lemma:generallinearop}. $\Box$

We may now apply Lemma \ref{lemma:generallinearop} to the rescaled low frequency equation. \\

\begin{proposition}
\label{prop:solutiontolow}
Let $a > 1/2$ and $0 < r < 1$. Then there exists 
$0< \alpha_2\le\alpha_1$ such that for all $\alpha\in (0,\alpha_2)$, there exists an even (symmetric) solution $\widehat{E_{_{\rm lo}}}$ to (\ref{eqn:lowrescaled}) which satisfies
\begin{align}
 \left\| \widehat{E_{_{\rm lo}}} \right\|_{L^{2,a}(\mathbb{R})} \lesssim \alpha^{2J + 2} . \label{eqn:philoest}
\end{align}
Furthermore, we have that $\widehat{E_{_{\rm lo}}} = \chi_{_{\rm lo}} \widehat{E_{_{\rm lo}}}$; that is, $\widehat{E_{_{\rm lo}}}(Q)$ is supported on $ Q \in [ - \alpha^{r -1}, \alpha^{ r - 1} ]$.
\end{proposition} \\

\noindent {\bf Proof of Proposition \ref{prop:solutiontolow}:} It suffices to show that the hypotheses to Lemma \ref{lemma:generallinearop} are satisfied by equation (\ref{eqn:lowrescaled}) for $a > 1/2$. By Proposition \ref{prop:Lplus}  $\widetilde{L_+}:L^{2,a}_{_{\rm even}}(\mathbb{R})\to L_{_{\rm even}}^{2,a-2}(\mathbb{R}) $ is an isomorphism. Moreover, by Proposition \ref{prop:rescale} the mapping $(\alpha,\widehat{E}_{\rm lo})\mapsto \mathcal{R}_{J,2}^{\sigma}[\alpha,\widehat{E}_{\rm lo}]$, maps $L^{2,a}_{_{\rm
even}}(\mathbb{R})$ to $L^{2,a-2}_{_{\rm even}}(\mathbb{R})$, and  is continuous at $(\alpha, \widehat{E_{_{\rm lo}}} ) = (0,0)$ .  Furthermore, by choosing $\alpha < \alpha_1$ the estimates (\ref{eqn:Hestimate}) and (\ref{eqn:Hdiffestimate}) on $\mathcal{R}_{J,2}^{\sigma}[ \alpha, \widehat{E_{_{\rm lo}}} ]$ hold. Hence,  hypotheses
(\ref{eqn:HIFTest}) and (\ref{eqn:HIFTestdiff}) of Lemma \ref{lemma:generallinearop} are satisfied. Lemma \ref{lemma:generallinearop} implies, for $0<\alpha<\alpha_2\le\alpha_2$, the existence of 
$\widehat{E_{_{\rm lo}}}$ satisfying the bound \eqref{eqn:philoest}. $\Box$

\subsection{Completion of the proof of Proposition \ref{prop:rigorousexpansion}}
\label{section:completion}
First, we summarize the results of sections \ref{subsect:lyapschm} through \ref{section:solutionlow}. Proposition \ref{prop:ifthigh} guarantees that a unique solution $\widehat{E_{_{\rm hi}}}[\alpha, \Gamma] \in L_{_{\rm even}}^{2,a}(\mathbb{R})$ exists to the high frequency equation \eqref{eqn:high} for any $\Gamma \in L_{_{\rm even}}^{2,a}(\mathbb{R})$ and $\alpha > 0 $ sufficiently small:
\begin{align}
& \left[ 1 + M_{\alpha}(Q) \right] \  \widehat{E_{_{\rm hi}}} \left[ \alpha, \Gamma \right] ( Q ) - \chi_{_{\rm hi}}( Q )  \ \chi_{_{\mathcal{B}_{_{\alpha}}}}(Q) \  \frac{3}{4 \pi^2}   \left( \widetilde{\psi_1} * \widetilde{\psi_1} * \Gamma (Q) + \widetilde{\psi_1} * \widetilde{\psi_1} * \widehat{E_{_{\rm hi}}}  \left[ \alpha, \Gamma \right]  (Q) \right) & \nonumber \\
& \hspace{7cm} = \chi_{_{\rm hi}}( Q ) \ \mathcal{R}_{J,1}^{\sigma} \left[ \alpha, \Gamma + \widehat{E_{_{\rm hi}}} \left[ \alpha, \Gamma \right] \right] ( Q).  \label{eqn:high2}
\end{align}
with the bound
\begin{align}
\left\| \widehat{E_{_{\rm hi}}}[\alpha, \Gamma] \right\|_{L^{2,a}(\mathbb{R})} \lesssim \alpha^{2 - 2r} \left\| \Gamma \right\|_{L^{2,a}(\mathbb{R})} + e^{- C \alpha^{r-1}}, \quad \quad 0 < r < 1. \label{eqn:hiestcomplete}
\end{align}
We apply this result with $\Gamma=\widehat{E_{_{\rm lo}}}\in L_{_{\rm even}}^{2,a}(\mathbb{R})$ to obtain   $\widehat{E_{_{\rm hi}}}[\alpha, \widehat{E_{_{\rm lo}}}] \in L_{_{\rm even}}^{2,a}(\mathbb{R})$.
 Thus,  (\ref{eqn:low}) may be rewritten as a closed equation (\ref{eqn:closed}) for  $\widehat{E_{_{\rm lo}}}$:
\begin{align}
&\left[ 1 + M_{\alpha}(Q) \right] \  \widehat{E_{_{\rm lo}}} ( Q ) - \chi_{_{\rm lo}}( Q )   \frac{3}{4 \pi^2}   \widetilde{\psi_1} * \widetilde{\psi_1} * \widehat{E_{_{\rm lo}}}  (Q)  & \nonumber \\ 
&  \quad  = \chi_{_{\rm lo}}( Q ) \ \mathcal{R}_{J,1}^{\sigma} \left[ \alpha, \widehat{E_{_{\rm lo}}} + \widehat{E_{_{\rm hi}}} [ \alpha, \widehat{E_{_{\rm lo}}} ] \right] ( Q) +   \chi_{_{\rm lo}}( Q )   \frac{3}{4 \pi^2}   \widetilde{\psi_1} * \widetilde{\psi_1} * \left( \widehat{E_{_{\rm hi}}}  [ \alpha, \widehat{E_{_{\rm lo}}} ] \right) (Q) . \label{eqn:closed2}
\end{align}
By  Proposition \ref{prop:solutiontolow} there exists for all $0 < \alpha < \alpha_2$, with $\alpha_2$ sufficiently small, and any $r\in(0,1)$  a unique solution $\widehat{E_{_{\rm lo}}} \in L_{_{\rm even}}^{2,a}(\mathbb{R})$ of (\ref{eqn:lowrescaled}):
\begin{align}
\widetilde{L_+} \widehat{E_{_{\rm lo}}}(Q) = \mathcal{R}_{J,2}^{\sigma}[\alpha, \widehat{E_{_{\rm lo}}}](Q), \label{eqn:lowrescaled2}
\end{align}
with the bound
\begin{align}
\left\| \widehat{E_{_{\rm lo}}} \right\|_{L^{2,a}(\mathbb{R})} \lesssim \alpha^{2J + 2}. \label{eqn:loestcomplete}
\end{align}
By Proposition \ref{prop:rescale}, equation \eqref{eqn:lowrescaled2} is equivalent to \eqref{eqn:closed2}. By Proposition \ref{prop:highlow}, adding together equations (\ref{eqn:closed2}) and (\ref{eqn:high2}) (with $\Gamma(q) = \widehat{E_{_{\rm lo}}}(q)$) gives equation \eqref{eqn:Eeqn} for 
\begin{align}
&\widehat{E}( Q )\ =\  \widehat{E_J^{\sigma, \alpha}}( Q )\ =\ \widehat{E_{_{\rm lo}}}(Q) + \widehat{E_{_{\rm hi}}} \left[ \alpha, \widehat{E_{_{\rm lo}}} \right] (Q), \qquad \qquad \widehat{E}(Q) = \chi_{_{\mathcal{B}_{_{\alpha}}}} ( Q ) \ \widehat{E} (Q), \nonumber\\
&\left[ 1 + M_{\alpha}( Q) \right] \ \widehat{E}( Q ) - \chi_{_{\mathcal{B}_{_{\alpha}}}}( Q)  \  \left( \frac{1}{2 \pi} \right)^2
\widetilde{\psi_1} * \widetilde{\psi_1} * \widehat{E}( Q ) = \mathcal{R}_{J,1}^{\sigma} \left[ \alpha, \widehat{E}\right]( Q )\ .   \label{eqn:Eeqnfinal}
\end{align}

By Proposition \ref{prop:ifthigh}, $\widehat{E_{_{\rm lo}}} \in L_{_{\rm even}}^{2,a}(\mathbb{R})$ implies that $\widehat{E_{_{\rm hi}}} [\alpha, \widehat{E_{_{\rm lo}}}] \in L_{_{\rm even}}^{2,a}(\mathbb{R})$ and therefore $\widehat{E} \in L_{_{\rm even}}^{2,a}(\mathbb{R})$. Furthermore, by \eqref{eqn:hiestcomplete} and \eqref{eqn:loestcomplete}, we have the bound
\begin{align}
\left\| \  \widehat{E} \  \right\|_{L^{2,a}(\mathbb{R})} \lesssim \left\| \widehat{E_{_{\rm lo}}} \right\|_{L^{2,a}(\mathbb{R})} + \left\| \widehat{E_{_{\rm hi}}} [ \alpha, \widehat{E_{_{\rm lo}}} ]  \right\|_{L^{2,a}(\mathbb{R})} \lesssim \left( 1 + \alpha^{2 - 2r} \right) \left\| \widehat{E_{_{\rm lo}}} \right\|_{L^{2,a}(\mathbb{R})} + \mathcal{O}(\alpha^{\infty})  
\lesssim \alpha^{2J + 2}, \label{eqn:phiestfinalr}
\end{align}
for $0 < \alpha < \alpha_2$ with $\alpha_2$ sufficiently small and $0 < r < 1$. Finally, from  \eqref{eqn:rigorousexpansion} we have
\begin{align}
\widehat{\Phi^{\sigma}}( Q) = \sum_{j = 0}^J \ \alpha^{2j} \  \chi_{_{\mathcal{B}_{_{\alpha}}}}(Q) \ F_j \left[ \widetilde{\psi_1} \right] (Q)  + \widehat{E_J^{\alpha, \sigma}}(Q) \qquad \in L_{_{\rm even}}^{2,a}(\mathbb{R}),  
\label{Phi-sigma}\end{align}
where $F_j [ \widetilde{\psi_1} ] = F_j \in L_{_{\rm even}}^{2,a}(\mathbb{R})$ are given in Proposition \ref{prop:orderj}.
\medskip

To complete the proof of Proposition \ref{prop:rigorousexpansion}, we show that $\widehat{\Phi^{\sigma}}$ is real-valued. By Proposition \ref{prop:orderj},  $F_j [ \widetilde{\psi_1} ], \ \ j = 0, \dots, J$ are real-valued. It therefore suffices to show that $\widehat{E_J^{\alpha, \sigma}}(Q)$ is real-valued. 

Recall equation \eqref{eqn:Eeqnfinal} for $\widehat{E} = \widehat{E_J^{\alpha, \sigma}}$, which has inhomogeneous forcing $\mathcal{D}^{\sigma,\alpha}[S_J^{\alpha}]$ contained in $\mathcal{R}_{J,1}^{\sigma} [ \alpha, \widehat{E } ]$ (here, the operator $\mathcal{D}^{\sigma,\alpha}$ is defined in \eqref{eqn:Phieqn}) . Note that since $S_J^{\sigma} = \sum_{j = 0}^J \ \alpha^{2j} \  \chi_{_{\mathcal{B}_{_{\alpha}}}} \ F_j [ \widetilde{\psi_1} ]$ is real-valued and since $e^{2 m \pi i \sigma} = \pm 1$ for $\sigma \in \{0, 1/2 \}$, the forcing $\mathcal{D}^{\sigma,\alpha}[S_J^{\alpha}]$ is also real-valued. Now define
\begin{align}
\widehat{E^{\sigma}_{_{\rm im}}} (Q) \equiv \widehat{E}(Q) - \overline{ \widehat{E}}(Q) = 2 \ i \   {\rm Im } \  \widehat{E^{\sigma}} (Q). 
\end{align}
Subtracting the complex conjugate of equation \eqref{eqn:Eeqnfinal} for $\widehat{E}$ from itself then gives the linear equation for $\widehat{E_{_{\rm im}}}$:
\begin{align}
\left[ 1 + M_{\alpha}( Q) \right] \ \widehat{E_{_{\rm im}}}( Q ) - \chi_{_{\mathcal{B}_{_{\alpha}}}}( Q)  \  \left( \frac{1}{2 \pi} \right)^2
\widetilde{\psi_1} * \widetilde{\psi_1} * \widehat{E_{_{\rm im}}} ( Q) = \mathcal{R}_{_{J, {\rm diff}}}^{\sigma} \left[ \alpha, \widehat{E_{_{\rm im}}} \right] (Q),  \label{eqn:Eim}
\end{align}
Since $\mathcal{D}^{\sigma,\alpha}[S_J^{\alpha}]$ is real-valued,  $\mathcal{R}_{_{J, {\rm diff}}}^{\sigma} $ contains no inhomogeneous forcing term. 
As such, a Lyapunov-Schmidt strategy applied to \eqref{eqn:Eim} yields, for some $0 < r < 1$, 
\begin{align}
& \left\| \widehat{E_{_{\rm im,  hi}}} \right\|_{L^{2,a}(\mathbb{R})} \lesssim \ \alpha^{2 - 2r} \ \left\| \widehat{E_{_{\rm im,  lo}}} \right\|_{L^{2,a}(\mathbb{R})}, \\
& \left\| \widehat{E_{_{\rm im,  lo}}} \right\|_{L^{2,a}(\mathbb{R})}  \lesssim ( \alpha^2 + \alpha^{2 - 2r} )  \left( \left\| \widehat{E_{_{\rm im, lo}}} \right\|_{L^{2,a}(\mathbb{R})}  + \left\| \widehat{E_{_{\rm im,  hi}}} \right\|_{L^{2,a}(\mathbb{R})} \right), \nonumber \\
\Longleftrightarrow \qquad & (1 - C \ [ \alpha^2 + \alpha^{2 - 2r}] ) \left\| \widehat{E_{_{\rm im,  lo}}} \right\|_{L^{2,a}(\mathbb{R})}  \leq 0.
\end{align}
 Taking $\alpha$ small enough that $ 1 - C  \ [ \alpha^2 + \alpha^{2 - 2r} ] > 1/2$ implies that $\widehat{E_{_{\rm im,  lo}}} = 0$. Therefore, $\widehat{E}$ is real-valued, which implies that $\widehat{\Phi^{\sigma}}$ is real-valued. This completes the proof of Proposition \ref{prop:rigorousexpansion}. 
$\Box$ \\

\subsection{Completion of the proof of Theorem \ref{th:main}} \label{section:mainproofend}

 Above we solved for, $\alpha\mapsto\widehat{\Phi^{\sigma,\alpha}}(Q)$,  the discrete Fourier transform of the on- and off-site standing waves as a function of  the scaled variable $Q$, restricted to the scaled Brillouin zone, $\mathcal{B}_{_\alpha}=[-\pi/\alpha,\pi/\alpha]$. To complete the proof we use this to construct $\widehat{\phi^{\sigma,\alpha}}(q)$, defined on
  $\mathcal{B}=[-\pi,\pi]$. From \eqref{Phi-sigma} we have
  \begin{align}
& \widehat{\phi^{\sigma}}(q) = \chi_{_{\mathcal{B}}}(q)  \widehat{\phi^{\sigma}}(q) = \sum_{j = 0}^{J} \ \alpha^{2j} \ \chi_{_{\mathcal{B}}}(q) \  F_j \left[ \widetilde{\psi_1} \right] \left( \frac{q}{\alpha} \right) + \widehat{E_J^{\alpha, \sigma}} \left( \frac{q}{\alpha} \right), \nonumber \\
& \left\| F_j \left[ \widetilde{\psi_1} \right] \left( \frac{\cdot}{\alpha} \right) \right\|_{L^{2,a}(\mathbb{R})} \lesssim \alpha^{1/2}, \qquad  \left\| \widehat{E_J^{\alpha, \sigma}} \left( \frac{\cdot}{\alpha} \right) \right\|_{L^{2,a}(\mathbb{R})} \lesssim \alpha^{2J + 5/2}\ .
\label{FjEJ-bounds}
\end{align}
The bounds \eqref{FjEJ-bounds} follow since $F_j\left[ \widetilde{\psi_1} \right](Q)$ and
$\widehat{E_J^{\alpha, \sigma}}(Q) $ have order one $L^{2,a}(\mathbb{R}_Q)$ norm, and 
using the general bound on $q\mapsto f(q/\alpha)$ in $L^{2,a}(\mathbb{R}_q)$:
\begin{align}
  \Big\| f\left( \frac{q}{\alpha} \right) \Big\|^2_{L^{2,a}(\mathbb{R}_q)} & = 
  \int_{\mathbb{R}} \ \left(1 + |q|^2 \right)^a \ \Big|f\left( \frac{q}{\alpha} \right)\Big|^2 \ dq \leq \int_{\mathbb{R}} \ \left(1 + \frac{|q|^2}{\alpha^2} \right)^a \ \Big|f\left( \frac{q}{\alpha} \right)\Big|^2 \ dq \nonumber \\
& = \alpha \ \int_{\mathbb{R}} \ \left(1 + |Q|^2  \right)^a \ \Big|f\left( Q \right)\Big|^2 \ dQ = \alpha \left\| f\right\|^2_{L^{2,a}(\mathbb{R}_Q)}. \label{eqn:generalrescl2a}
\end{align}

Next,  the ($2\pi-$ periodic in $q$)  discrete Fourier transform of $\alpha\mapsto\{G_n^{\sigma,\alpha}\}_{n\in\mathbb{Z}}$ is 
 $\widehat{G^{\sigma,\alpha}}(q)=e^{-i\sigma q} \widehat{K^{\sigma,\alpha}}(q)$, where
  \begin{align}
\widehat{K^{\sigma,\alpha}}(q) = \sum_{m \in \mathbb{Z}} \ \chi_{_{\mathcal{B}}}(q - 2 m \pi) \  \widehat{\phi^{\sigma,\alpha}}(q - 2 m \pi) \ e^{2 m \pi i \sigma}, \label{eqn:Kdef2}
\end{align} 
This implies the expansion on the Brillouin zone $q \in \mathcal{B} = [- \pi, \pi]$:
\begin{align}
\widehat{G^{\alpha, \sigma}}(q) = e^{- i q \sigma}  \ \widehat{\phi^{\sigma}}(q) = e^{- i q \sigma} \left(\sum_{j = 0}^{J} \ \alpha^{2j} \ \chi_{_{\mathcal{B}}}(q) \  F_j \left[ \widetilde{\psi_1} \right] \left( \frac{q}{\alpha} \right) + \widehat{E_J^{\alpha, \sigma}} \left( \frac{q}{\alpha} \right) \right), \quad \quad q \in \mathcal{B} . \label{eqn:GonB}
\end{align}
Define
\begin{align}
 \mathcal{G}_j[ \widetilde{\psi_1} ](n) = \mathcal{F}_{_D}^{-1} \left[ \ F_j \left[ \widetilde{\psi_1} \right] \left( \frac{q}{\alpha} \right) \right]_n, \qquad {\rm and} \qquad \mathcal{E}^{\alpha, J, \sigma}_n \equiv \mathcal{F}^{-1}_{_D} \left[ e^{- i q  \sigma} \widehat{E_J^{\alpha, \sigma}}  \left( \frac{q}{\alpha} \right) \right]_n. \label{eqn:transdefs}
\end{align}
Therefore, for any $\sigma$, in particular $\sigma = 0, 1/2$,
\begin{align}
& \mathcal{G}_j[ \widetilde{\psi_1} ](n - \sigma) = \mathcal{F}_{_D}^{-1} \left[ e^{- i q \sigma} \ F_j \left[ \widetilde{\psi_1} \right] \left( \frac{q}{\alpha} \right) \right]_n.
\end{align}
 Applying the inverse discrete Fourier transform \eqref{eqn:invDFT} to $\widehat{G^{\alpha, \sigma}}(q)$ in \eqref{eqn:GonB} gives the branches of on-site ($\sigma=0$) and off-site ($\sigma=1/2$) discrete solitary waves \eqref{eqn:solutions}. We use the Plancherel identity with the bounds \eqref{FjEJ-bounds} to get
 \begin{align}
& \left\|  \mathcal{G}_j [ \widetilde{\psi_1}] (n - \sigma) \right\|_{l^2(\mathbb{Z}_n)} = \frac{1}{(2 \pi)^2} \  \left\| e^{- i q \sigma} F_j \left[ \widetilde{\psi_1} \right] \left( \frac{q}{\alpha} \right) \right\|_{L^2(\mathcal{B}; dq)} \lesssim  \left\|  F_j \left[ \widetilde{\psi_1} \right] \left( \frac{q}{\alpha} \right) \right\|_{L^{2,a}(\mathbb{R}_q)} \lesssim \alpha^{1/2}, \nonumber \\
& \left\|  \mathcal{E}^{\alpha, J, \sigma}  \right\|_{l^2(\mathbb{Z})} = \frac{1}{(2 \pi)^2} \ \left\| e^{- i q \sigma} E_J^{\alpha, \sigma} \left( \frac{q}{\alpha} \right) \right\|_{L^2(\mathcal{B};dq)} \lesssim \left\| e^{- i q \sigma} E_J^{\alpha, \sigma} \left( \frac{q}{\alpha} \right) \right\|_{L^{2,a}(\mathbb{R}_q)}  \lesssim \alpha^{2J + 5/2}. 
\end{align}
\medskip

To complete the proof of Theorem \ref{th:main},  we show that the solitary wave $G^{\sigma, \alpha}_n = \mathcal{F}_{_D}^{-1} [ \widehat{G^{\sigma, \alpha}} ]_n$ corresponds for $\sigma = 0$ to a real-valued, on-site symmetric solitary wave and corresponds for $\sigma = 1/2$ to a real-valued, off-site symmetric solitary wave. By Proposition \ref{prop:off-on}, it suffices to show that $\widehat{K^{\sigma}}(q)$ is symmetric (even) and real-valued. Proposition \ref{prop:rigorousexpansion} gives that $\widehat{\Phi^{\sigma}} \in L_{_{\rm even}}^{2,a}(\mathbb{R}_Q)$ is real-valued, which implies by \ref{eqn:generalrescl2a} that its rescaling $\widehat{\phi^{\sigma}} \in L_{_{\rm even}}^{2,a}(\mathbb{R}_q)$ is real-valued. Since $e^{2 m \pi i \sigma} = \pm 1$ for $\sigma \in \{0, 1/2 \}$, $\widehat{K^{\sigma}}$ as given in \eqref{eqn:Kdef2} will be also be even and real-valued. This completes the proof of Theorem \ref{th:main}.

\section{Exponential smallness of the Peierls-Nabarro barrier; Proof of Theorem \ref{th:PN} for $d = 1$} \label{section:PNbarrier}
We now prove Theorem \ref{th:PN} for dimension $d = 1$ in sections \ref{section:PNbarrier} through \ref{section:diffequation}. Recall the definitions
\begin{align}
& \mathcal{N}[{G} ] = \| {G} \|^2_{l^2(\mathbb{Z})} = \sum_{n \in \mathbb{Z}} |G_n|^2,   \nonumber \\
& \mathcal{H}[ {G} ] = \sum_{n \in \mathbb{Z}} | G_{n + 1} - G_n |^2 - \frac12 |G_n|^4  = \| \delta {G} \|_{l^2(\mathbb{Z})}^2 - \frac12 |G_n|^4.
\end{align}
By Theorem \ref{th:main}, for $\alpha$ sufficiently small, there exist solutions ${G}^{\alpha, {\rm on}} = \{G_n^{\alpha, {\rm on}} \}_{n \in \mathbb{Z}} $ and ${G}^{\alpha, {\rm off}} = \{G_n^{\alpha, {\rm off}} \}_{n \in \mathbb{Z}}$ to DNLS:
$
- \alpha^2 G_n^{\alpha, \sigma} = - (\delta^2 G)_n - (G_n)^3. 
$
We shall prove that there exists a constant $C > 0$ such that for $\alpha$ sufficiently small, 
\begin{align}
\mathbf{PN \ Barrier \ for \ } d = 1: \hspace{1cm} & \bigg| \mathcal{N} [{ G^{\rm off}}] - \mathcal{N} [{G^{\rm on}} ] \bigg| \lesssim  \alpha \ e^{- C/\alpha} \hspace{2cm} \\
& \bigg| \mathcal{H} [{G^{\rm off}}] - \mathcal{H} [{G^{\rm on}} ] \bigg| \lesssim \alpha \ e^{- C/\alpha} . 
\end{align}
The following identity allows us to equate the nonlinear term in $\mathcal{H}$ with terms involving $ \|{G} \|^2_{l^2(\mathbb{Z})} $ and $ \| \delta{G} \|^2_{l^2(\mathbb{Z})}$.

\begin{proposition} \label{prop:virial}
Suppose that ${G} = \{G_n \}_{n \in \mathbb{Z}} $ solves DNLS:
$
- \alpha^2 G_n = - (\delta^2 G)_n - (G_n)^3.  
$
Then ${G}$ satisfies
\begin{align}
\alpha^2 \| {G} \|_{l^2(\mathbb{Z})}^2  + \| \delta {G} \|_{l^2(\mathbb{Z})}^2  = \sum_{n \in \mathbb{Z}} |G_n|^4 .
\end{align}
\end{proposition}

\noindent {\bf Proof of Proposition \ref{prop:virial}:} Multiplying  DNLS by $G_n$ and summing over $n \in \mathbb{Z}$ gives the result. $\Box$ \\

Now recall from \eqref{eqn:ansatz3} that
\begin{align}
\widehat{G^{\sigma}}(q) = \mathcal{F}_{_D} \left[ {G^{\sigma}} \right](q) = e^{- i q \sigma} \ \widehat{K^{\sigma}}(q), \quad \quad q \in \mathbb{R}, \ \sigma \in \{0, 1/2 \},
\end{align}
where $\widehat{K^{\sigma}}(q)$ is even, real-valued, and $2 \pi \sigma $- pseudoperiodic, and where $  \chi_{_{\mathcal{B}}}(q) \  \widehat{K^{\sigma}}(q) = \widehat{\phi^{\sigma}}(q).$


Next, observe that by the Plancherel identity and since $\widehat{\phi^{\sigma}}$ is real-valued, 
\begin{align}
& 2 \pi \ \bigg| \mathcal{N} [{G^{\rm off}}] - \mathcal{N} [{G^{\rm on}} ] \bigg| = 2 \pi \ \bigg| \| 
{ G^{\rm off}} \|^2_{l^2(\mathbb{Z})} - \| {G^{\rm on}} \|^2_{l^2(\mathbb{Z})} \bigg| = \bigg| \| \widehat{G^{\rm off}} \|^2_{L^2(\mathcal{B})} -  \| \widehat{G^{\rm on}} \|^2_{L^2(\mathcal{B})}  \bigg| \nonumber \\
&= \bigg|  \| \widehat{K^{\rm off}} \ \|^2_{L^2(\mathcal{B})} -  \| \widehat{K^{\rm on}} \|^2_{L^2(\mathcal{B})} \bigg|  =  \bigg|  \| \widehat{\phi^{\rm off}} \ \|^2_{L^2(\mathcal{B})} -  \| \widehat{\phi^{\rm on}} \|^2_{L^2(\mathcal{B})} \bigg|   \leq \ \left\| \widehat{\phi^{\rm off}} + \widehat{\phi^{\rm on}}  \right\|_{L^2(\mathcal{B})} \  \left\| \widehat{\phi^{\rm off}} - \widehat{\phi^{\rm on}}  \right\|_{L^2(\mathcal{B})}. \label{eqn:planchdif1}
\end{align}
Similarly, by Lemma \ref{lemma:discderivtransform} and the Plancherel identity,
\begin{align}
& 2 \pi \ \bigg| \left( \sum_{n \in \mathbb{Z}} |G^{\rm on}_{m + n} - G^{\rm on}_n |^2 - |G^{\rm on}_{m + n} - G^{\rm on}_n |^2 \right) \bigg|
= 2 \pi  \ \bigg|   \| \delta {G^{\rm on}} \|^2_{l^2(\mathbb{Z})} - \| \delta {G^{\rm off}} \|^2_{l^2(\mathbb{Z})}   \bigg| \nonumber \\
& =  \bigg|   \left\|   \ \sin(m \cdot/2) \  \widehat{G^{\rm on}} \right\|^2_{L^2(\mathcal{B})}  -  \left\| \ \sin(m \cdot/2) \ \widehat{G^{\rm on}} \right\|^2_{L^2(\mathcal{B})}  \bigg| \nonumber \\
& =  \bigg|   \left\|   \ \sin(m \cdot/2) \  \widehat{\phi^{\rm on}} \right\|^2_{L^2(\mathcal{B})}  -  \left\| \ \sin(m \cdot/2) \ \widehat{\phi^{\rm on}} \right\|^2_{L^2(\mathcal{B})}   \bigg|  \leq \ \left\| \widehat{\phi^{\rm off}} + \widehat{\phi^{\rm on}}   \right\|_{L^2(\mathcal{B})}  \ \left\| \widehat{\phi^{\rm off}} - \widehat{\phi^{\rm on}}  \right\|_{L^2(\mathcal{B})}  .
\label{eqn:planchdif2}
\end{align}
Note also since $\widehat{\phi^{\sigma}}(q) = \chi_{_{\mathcal{B}}}(q) \widehat{\phi^{\sigma}}(q)  =  \widehat{\Phi^{\sigma}}(q/\alpha) = \widehat{\Phi^{\sigma}}(Q)$ , 
\begin{align}
\left\| \widehat{\phi^{\rm off}} - \widehat{\phi^{\rm on}}  \right\|_{L^2(\mathcal{B})} = \left\| \widehat{\phi^{\rm off}} - \widehat{\phi^{\rm on}}  \right\|_{L^2(\mathbb{R})} = \alpha^{1/2} \ \left\| \widehat{\Phi^{\rm off}} - \widehat{\Phi^{\rm on}}  \right\|_{L^2(\mathbb{R})} \lesssim  \alpha^{1/2} \ \left\| \widehat{\Phi^{\rm off}} - \widehat{\Phi^{\rm on}}  \right\|_{L^{2,a}(\mathbb{R})}. \label{eqn:planchdiff3}
\end{align}
Proposition \ref{prop:virial}, the bound $ \| \widehat{\phi^{\rm off}} + \widehat{\phi^{\rm on}}   \|_{L^2(\mathcal{B})}  \lesssim \alpha^{1/2}$, and equations \eqref{eqn:planchdif1}, \eqref{eqn:planchdif2}, and \eqref{eqn:planchdiff3} give
\begin{align}
& \bigg| \mathcal{N} [{G^{\rm off}}] - \mathcal{N} [{G^{\rm on}} ] \bigg| + 
\bigg| \mathcal{H} [{ G^{\rm off}}] - \mathcal{H} [{G^{\rm on}} ] \bigg| \lesssim  \alpha \  \left\| \widehat{\Phi^{\rm off}} - \widehat{\Phi^{\rm on}}  \right\|_{L^{2,a}(\mathbb{R})}, \label{eqn:Nest} 
\end{align}

We shall prove that  $  \widehat{\Phi^{\rm off}}(Q) -  \widehat{\Phi^{\rm on}}( Q ) $ is exponentially small in $L^{2,a}(\mathbb{R})$. 
\medskip
\begin{proposition}
\label{prop:expdiffEonoff} Let $\alpha_0  > 0$ be that prescribed in Proposition \ref{prop:rigorousexpansion}. Then for $0 < \alpha < \alpha_0$, there exists a constant $C > 0$ such that
\begin{align}
  \left\| \widehat{\Phi^{\rm off}} - \widehat{\Phi^{\rm on}} \right\|_{L^{2,a}(\mathbb{R})} \lesssim e^{- C / \alpha} . \label{eqn:diffexpo}
\end{align}
\end{proposition}

\medskip
We prove Proposition \ref{prop:expdiffEonoff} in the subsequent section. Theorem \ref{th:PN} follows directly from Proposition \ref{prop:expdiffEonoff} and \eqref{eqn:Nest}. $\Box$ \\
 
\subsection{Estimation of the difference $ \widehat{\Phi^{\rm off}} - \widehat{\Phi^{\rm on}}$} \label{section:diffequation}

We embark on the proof of Proposition \ref{prop:expdiffEonoff}. Recall from Proposition \ref{prop:rigorousexpansion} that $\widehat{\Phi^{\sigma, \alpha}}  \in L^{2,a}(\mathbb{R}) $ is well-defined, $ \| \widehat{\Phi^{\sigma,\alpha}} \|_{L^{2,a}(\mathbb{R}_Q)} \lesssim 1$, for $\alpha$ sufficiently small and satisfies equation \eqref{eqn:Phieqn}. Note that $\widehat{\Phi^{\sigma,\alpha}}$ is supported on $\mathcal{B}_{_{\alpha}} = [ - \frac{\pi}{\alpha}, \frac{\pi}{\alpha}]$, an interval which grows as $\alpha \downarrow 0$. We begin by proving a uniform decay bound for $\widehat{\Phi^{\sigma, \alpha}}$. 
\medskip
\begin{proposition} \label{prop:Phi} 
Let $0  < \alpha < \alpha_2$. Then there exist constants $C_1, C_2 > 0$ such that $\widehat{\Phi^{\sigma, \alpha}}$ satisfies $ \| e^{C_1 |Q|} \widehat{\Phi^{\sigma,\alpha}} \|_{L^{2,a}(\mathbb{R}_Q)} \leq C_2 \ \|  \widehat{\Phi^{\sigma,\alpha}} \|_{L^{2,a}(\mathbb{R}_Q)}$. 
\end{proposition}

\medskip
\noindent {\bf Proof of Proposition \ref{prop:Phi}:} We apply Lemma \ref{lemma:expogeneral} from the appendix to equation \eqref{eqn:Phieqn} for $\widehat{\Phi^{\sigma, \alpha}}$. To see that the hypotheses of the lemma are satisfied, it suffices to observe that
\begin{align}
M_{\alpha}( Q ) = \frac{4}{\alpha^2} \sin^2 \left( \frac{Q \alpha}{2} \right) \geq  \frac{4 }{  \pi^2} \ | Q |^2, \qquad Q \in \mathcal{B}_{_{\alpha}}.
\end{align}
and for $ m \in \{-1,0,1\}$ and $Q \in \mathcal{B}_{_{\alpha}},$  $
|Q| \leq |Q - 2 m \pi / \alpha|. $  $\Box$

\bigskip
Next, we derive the equation for $\widehat{\Phi^{^{\rm diff}}} = \widehat{\Phi^{\rm off}} - \widehat{\Phi^{\rm on}}$.

\begin{proposition}
\label{prop:ediffeqn}
Let $0 < \alpha < \alpha_0$. Then $\widehat{\Phi^{^{\rm diff}}}(q) = \widehat{\Phi^{\rm off}}(q) - \widehat{\Phi^{\rm on}}(q)$ solves the following \emph{linear} equation:
\begin{align}
& \left[ 1 + M_{\alpha}(Q) \right] \ \widehat{\Phi^{^{\rm diff}}}(q) - \chi_{_{\mathcal{B}_{_{\alpha}}}} (Q)  \frac{1}{ 4 \pi^2} \bigg( \widehat{\Phi^{\rm off}} *  \widehat{\Phi^{\rm off}} * \widehat{\Phi^{^{\rm diff}}}   (Q)  \nonumber \\
& + \widehat{\Phi^{\rm on}} *  \widehat{\Phi^{\rm off}} *  \widehat{\Phi^{^{\rm diff}}}  (Q) + \widehat{\Phi^{\rm on}} *    \widehat{\Phi^{\rm on}} * \widehat{\Phi^{^{\rm diff}}}  (Q) =  R_{_{\rm diff}} \left[ \widehat{\Phi^{\rm off }}, \widehat{\Phi^{\rm on }} \right]( Q ), \label{eqn:Ediff}
\end{align}
where the inhomogeneous right-hand side is given by
\begin{align}
 R_{_{\rm diff}} \left[ \widehat{\Phi^{\rm off }}, \widehat{\Phi^{\rm on }} \right]( Q ) = -
 \chi_{_{\mathcal{B}_{_{\alpha}}}}( Q ) \frac{1}{4 \pi^2} \sum_{m = \pm 1} \bigg(  & \ \widehat{\Phi^{\rm off}}  * \widehat{\Phi^{\rm off}} * \widehat{\Phi^{\rm off}}  ( Q - 2 m \pi / \alpha) \nonumber \\
& + \ \widehat{\Phi^{\rm on}} *  \widehat{\Phi^{\rm on}}  * \widehat{\Phi^{\rm on}}   ( Q - 2 m \pi / \alpha)  \bigg), \label{eqn:R3def}
\end{align}
with
\begin{align}
\left\|  R_{_{\rm diff}} \left[ \widehat{\Phi^{\rm off }}, \widehat{\Phi^{\rm on }} \right] \right\|_{L^{2,a}(\mathbb{R})}  \lesssim e^{- C \pi / \alpha}. \label{eqn:residdiffest}
\end{align}
\end{proposition}

\medskip

\noindent {\bf Proof of Proposition \ref{prop:ediffeqn} :} We subtract equation \eqref{eqn:Phieqn} for $\sigma = 0$ from the same equation for $\sigma = 1/2$. To estimate the $m = \pm 1$ terms in \eqref{eqn:R3def}, we apply Lemma \ref{lemma:expconvo}  along with Proposition \ref{prop:Phi}. This gives for $m = \pm 1$, 
\begin{align}
& \left\| \chi_{_{\mathcal{B}_{_{\alpha}}}} \ \widehat{\Phi^{\sigma}} * \widehat{\Phi^{\sigma}} * \widehat{\Phi^{\sigma}} (\cdot - 2 m \pi /\alpha) \right\|_{L^{2,a}(\mathbb{R})} \lesssim  e^{- C \pi / \alpha}.  
 \end{align}
which gives \eqref{eqn:residdiffest}. $\Box$ \\
\medskip

We now use Proposition \ref{prop:ediffeqn} to prove the exponential bound \eqref{eqn:diffexpo} on $ \widehat{\Phi^{^{\rm diff}}}$. We use an argument analogous to that used in the proof of Proposition \ref{prop:rigorousexpansion}. Here, we only summarize the argument since the details are quite familiar. Introduce
\begin{align}
\widehat{\Phi^{\rm diff}_{_{\rm lo}}}(q) \equiv  \chi_{_{\rm lo}}(q) \ \widehat{\Phi^{^{\rm diff}}}(q), \quad \quad {\rm and } \quad \quad \widehat{\Phi^{\rm diff}_{_{\rm hi}}}(q) \equiv  \chi_{_{\rm hi}}(q) \ \widehat{\Phi^{^{\rm diff}}}(q).
\end{align}
Estimation of $\widehat{\Phi^{\rm diff}_{_{\rm hi}}}$ gives
\begin{align}
\left\| \widehat{\Phi^{^{\rm diff}}_{_{\rm hi}}} \right\|_{L^{2,a}(\mathbb{R})}  \lesssim \alpha^{2 - 2r} \ e^{ - C / \alpha} + \ \alpha^{2 - 2r} \  \left\| \widehat{\Phi^{^{\rm diff}}_{_{\rm lo}}} \right\|_{L^{2,a}(\mathbb{R})} , \label{eqn:highestdiff}
\end{align}
$\widehat{\Phi^{^{\rm diff}}_{_{\rm lo}}} $ satisfies an inhomogeneous equation forced by  $\chi_{_{\rm lo}} \ R_{_{\rm diff}} \left[ \widehat{\Phi^{\rm off }}, \widehat{\Phi^{\rm on }} \right] $ which satisfies the exponential bound \eqref{eqn:residdiffest}. A simple bootstrap argument using \eqref{eqn:highestdiff} and the proved bounds on $\widehat{\Phi^{\rm off }}$ and $\widehat{\Phi^{\rm on }}$ gives
$ 
\left\| \widehat{\Phi^{^{\rm diff}}_{_{\rm lo}}} \right\|_{L^{2,a}(\mathbb{R})} \lesssim  e^{- C \pi / \alpha},
$ 
for $\alpha$ sufficiently small, which dominates $\widehat{\Phi^{^{\rm diff}}}$. This completes the proof of Proposition \ref{prop:expdiffEonoff}. $\Box$
\medskip

\section{Extension of our analysis to dimensions $d = 2, 3$}
\label{section:higherdim}
In Sections \ref{section:1dproof} through \ref{section:PNbarrier} we proved Theorems \ref{th:main} and \ref{th:PN}, concerning the bifurcation of discrete solitary waves and a bound on the PN-barrier in dimension $d=1$. In this section, we show how to adapt the the proof  to dimensions $d = 2$ and $d = 3$ . This will then prove Theorem \ref{th:maingeneral}. For many of the details, we refer to the proofs in one dimension and emphasize those aspects in which the spatial dimension, $d$, appears explicitly. In particular, the proof of Theorem \ref{th:PN} for $d = 2, 3$  concerning the PN-barrier bound follows Section \ref{section:PNbarrier} using the $d$-dimensional scalings of the discrete solitary waves in this section.

We begin with a generalization of Definition \ref{defn:onoff}, concerning the different centerings of discrete solitary standing waves. \\

 \begin{definition} \label{defn:sigmacentered}
Let ${G} = \{ G_n \}_{n \in \mathbb{Z}^d}$ be a solution to equation \eqref{eqn:tiDNLS2general} and let $\sigma \in \{ 0 , 1/2 \}^d$. We say that ${G}$ is $\sigma$- centered if it is symmetric about the point $ \sigma$ in space. That is, for each spatial component $k = 1, \dots, d$, we have $G_n = G_m^{(k)}$ where $m^{(k)} = (n_1, \dots, n_k  + 2 \sigma_k, \dots, n_d)^T \in \mathbb{Z}^d$. Note that this definition is consistent with its one-dimensional analogue given in Definition \ref{defn:onoff}.
\end{definition} \\

Applying the $d$-dimensional discrete Fourier transform  to DNLS, we obtain 
\begin{align}
  \widehat{DNLS}[\widehat{G^{\alpha}}](q) & \equiv \big[ \alpha^2 +  M(q) \big] \widehat{G^{\alpha}}(q) - \left(\frac{1}{2 \pi} \right)^{2d} \widehat{G^{\alpha}} *_{_1} \widehat{G^{\alpha}} *_{_1} \widehat{G^{\alpha}}(q) = 0, \\
 & \widehat{G^{\alpha}}(q + 2\pi e^{(k)}) = \widehat{G^{\alpha}}(q). 
\label{eqn:tiDNLS2realfouriergeneral}
\end{align}
Here, $e^{(k)}$ is the unit vector in the $k$th coordinate direction, $q_k=e^{(k)}\cdot q$ and 
\begin{align}
M(q) = 4 \sum_{k = 1}^d \sin^2(q_k/2). \label{eqn:Mdefgeneral}
\end{align}
Note that $\widehat{G^{\alpha}}$ is periodically tiled over $d$-cubes of volume $(2 \pi)^d$ in $\mathbb{R}^d$.
\medskip
\begin{proposition} \label{prop:sigmacentered}
If ${G} = \{ G_n \}_{n \in \mathbb{Z}^d}$ is real and $\sigma$-centered, then $\widehat{G}(q) = e^{- i q \cdot \sigma} \widehat{K}(q)$, where $\widehat{K}(q)$ is real and symmetric. Conversely, if $\widehat{G}(q) = e^{- i q \cdot \sigma} \widehat{K}(q)$, where $\widehat{K}(q)$ is real-valued and symmetric, then $\mathcal{F}^{-1}_{_D}[ \widehat{G} ]$ is real and $\sigma$-centered. This is consistent with the one-dimensional analogue given in Proposition \ref{prop:off-on}.
\end{proposition}

\medskip

We take $\sigma = \{ \sigma_k \}_{k = 1, \ldots , d}$ where $ \sigma_k \in \{ 0, 1/2 \}$ and seek $\widehat{G}(q)$ in the form
\begin{align}
\widehat{G^{\sigma}}(q)\ =\ e^{-i\sigma \cdot q}\ \widehat{K^{\sigma}}(q), \hspace{1cm} \widehat{K^{\sigma}}(q) = \widehat{K^{\sigma}}(-q),  \hspace{1cm} \ \overline{\widehat{K^{\sigma}}(q)}=\widehat{K^{\sigma}}(q),  \label{Gaqgeneral}
\end{align}
By Lemma \ref{lemma:producttransform}, the periodicity of  $\widehat{G}(q)$ and \eqref{Gaqgeneral}  we have
\begin{align}
      &\big[ \alpha^2 +  M(q) \big] \  \widehat{K^\sigma}(q) - \left(\frac{1}{2 \pi} \right)^{2d} \ \widehat{K^\sigma} *_{_1} \widehat{K^\sigma} *_{_1} \widehat{K^\sigma}(q) = 0, \quad q \in \mathbb{R}^d,  \label{Heqngeneral}\\
& \widehat{K^{\sigma}} \left( q+2\pi e^{(k)} \right) = e^{2\pi i\sigma \cdot e^{(k)}} \ \widehat{K^{\sigma}}(q), \quad k = 1, \dots, d. 
\label{Hpseudogeneral}\end{align}
The ``Bloch'' phase factor, $e^{2\pi i\sigma \cdot e^{(k)}}$,  in \eqref{Hpseudogeneral} is equal to $\pm1$.
\bigskip

\begin{lemma}\label{pisigma-pseudogeneral}
Let $A(q)$, defined on $\mathbb{R}^d$, be $2\pi\sigma-$ pseudo-periodic (condition \eqref{Hpseudogeneral}). Then,
 $A(q)$ is completely determined by its values on $\mathcal{B}=[-\pi,\pi]^d$ and has the representation:
 \begin{align}
A(q) =  \sum_{m \in \mathbb{Z}^d} \chi_{_{\mathcal{B}}}(q - 2 m \pi) A(q - 2  m \pi) e^{2 \pi i m \cdot \sigma}. \quad \Box \label{eqn:Atilinggeneral}
\end{align}
\end{lemma}

By Lemma \ref{pisigma-pseudogeneral} we may express $\widehat{K^{\sigma}}(q)$, for any $q \in \mathbb{R}^d$, explicitly in terms of its values on $q \in \mathcal{B}$.  In particular,  we set
$ 
\widehat{K^{\sigma}}(q) = \widehat{\phi^{\sigma}}(q) $ for $q \in \mathcal{B},
$ 
and extend $\widehat{\phi^{\sigma}}(q)$ to $\mathbb{R}^d$ to get:
\begin{align}
& \widehat{K^\sigma}(q) \equiv \sum_{m \in \mathbb{Z}^d} \chi_{_{\mathcal{B}}}(q - 2 m \pi) \  \widehat{\phi^{ \sigma}} (q - 2 m \pi) \ e^{2   \pi i m \cdot \sigma}, \qquad   {\rm and}\ \qquad \widehat{G^\sigma}(q) = e^{-i\sigma \cdot q} \ \widehat{K^\sigma}(q)  \label{eqn:ansatz3general}.
\end{align}
Note that $\widehat{\phi^{\sigma}}(q - 2 m \pi) = \chi_{_{\mathcal{B}}}(q - 2 m \pi) \ \widehat{\phi^{\sigma}}(q - 2 m \pi)$ is supported on $ \{ q: q \in 2 m \pi + \mathcal{B}   \}$. Therefore,
\begin{align}
& \chi_{_{\mathcal{B}}}(q)  \ \widehat{K^{\sigma}}(q) = \widehat{\phi^{\sigma}}(q), \qquad {\rm and} \qquad \chi_{_{\mathcal{B}}}(q) \ \widehat{G^{\sigma}}(q) = e^{- i q \cdot \sigma} \ \widehat{\phi^{\sigma}}(q). \label{eqn:decomp22general}
\end{align}
Equation \eqref{eqn:ansatz3general} encodes the required $2\pi-$ periodicity of
 $\widehat{G^\sigma}(q)$ on all $\mathbb{R}^d$, the $2\pi\sigma-$ pseudo-periodicity of $\widehat{K^\sigma}(q)$. Furthermore,
$ 
 \widehat{G^\sigma}(q)$ and $\widehat{K^\sigma}(q)$ are completely specified by $\widehat{\phi^\sigma}(q)$ for $q\in\mathcal{B}$.

 With a view toward obtaining an equation determining $\widehat{\phi^\sigma}(q)$ for $ q\in\mathcal{B}$, we require a lemma to simplify the convolution terms in \eqref{Heqngeneral}.  

\begin{lemma}
\label{lemma:decompgeneral}
Let  $\widehat{A}, \widehat{B}, \widehat{C}$ be bounded $2\pi\sigma-$ pseudo-periodic.
Furthermore, let $\widehat{C}$ have the form:
\begin{align}
\widehat{C}(q) \equiv \sum_{m \in \mathbb{Z}^d} e^{2 \pi i m \cdot \sigma} \chi_{_{\mathcal{B}}}(q - 2 m \pi) \ \widehat{C}(q - 2 m \pi).
\end{align}
Then
\begin{align}
\chi_{_{\mathcal{B}}}(q) \widehat{A} *_{_1} \widehat{B} *_{_1} \widehat{C} (q) = \chi_{_{\mathcal{B}}}(q) \sum_{m \in \{ -1, 0, 1 \}^d } e^{2 \pi i m \cdot \sigma} \ \widehat{A} *_{_1} \left[\ \widehat{B} *_{_1}  \left( \chi_{_{\mathcal{B}}} \widehat{C} \right)\right]  ( q - 2 m \pi). \label{eqn:decomp2general}
\end{align}
\end{lemma}
This lemma is a simple generalization of Lemma \ref{lemma:decomp}; the proof is nearly identical. 
Applying Lemma \ref{lemma:decompgeneral} and \eqref{eqn:ansatz3general}, we have:

\begin{proposition} \label{prop:phieqngeneral}
Equation \eqref{Heqngeneral} for $\widehat{K^{\sigma}}$ on $q \in \mathbb{R}^d$ is equivalent to the following equation for $\widehat{\phi^{\sigma}}$ on $q \in \mathcal{B}$: 
\begin{align}
  [\alpha^2 + M(q)] \ \widehat{\phi^{\sigma}}(q) & - \chi_{_{\mathcal{B}}}(q) \  \left( \frac{1}{2 \pi} \right)^{2d}
 \ \left( \widehat{\phi^{\sigma}} *  \widehat{\phi^{\sigma}} *  \widehat{\phi^{\sigma}}  \right) (q) \nonumber \\
 & - \chi_{_{\mathcal{B}}}(q) \  \left( \frac{1}{2 \pi} \right)^{2d} \sum_{ \substack{m \in \{-1,0,1\}^d \\ m \neq 0} } e^{2 \pi i m \cdot \sigma}  \left( \widehat{\phi^{\sigma}} *  \widehat{\phi^{\sigma}} *  \widehat{\phi^{\sigma}}  \right) (q - 2 m \pi) = 0,
\label{eqn:phieqngeneral}
\end{align}
where $M(q) \widehat{G}(q) = 4 \sum_{j = 1}^d 4 \sin^2 (q_j /2) \widehat{G}(q) = \widehat{ \delta^2 G }(q)$; see \eqref{eqn:Mdefgeneral}.
\end{proposition}

Note now that the convolutions are taken over $\mathbb{R}^d$.

\subsection{Rescaled equation for $\widehat{\phi^{\sigma}}$} \label{section:rescaledphigeneral}
In analogy with the one-dimensional case discussed in Section \ref{section:strategy}, we expect 
\begin{align}
\widehat{\phi^{\sigma}}(q) \sim \widetilde{\psi_{\alpha^2}}(q) =  \alpha^{1 - d} \ \widetilde{\psi_1} \left( \frac{q}{\alpha} \right), \quad \quad \alpha \ll 1, 
\end{align}
(see Proposition \ref{prop:Psi}). To anticipate this leading order behavior, we study \eqref{eqn:phieqngeneral} by a rescaling which will make explicit the connection, for $\alpha\downarrow0$, between
DNLS and the continuum (NLS) limit. With the goal of obtaining an asymptotic expansion for $\widehat{\phi^{\sigma}}$ as a functional of $\widetilde{\psi_1}$ in powers of $\alpha$, we introduce:
\begin{align}
& \textrm{\rm Rescaled momentum:} \quad \quad & Q \equiv q/\alpha, \qquad Q_k = q_k/\alpha\ \ \textrm{and therefore}\ \
Q\in \mathcal{B}_{\alpha} = [-\pi/\alpha,\pi/\alpha]^d,  \nonumber \\
& \textrm{\rm Rescaled projection:} \quad \quad  &  \chi_{_{\mathcal{B}_{_{\alpha}}}}(Q) \equiv \chi_{_{\mathcal{B}}}(Q \alpha) = \chi_{_{ \left[ - \frac{\pi}{\alpha}, \frac{\pi}{\alpha} \right]^d }}(Q), 
 \nonumber \\
& \textrm{\rm Rescaled wave:} \quad \quad  &  \widehat{\Phi^{\sigma}}(Q) \equiv \alpha^{d - 1} \ \widehat{\phi^{\sigma}}(Q \alpha) = \alpha^{d - 1} \  \widehat{\phi^{\sigma}}( q), \nonumber \\
& \textrm{\rm Rescaled symbol:} \quad \quad  & M_{\alpha}(Q) \equiv \frac{1}{\alpha^2} M(Q \alpha) =  \frac{4}{\alpha^2}  \sum_{k = 1}^d \sin^2 \left( \frac{ Q_k \alpha}{2} \right). \label{eqn:M2defgeneral}
\end{align}

The following proposition is a formulation of Proposition \ref{prop:phieqn}  in terms of functions of the rescaled quasi-momentum, $Q$:

\begin{proposition}
\label{prop:Phieqngeneral}
Equation \eqref{Heqn} for $\widehat{K^{\sigma}}(q)$ on $q \in \mathbb{R}$ is equivalent to the following equation for  $\widehat{\Phi^{\sigma}}(Q)=\chi_{_{\mathcal{B}_\alpha}}(q)\ \widehat{\Phi^{\sigma}}(Q)$, compactly supported on $\mathcal{B}_\alpha=[-\pi/\alpha,\pi/\alpha]$: 
\begin{align}
&\mathcal{D}^{\sigma,\alpha}[\widehat{\Phi^{\sigma}}](Q)\ \equiv\ [1 + M_\alpha(Q)] \ \widehat{\Phi^{\sigma}}(Q) - \frac{\chi_{_{\mathcal{B}_\alpha}}(Q)}{ (2 \pi)^d}  
\  \left(\ \widehat{\Phi^{\sigma}} *  \widehat{\Phi^{\sigma}} *  \widehat{\Phi^{\sigma}}\ \right) (Q) 
\ +\ R_1^{\sigma} [ \widehat{\Phi^{\sigma}} ] (Q) = 0\ ,
\label{eqn:Phieqngeneral}
\end{align}
where $R_1^{\sigma} [ \widehat{\Phi^{\sigma}} ]$ contains the $\pm 1$-sideband contributions:
\begin{align}
R_1^{\sigma} [ \widehat{\Phi^{\sigma}} ] (Q) &\equiv  -  \frac{\chi_{_{\mathcal{B}_{_{\alpha}}}}( Q)}{ (2 \pi)^d } \ \sum_{ \substack{ m \in \{-1, 0, 1\}^d \\ m \neq 0} } \ e^{2 m \pi i \sigma} \  \left(\ \widehat{\Phi^{\sigma}} *  \widehat{\Phi^{\sigma}} *  \widehat{\Phi^{\sigma}}\ \right)  ( Q - 2 m \pi / \alpha), \label{eqn:R1defgeneral}
\end{align}
and 
$M_\alpha(Q)=\ \frac{4}{\alpha^2} \sum_{j = 1}^d  \sin^2(\frac{\alpha Q_j}{2})$; see \eqref{eqn:M2defgeneral}. 
\end{proposition}

\noindent To prove  Proposition \ref{prop:Phieqngeneral} we need to re-express the convolutions in \eqref{eqn:phieqngeneral} in terms of $\widehat{\Phi^{\sigma}}(Q)$. 
For this we use the following lemma, which is proved by change of variables and generalizes Lemma \ref{lemma:convresc}. 
\begin{lemma}
\label{lemma:convrescgeneral}
Suppose that $\widehat{a}(q) = \widehat{A}(Q)$, $\widehat{b}(q) = \widehat{B}(Q)$, and $\widehat{c}(q) = \widehat{C}(Q)$, where $Q=q/\alpha$.  Then
\begin{align}
\left( \widehat{a} * \widehat{b} * \widehat{c} \right)  (q)  = \alpha^{2d} \ \left( \widehat{A} * \widehat{B} * \widehat{C} \right) (Q) .
\end{align}
\end{lemma}
 
Applying the rescalings \eqref{eqn:M2defgeneral} and Lemma \ref{lemma:convrescgeneral} to \eqref{eqn:phieqngeneral} and  then dividing by $\alpha^{3-d}$, we obtain \eqref{eqn:Phieqngeneral}.\ $\Box$ \\

%

We will again formally Taylor expand, for $\alpha \ll 1$ and $Q \in \mathbb{R}^d$ fixed, 
\begin{align}
M_{\alpha}(Q) = \frac{4}{\alpha^2} \sum_{k = 1}^d \sin^2 \left( \frac{Q_k \alpha}{2} \right) & = 2 \sum_{k = 1}^d \sum_{j  = 0}^{\infty} \frac{  \alpha^{2j}  \ (-1)^{j} \  |Q_k |^{2j + 2} }{ (2j + 2)!} \nonumber \\
& = \sum_{k = 1}^d \ |Q_k|^2 - \frac{\alpha^2 \ | Q _k |^4}{12} + \frac{ \alpha^4 \ |Q _k|^6}{360} + \mathcal{O} \left( \alpha^6 \ |Q_k |^8 \right) \nonumber \\
& = | Q|^2 - \sum_{k= 1}^d   \frac{\alpha^2 \ | Q _k |^4}{12} - \frac{ \alpha^4 \ |Q _k|^6}{360} + \mathcal{O} \left( \alpha^6 \ |Q_k |^8 \right).
\end{align}
Using truncations of the expansion of $M_\alpha(Q)$ we shall, for any $J=0,1,2,\dots$,  construct $\widehat{\Phi}^\sigma(Q)$ in the form of a finite expansion in power of $\alpha^{2j},\  j=0,\dots, J$, with an error term of 
which is of order $\alpha^{2J+2}$
 plus a corrector of higher order. For each $J$, the polynomial expansion in $\alpha^2$ is {\sl independent of $\sigma$}. The construction is summarized in the following:\\

\begin{proposition} \label{prop:rigorousexpansiongeneral}
Fix $J \geq 0$, $a > 1/2$, and $\sigma \in \{ 0, 1/2 \}^d $. Then there exist a constant $\alpha_0 = \alpha_0[a, J, \sigma] > 0$, and  $J$ mappings $F_j: L_{_{\rm even}}^{2,a}(\mathbb{R}^d ) \rightarrow L_{_{\rm even}}^{2,a}(\mathbb{R}^d ), \ j = 0, \dots, J$, and a unique, real-valued function $\widehat{E_J^{\alpha, \sigma}} \in L_{_{\rm even}}^{2,a}(\mathbb{R}^d )$ such that for all $0 < \alpha < \alpha_0$, 
\begin{align}
\widehat{\Phi^{\sigma}}( Q)\ =\ \chi_{_{\mathcal{B}_{_{\alpha}}}}(Q) \ \widetilde{\psi_1}(Q)\ +\ \sum_{j = 1}^J \ \alpha^{2j} \  \chi_{_{\mathcal{B}_{_{\alpha}}}}(Q) \ F_j \left[ \widetilde{\psi_1} \right] (Q)  + \widehat{E_J^{\alpha, \sigma}}(Q), \label{eqn:rigorousexpansiongeneral}
\end{align}
solves equation \eqref{eqn:Phieqngeneral}
with the error bound:
\begin{align}
\left\| \widehat{E_J^{\alpha, \sigma}} \right\|_{L^{2,a}(\mathbb{R}^d )} \lesssim \alpha^{2J + 2}
\end{align}
$F_j,\ j\ge1$, defined in Proposition \ref{prop:orderjgeneral} below,  is independent of $\sigma$ and $\alpha$, and 
$
 \widehat{E_J^{\alpha, \sigma}}( Q) = \chi_{_{\mathcal{B}_{_{\alpha}}}}(Q) \ \widehat{E_J^{\alpha, \sigma}}( Q).
$
\end{proposition}
 \medskip
 
\begin{remark}
By Proposition \ref{prop:rigorousexpansiongeneral}, since the polynomial expansion in $\alpha^2$ is completely determined by $\widetilde{\psi_1}(Q)$, 
$\widehat{\Phi^{\sigma}}(Q)$ is completely specified once we have constructed  $\widehat{E_J^{\alpha, \sigma}}(Q)$ for $Q \in \mathbb{R}^d$.
And, in turn by the rescalings $Q = q/\alpha$ and $\widehat{\Phi^{\sigma}}(Q) = \widehat{\phi^{\sigma}}(Q \alpha) = \widehat{\phi^{\sigma}}(q)$,  
$\widehat{G^\sigma}(q)$ and $\widehat{K^\sigma}(q)$ are completely specified by $\widehat{E_J^{\alpha, \sigma}}(q/\alpha)$ for $q \in \mathbb{R}^d$.
Therefore, Proposition \ref{prop:rigorousexpansiongeneral} completely characterizes $\widehat{G^{\sigma}}(q)$. \end{remark}
\medskip

The formal asymptotic analysis in Section \ref{section:formalexpansion} is easily generalized to dimension $d \geq 1$ and $L^{2,a}(\mathbb{R}^d)$ for $a > d/2$ using Propositions \ref{prop:Psi} and \ref{prop:Lplus}, and \eqref{eqn:algebra0}. We therefore construct and characterize $F_j \left[ \widetilde{\psi_1} \right]$ via the following proposition, which itself is a generalization of Proposition \ref{prop:orderj}. \\

\begin{proposition} \label{prop:orderjgeneral}
Let $ j \geq 1$. The equation for $F_j$ at order $\mathcal{O}(\alpha^{2j})$, independent of $\alpha$ and $\sigma$, is given by
\begin{flalign}
 & \mathcal{O}(\alpha^{2 j}) {\bf \ equation:} &  \widetilde{L_+} \ F_j(Q) & = 2 \sum_{l = 1}^d \sum_{k = 0}^{j-1} \frac{  (-1)^{k -  j + 1} \  | Q_l |^{2j - 2k + 2} \ F_k (Q) }{ (2j - 2k + 2)!}, & \nonumber \\
& & & + \frac{1}{( 2 \pi)^{2d}} \sum_{ \substack{ k + l + z = j \\ 0 \leq k, l, z < j  }} \ F_k * F_l * F_z (Q) \equiv H_j \left[F_0, \dots, F_{j-1} \right](Q) , & \label{eqn:orderjgeneral}
 \end{flalign}
 and has the unique solution 
 \begin{align}
 F_j & = \left( \widetilde{L_+} \right)^{-1} \bigg( H_j \left[ F_0 , \dots, F_{j-1} \right] \bigg) \in L_{_{\rm even}}^{2,a}(\mathbb{R}^d;dQ). 
 \end{align}
 Furthermore, $F_j$ is real-valued and $e^{C_j |Q| } \ F_j(Q) \in L^{2,a}(\mathbb{R}^d;dQ)$, where
 $C_j \equiv C_0 \left( \frac{1}{2} + \frac{1}{2^{j+1}} \right) \geq \frac{C_0}{2}$ and $C_0>0$ is as in \eqref{C0-def}.   \\
\end{proposition}

The proof closely follows that of Proposition \ref{prop:orderj}. The remainder of the proof of Proposition \ref{prop:rigorousexpansiongeneral} follows that of Proposition \ref{prop:rigorousexpansion} for the 1-d case given in sections \ref{section:rigorousexpansionproof} and \ref{section:rescalinglow0}.

\subsection{Completion of the proofs of Theorems \ref{th:maingeneral} and \ref{th:PN} for $d = 1,2,3$} \label{section:mainproofend}

 Above we solved for, $\alpha\mapsto\widehat{\Phi^{\sigma,\alpha}}(Q)$,  the discrete Fourier transform of the on- and off-site standing waves as a function of  the scaled variable $Q$, restricted to the scaled Brillouin zone, $\mathcal{B}_{_\alpha}=[-\pi/\alpha,\pi/\alpha]^d$. To complete the proof we use this to construct $\widehat{\phi^{\sigma,\alpha}}(q)$, defined on
  $\mathcal{B}=[-\pi,\pi]^d$. From \eqref{eqn:rigorousexpansiongeneral} and the scaling $ \widehat{\phi^{\sigma}}(q) = \alpha^{1 - d} \ \widehat{\Phi}(q/\alpha)$, we have
  \begin{align}
& \widehat{\phi^{\sigma}}(q) = \chi_{_{\mathcal{B}}}(q)  \widehat{\phi^{\sigma}}(q) = \sum_{j = 0}^{J} \ \alpha^{2j + 1 - d} \ \chi_{_{\mathcal{B}}}(q) \  F_j \left[ \widetilde{\psi_1} \right] \left( \frac{q}{\alpha} \right) + \alpha^{1  - d} \ \widehat{E_J^{\alpha, \sigma}} \left( \frac{q}{\alpha} \right), \nonumber \\
& \left\| F_j \left[ \widetilde{\psi_1} \right] \left( \frac{\cdot}{\alpha} \right) \right\|_{L^{2,a}(\mathbb{R}^d)} \lesssim \alpha^{d/2}, \qquad  \left\| \widehat{E_J^{\alpha, \sigma}} \left( \frac{\cdot}{\alpha} \right) \right\|_{L^{2,a}(\mathbb{R}^d)} \lesssim \alpha^{2J + 2 + d/2}\ .
\label{FjEJ-boundsgeneral}
\end{align}
The bounds \eqref{FjEJ-boundsgeneral} follow since $F_j\left[ \widetilde{\psi_1} \right](Q)$ and
$\widehat{E_J^{\alpha, \sigma}}(Q) $ have order one $L^{2,a}(\mathbb{R}^d_Q)$ norm, and 
using the general bound on $q\mapsto f(q/\alpha)$ in $L^{2,a}(\mathbb{R}^d_q)$:
\begin{align}
  \Big\| f\left( \frac{q}{\alpha} \right) \Big\|^2_{L^{2,a}(\mathbb{R}^d_q)} & = 
  \int_{\mathbb{R}^d} \ \left(1 + |q|^2 \right)^a \ \Big|f\left( \frac{q}{\alpha} \right)\Big|^2 \ dq \leq \int_{\mathbb{R}^d} \ \left(1 + \frac{|q|^2}{\alpha^2} \right)^a \ \Big|f\left( \frac{q}{\alpha} \right)\Big|^2 \ dq \nonumber \\
& = \alpha^d \ \int_{\mathbb{R}^d} \ \left(1 + |Q|^2  \right)^a \ \Big|f\left( Q \right)\Big|^2 \ dQ = \alpha^d \left\| f\right\|^2_{L^{2,a}(\mathbb{R}^d_Q)}. \label{eqn:generalrescl2ageneral}
\end{align}

Next,  the ($2\pi-$ periodic in $q$)  discrete Fourier transform of $\alpha\mapsto\{G_n^{\sigma,\alpha}\}_{n\in\mathbb{Z}^d}$ is 
 $\widehat{G^{\sigma,\alpha}}(q)=e^{-i\sigma \cdot q} \widehat{K^{\sigma,\alpha}}(q)$, where
  \begin{align}
\widehat{K^{\sigma,\alpha}}(q) = \sum_{m \in \mathbb{Z}^d} \ \chi_{_{\mathcal{B}}}(q - 2 m \pi) \  \widehat{\phi^{\sigma,\alpha}}(q - 2 m \pi) \ e^{2 \pi i m \cdot \sigma}, \label{eqn:Kdef2general}
\end{align} 
This implies the expansion on the Brillouin zone $q \in \mathcal{B} = [- \pi, \pi]^d$:
\begin{align}
\widehat{G^{\alpha, \sigma}}(q) = e^{- i q  \cdot \sigma}  \ \widehat{\phi^{\sigma}}(q) = e^{- i q \cdot  \sigma} \left(\sum_{j = 0}^{J} \ \alpha^{2j + 1 - d} \ \chi_{_{\mathcal{B}}}(q) \  F_j \left[ \widetilde{\psi_1} \right] \left( \frac{q}{\alpha} \right) + \alpha^{1 - d} \ \widehat{E_J^{\alpha, \sigma}} \left( \frac{q}{\alpha} \right) \right), \quad \quad q \in \mathcal{B} . \label{eqn:GonBgeneral}
\end{align}
Define  
\begin{align}
 \mathcal{G}_j[ \widetilde{\psi_1} ](n) = \alpha^{1 - d} \ \mathcal{F}_{_D}^{-1} \left[ \ F_j \left[ \widetilde{\psi_1} \right] \left( \frac{q}{\alpha} \right) \right]_n, \qquad {\rm and} \qquad \mathcal{E}^{\alpha, J, \sigma}_n \equiv \alpha^{1 - d} \ \mathcal{F}^{-1}_{_D} \left[ e^{- i q  \cdot \sigma} \widehat{E_J^{\alpha, \sigma}}  \left( \frac{q}{\alpha} \right) \right]_n. \label{eqn:transdefsgeneral}
\end{align}
Therefore, for any $\sigma$, in particular $\sigma = \{0, 1/2\} ^d$,
\begin{align}
& \mathcal{G}_j[ \widetilde{\psi_1} ](n - \sigma) = \mathcal{F}_{_D}^{-1} \left[ e^{- i q \cdot \sigma} \ F_j \left[ \widetilde{\psi_1} \right] \left( \frac{q}{\alpha} \right) \right]_n.
\end{align}
 Applying the inverse discrete Fourier transform \eqref{eqn:invDFT} to $\widehat{G^{\alpha, \sigma}}(q)$ in \eqref{eqn:GonBgeneral} gives the branches of vertex-, bond-, face-, and cell-centered discrete solitary waves \eqref{eqn:solutionsgeneral}. We use the Plancherel identity with the bounds \eqref{FjEJ-boundsgeneral} to get
 \begin{align}
& \left\|  \mathcal{G}_j [ \widetilde{\psi_1}] (n - \sigma) \right\|_{l^2(\mathbb{Z}^d_n)} = \frac{\alpha^{1 -d } }{(2 \pi)^2} \  \left\| e^{- i q \sigma} F_j \left[ \widetilde{\psi_1} \right] \left( \frac{q}{\alpha} \right) \right\|_{L^2(\mathcal{B}; dq)} \lesssim \alpha^{1 - d} \  \left\|  F_j \left[ \widetilde{\psi_1} \right] \left( \frac{q}{\alpha} \right) \right\|_{L^{2,a}(\mathbb{R}^d_q)} \lesssim \alpha^{1 - d/2}, \nonumber \\
& \left\|  \mathcal{E}^{\alpha, J, \sigma}  \right\|_{l^2(\mathbb{Z}^d)} = \frac{\alpha^{1-d}}{(2 \pi)^2} \ \left\| e^{- i q \cdot \sigma} E_J^{\alpha, \sigma} \left( \frac{q}{\alpha} \right) \right\|_{L^2(\mathcal{B};dq)} \lesssim \alpha^{1 - d} \left\| e^{- i q \cdot \sigma} E_J^{\alpha, \sigma} \left( \frac{q}{\alpha} \right) \right\|_{L^{2,a}(\mathbb{R}^d_q)}  \lesssim \alpha^{2J + 3 - d/2}. 
\end{align}
\medskip

\appendix

\section{Properties of the Discrete Fourier Transform}
\label{subsection:DFT}
In this section, we prove some general properties of the discrete Fourier transform as defined in (\ref{eqn:DFT}).

\begin{lemma}
\label{lemma:discderivtransform}
For any function $u = \{ u_n \}_{n \in \mathbb{Z}^d} \in l^1(\mathbb{Z}^d) \cap l^2(\mathbb{Z}^d)$,
\begin{align}
\widehat{(\delta_j u)}(q) = \left( e^{i q} - 1 \right) \widehat{u}(q),
\end{align}
and
\begin{align}
\widehat{(\delta^2 u) }(q) = \mathcal{F}_{_D}[ (\delta^2 u) ](q) = - 4 \sum_{j = 1}^{d} \sin^2(q_j/2) \widehat{u}(q),
\end{align}
where $q_j$ is the $j$th component of $q \in \mathbb{R}^d$.
\end{lemma}
\medskip

\begin{lemma}
\label{lemma:producttransform}
For any two functions $u, v \in l^1(\mathbb{Z}^d) \cap l^2(\mathbb{Z}^d)$, and with their product given by $u \cdot v = \{ u_n v_n \}_{n \in \mathbb{Z}^d}, $
\begin{align}
\mathcal{F}_{_D} [ u \cdot v ](q) = \widehat{u \cdot v}(q) = (2 \pi)^{-d} \ \left( \widehat{u} *_{_1} \widehat{v} \right) (q).
\end{align}
where the periodic convolution $*_{_1}$ is defined in \eqref{eqn:periodicconvolution}.
\end{lemma}

\medskip

\begin{lemma}
\label{lemma:exponentials}
Let  $\widehat{u}$ and $ \widehat{v}$ be \ $L^1_{_{\rm loc}}(\mathbb{R}^d)$ functions. Then if $C$ is any constant, we have
\begin{align}
\left( e^{i C (\cdot)} \widehat{u} \right) *_{_1} \left( e^{i C (\cdot)} \widehat{v} \right)(q) = e^{i C q} \left( \widehat{u} *_{_1} \widehat{v}\right)(q).
\end{align}
\end{lemma}

\medskip

\begin{lemma}
\label{lemma:commutativity}
(Commutativity) For $2 \pi$-periodic functions $\widehat{u}\ ,\ \widehat{v} \in L^1_{_{\rm loc}}(\mathbb{R}^d)$, we have
\begin{align}
\widehat{u} *_{_1} \widehat{v}(q) = \widehat{v} *_{_1} \widehat{u}(q).
\end{align}
\end{lemma}

\medskip

\begin{lemma}
\label{lemma:commutativity2}
For three functions $\widehat{u}, \widehat{v}, \widehat{w} \in L^1_{_{\rm loc}}(\mathcal{\mathbb{R}}^d)$, we have
\begin{align}
\widehat{u} *_{_1} \left[ \widehat{v} *_{_1} \widehat{w} \right] (q) = \widehat{v} *_{_1} \left[ \widehat{u} *_{_1} \widehat{w} \right](q).
\end{align}
Note that $\widehat{u}, \widehat{v}$, and $\widehat{w}$ need not be periodic.
\end{lemma}

\noindent {\bf Proof of Lemma \ref{lemma:commutativity2}:} By a simple application of Fubini's theorem, we have
\begin{align}
\widehat{u} *_{_1} \left[ \widehat{v} *_{_1} \widehat{w} \right](q) & = \int_{\mathcal{B}} \int_{\mathcal{B}} \widehat{u}(\xi) \widehat{v}(\zeta) \widehat{w}(q - \xi - \zeta) d\xi d\zeta \nonumber \\
& = \int_{\mathcal{B}} \int_{\mathcal{B}} \widehat{u}(\xi) \widehat{v}(\zeta) \widehat{w}(q - \xi - \zeta) d\zeta d\xi
= \widehat{v} *_{_1} \left[ \widehat{u} *_{_1} \widehat{w} \right](q).\ \ \Box
\end{align}
 \\

\begin{lemma}
\label{lemma:evenconvo}
Suppose that $\widehat{u}$ and $\widehat{v}$ are even functions. Then $\widehat{u} *_{_{\alpha}} \widehat{v}$ is also even. That is, for any $\tau_j = \pm 1$, $j = 1,\ \dots \,d$, we have $\widehat{u} *_{_{\alpha}} \widehat{v}(\tau_1 q_1, \ \dots \, \tau_d q_d) = \widehat{u} *_{_{\alpha}} \widehat{v}(q)$.
\end{lemma}
\medskip

We also state here an analogous result to Lemma \ref{lemma:evenconvo} for standard convolutions on the line.
\begin{corollary}
\label{cor:standardconvosym}
Suppose that $\widehat{u}$ and $\widehat{v}$ are even functions. Then $\widehat{u} * \widehat{v}$ is also even. That is, for any $\tau_j = \pm 1$, $j = 1,\ \dots \,d$, we have $\widehat{u} * \widehat{v}(\tau_1 \xi, \ \dots \, \tau_d q_d) = \widehat{u} * \widehat{v}(q)$.
\end{corollary}

\medskip

\section{Implicit Function Theorem} \label{IFT}
In this section, we state a variant of the implicit function theorem which is used in the proofs in Sections \ref{section:1dproof} and \ref{section:higherdim}. \\

\begin{theorem}
\label{th:IFT}
(Implicit Function Theorem) Assume the following hypotheses.
\begin{enumerate}
\item $X$, $Y$, and $Z$ are Banach spaces.

 \item The mapping
 \begin{align}
 & f:[0,1] \times X \times Y \rightarrow Z \nonumber \\
 & (\alpha, x , y ) \mapsto f(\alpha,x,y).
 \end{align}
 satisfies $f(0, 0, 0) = 0$ and is continuous at $(0,0,0)$.

\item For all $(\alpha, x) \in [0,1] \times X$, the mapping
\begin{align}
y \mapsto f(\alpha,x,y)
\end{align}
is Fr\'{e}chet differentiable which we denote $D_y f(\alpha, x,y): Y \rightarrow Z$. Furthermore, the mapping
 \begin{align}
(\alpha, x, y ) \mapsto  D_y f(\alpha, x,y)
 \end{align}
 is continuous at $(0, 0, 0)$.

 \item $D_y f(0, 0, 0)$ is an isomorphism of $Y$ onto $Z$.
\end{enumerate}
Then there exist $\alpha_0 , \delta, \kappa > 0$ such that for $(\alpha, x) \in [0,\alpha_0) \times B_{\delta}(0) $, there exists a unique map $y_*: [0,\alpha_0) \times B_{\delta}(0) \mapsto Y$ such that $y_*[\alpha, x] \quad \text{is well-defined on} \quad [0, \alpha_0) \times B_{\delta}(0)$ and
\begin{align}
& y_*[0,0] = 0, \hspace{4.15cm}
\| y_*[\alpha, x] \|_{Y} \leq \kappa,  & \\
& \underset{(\alpha,x) \rightarrow (0,0)}{\lim} y_*[\alpha, x] = y_*[0,0] = 0, \hspace{1cm} 
 f(\alpha, x,y_*[\alpha,x]) = 0. &
\end{align}
Suppose also that
\begin{itemize}
\item[5.] For all $(\alpha, y) \in [0,1] \times Y$, the mapping
\begin{align}
x \mapsto f(\alpha,x,y)
\end{align}
Fr\'{e}chet differentiable which we denote $D_x f(\alpha, x,y): Y \rightarrow Z$. Furthermore, the mapping
 \begin{align}
(\alpha, x, y ) \mapsto  D_x f(\alpha, x,y)
 \end{align}
 is continuous at $(0, 0, 0)$.
\end{itemize}
Then $D_x y_*[\alpha,x]: X \rightarrow Y $ exists and is continuous.
\end{theorem}
\medskip
\begin{remark}
The proof below follows that found in \cite{N:01}. However, we include the parameter $\alpha > 0$ explicitly and require only that $f(\alpha,x,z)$ and $D_x f(\alpha,x,y)$ be continuous at the origin.  \\
\end{remark}

\noindent {\bf Proof of Theorem \ref{th:IFT}:} Let $\mathcal{L} \equiv D_y f(0,0,0)$. Observe that $ f(\alpha, x,y) = 0$ is equivalent to the equation
\begin{align}
& \mathcal{L} y = \mathcal{L} y - f(\alpha, x,y) \equiv \mathcal{K}_1(\alpha, x,y), \\
& \mathcal{K}_1 : [0,1] \times X \times Y \rightarrow Z.
\end{align}
Since $\mathcal{L}: Y \rightarrow Z$ is an isomorphism, it must have a bounded inverse $\mathcal{L}^{-1}: Z \rightarrow Y$ by the bounded inverse theorem, such that we may define
\begin{align}
& y = y - \mathcal{L}^{-1} f(\alpha,x,y) = \mathcal{L}^{-1} \mathcal{K}_1(\alpha, x,y) \equiv \mathcal{K}_2(\alpha ,x,y), \\
& \mathcal{K}_2 : [0,1] \times X \times Y \rightarrow Y.
\end{align}
Note that
\begin{align}
f(0,0,0) = 0 \quad  \Longleftrightarrow \quad \mathcal{K}_2(0,0,0) = 0.
\end{align}
We seek a fixed point $y_*[\alpha,x]$ of $\mathcal{K}_2(\alpha, x,y)$. In particular, we want to show that
 \begin{align}
 & y \in  B_{\kappa}(0)  \mapsto \mathcal{K}_2(\alpha, x , y ) \in  B_{\kappa}(0) , 
 \end{align}
 is a contraction which maps the ball $B_{\kappa}(0)$ to itself for some $\kappa > 0$. 
 
 First we show that on a subset of $[0,1] \times X \times Y$ which is restricted sufficiently close to the origin, $G_2$ is contracting. That is, we may choose $(\alpha, x,y_1), (\alpha, x,y_2) \in [0,1] \times X \times Y$ small enough such that we have
 \begin{align}
 \| \mathcal{K}_2(\alpha , x,y_1) - \mathcal{K}_2(\alpha,x,y_2) \|_Y \leq C \| y_1 - y_2 \|_Y, \quad \quad  C < 1.
 \end{align}
 Observe that
\begin{align}
\mathcal{K}_1(\alpha , x,y_1) - \mathcal{K}_1(\alpha,x,y_2) = \mathcal{L} y_1 - \mathcal{L} y_2 - f(\alpha, x,y_1) + f(\alpha, x,y_2) \nonumber \\
= \mathcal{L} (y_1 - y_2) - \int_0^1 \left[ D_y f \big(\alpha, x, t y_1 + (1 - t) y_2 \big) dt \right] (y_1 - y_2) \nonumber \\
= \int_0^1 \left[ \mathcal{L} -  D_y f \big(\alpha, x, t y_1 + (1 - t) y_2 \big) dt \right] (y_1 - y_2) .
\end{align}
Since $D_y f(\alpha, x,y)$ is continuous at $(0,0,0)$ and since $\mathcal{L} = D_y f(0, 0,0)$, for any $\epsilon > 0$ we may choose some $\alpha_0, \delta, \kappa > 0 $ such that for $x \in B_{\delta}(0)$ and $y_1, y_2 \in B_{\kappa}(0)$, we have
\begin{align}
& \| \mathcal{L} -  D_y f \big(\alpha, x, t y_1 + (1 - t) y_2 \big) \|_{Y \rightarrow Z} \leq \epsilon, \\
  \Longrightarrow & \| \mathcal{K}_2(\alpha, x,y_1) - \mathcal{K}_2(\alpha, x,y_2) \|_Y \leq \| \mathcal{L}^{-1} \|_{Z \rightarrow Y} \ \| \mathcal{K}_1(\alpha, x,y_1) - \mathcal{K}_1(\alpha, x,y_2) \|_Z  \nonumber \\
& \quad \quad \quad \quad \quad  \quad \quad \quad \quad \quad   \leq
 \ \epsilon \ \| \mathcal{L}^{-1} \|_{Z \rightarrow Y} \| y_1 - y_2 \|_Y.
\end{align}
Thus, we choose $ 
\epsilon \leq \frac{1}{2} \left( \| \mathcal{L}^{-1} \|_{Z \rightarrow Y} \right)^{-1},
$
such that
$
\| \mathcal{K}_2(\alpha, x,y_1) - \mathcal{K}_2(\alpha, x,y_2) \|_Y \leq \frac{1}{2}  \| y_1 - y_2 \|_Y.  
$

Next, fix $\alpha \in [0, \alpha_0),$ $x \in B_{\delta}(0)$. We seek to show that for a small enough choice of $\alpha_0, \delta,$ and $\kappa$, the map $y \mapsto G_2(\alpha,x,y)$ sends the ball $B_{\kappa}(0)$ into itself, such that
$
\| y \|_Y \leq \kappa$ and therefore $ \| G_2(\alpha,x,y) \|_Y \leq \kappa.
$
By the continuity of $f$ and thus $\mathcal{K}_2$ at $(0,0,0)$, we may choose $\alpha_0, \delta$ small enough that
\begin{align}
\| \mathcal{K}_2(\alpha,x,0) - \mathcal{K}_2(0,0,0) \|_Y \leq \frac{1}{2} \kappa. \label{eqn:contraction2}
\end{align}
Therefore,
\begin{align}
\| \mathcal{K}_2(\alpha,x,y) \|_Y & \leq \| \mathcal{K}_2(\alpha,x,y) - \mathcal{K}_2(\alpha,x,0) \|_Y + \| \mathcal{K}_2(\alpha,x,0) \|_Y \nonumber \\
& = \| \mathcal{K}_2(\alpha,x,y) - \mathcal{K}_2(\alpha,x,0) \|_Y + \| \mathcal{K}_2(\alpha,x,0) - \mathcal{K}_2(0,0,0) \|_Y \nonumber \\
& \leq \frac{1}{2} \| y \|_Y  + \frac{1}{2} \kappa \leq \kappa. \label{eqn:contraction2}
\end{align}
Thus,  $\mathcal{K}_2: B_{\kappa}(0) \rightarrow B_{\kappa}(0)$ is a contraction for $\alpha \in [0, \alpha_0)$, $x \in B_{\delta}(0)$, and the Banach fixed-point theorem admits a unique fixed point $ y = y_*[\alpha,x]$ such that $y_*[0,0] = 0$ and
\begin{align}
\mathcal{K}_2(\alpha,x,y_*[\alpha,x]) = y_*[\alpha, x] \quad \Longleftrightarrow \quad f(\alpha, x,u[\alpha,x]) = 0.
\end{align}
Furthermore, by the contraction estimate $\| \mathcal{K}_2(\alpha, x,y_1) - \mathcal{K}_2(\alpha, x,y_2) \|_Y \leq \frac{1}{2}  \| y_1 - y_2 \|_Y $, 
\begin{align}
  \| y_*[\alpha, x] \|_Y & = \| \mathcal{K}_2(\alpha,x,y_*[\alpha,x]) \|_Y \nonumber \\
&  \leq \| \mathcal{K}_2(\alpha,x,y_*[\alpha,x]) -  \mathcal{K}_2(\alpha,x,0) \|_Y + \| \mathcal{K}_2(\alpha,x,0)  \|_Y \nonumber \\
&  \leq \frac{1}{2}  \| y_*[\alpha,x]  \|_Y + \| \mathcal{K}_2(\alpha,x,0)  \|_Y,  \nonumber \\
\Longleftrightarrow \quad & \| y_*[\alpha, x] \|_Y \leq 2 \| \mathcal{K}_2(\alpha,x,0)  \|_Y,
\end{align}
such that the continuity of $\mathcal{K}_2$ at the origin implies that
\begin{align}
\underset{(\alpha,x) \rightarrow (0,0)}{\lim} y_*[\alpha, x] = y_*[0,0] = \underset{(\alpha,x) \rightarrow (0,0)}{\lim} G(\alpha, x, 0) = 0. \qquad \qquad \Box
\end{align}

\section{Exponential decay of solitary waves in Fourier (momentum) space} \label{section:expogeneral} 
In this section, we prove a general lemma, Lemma \ref{lemma:expogeneral}, which is used to establish an exponential decay property for both the continuum NLS and DNLS solitary waves in Fourier space under a certain scaling.

We prove the exponential decay of solutions to an equation of the form,
\begin{align}
\left[ 1 + M(Q) \right] \ \Phi(Q) +  \chi_{_{A}}(q)  \ \sum_{k = 1}^K \     C_k \ \Phi * \Phi * \Phi (Q + \tau_k ) = 0.   \label{eqn:generaldecayeqn2}
\end{align}
where 
$K \geq 1$ is an integer, and $C_k \in \mathbb{R}, \ \tau_k \in \mathbb{R}^d$ are constants for $1 \leq k \leq K$, $A \subset \mathbb{R}^d$ and $M(Q)$ is a continuous function which satisfies, for some constant $D_M > 0$, 
\begin{align}
& M(Q) \geq D_M |Q|^{2},  \qquad {\rm and} \qquad  |Q| \leq |Q + \tau_k |, \qquad {\rm for} \qquad Q \in A. 
\end{align}
We shall apply our analysis of \eqref{eqn:generaldecayeqn2} to the following two examples:  \\
\begin{enumerate}

\item The Fourier transform of continuum NLS for $d = 1,2,3$ and frequency $\omega = -1$, \eqref{Psi-eqn} :
\begin{align}
 \left[ 1 + |Q|^2 \right] \  \widetilde{\psi_1}(Q) - \frac{1}{4 \pi^2} \widetilde{\psi_1} * \widetilde{\psi_1} * \widetilde{\psi_1}(Q)  = 0 , \qquad Q \in \mathbb{R}^d. \label{eqn:NLS1}
\end{align}
Here, $K = 1$, $C_1 \equiv -1/4 \pi^2$, $\tau_1 \equiv 0$, $A \equiv \mathbb{R}^d$, and $M(Q) \equiv |Q|^2$. 

\item The rescaled Fourier transform of DNLS localized to the Brillouin zone for $d = 1, 2, 3$ \eqref{eqn:Phieqngeneral} (see also \eqref{eqn:Phieqn} when $d = 1$):
\begin{align}
  \left[ 1 + M_{\alpha}(Q) \right] \  \widehat{\Phi}(Q) - \frac{\chi_{_{\mathcal{B}_{_{\alpha}}}}(Q)}{4 \pi^2} \sum_{m \in \{-1, 0, 1 \}^d} e^{2 \pi i \sigma \cdot m} \  \widehat{\Phi} * \widehat{\Phi} * \widehat{\Phi}(Q - 2 m \pi / \alpha) = 0 , \qquad Q \in \mathbb{R}^d. \label{eqn:DNLS1}
\end{align}
Here, $K = 3^d$, $C_k \equiv (-1/4 \pi^2) e^{2 \pi i \sigma \cdot m_k}$ for $m_k \in \{-1, 0, 1 \}^d$,  $\tau_k \equiv 2 m_k \pi / \alpha$, $A \equiv \mathcal{B}_{_{\alpha}}$, and $M(Q) \equiv (4/\alpha^2) \sin^2(Q \alpha/2)$. 
\end{enumerate}

\bigskip
Below we state and prove the weighted $L^{2,a}(\mathbb{R}^d)$ exponential decay estimate, Lemma \ref{lemma:expogeneral}, for solutions of \eqref{eqn:generaldecayeqn2}. The application of Lemma \ref{lemma:expogeneral} to equation \eqref{eqn:NLS1} is in Proposition \ref{prop:Psi} and to equation \eqref{eqn:DNLS1} in Proposition \ref{prop:Phi}. 

Recall from \eqref{eqn:algebra0} that for $a > d/2$, $L^{2,a}(\mathbb{R}^d)$ forms an algebra with the convolution operation such that for some constant $D_a > 0$, 
\begin{align}
\left\| \widehat{f}_1 * \widehat{f}_2 \right\|_{L^{2,a}(\mathbb{R}^d)} \leq D_a \left\| \widehat{f}_1  \right\|_{L^{2,a}(\mathbb{R}^d)} \ \left\| \widehat{f}_2 \right\|_{L^{2,a}(\mathbb{R}^d)} 
\end{align}
We follow an approach similar to that given in \cite{FL:10} and motivated by the approach to proving exponential decay estimates in \cite{BL:97}. In particular, we make use of the following identity.  \\

\begin{lemma} \label{lemma:abel} (Abel's identity \cite{R:68}) For any constants $c_1, c_2 \neq 0$, we have
\begin{align}
\sum_{x = 0}^{y}  \left( \begin{array}{l} y \\ x \end{array} \right) (x + c_1)^{x - 1} \ (y - x + c_2)^{y - x  - 1} = \frac{c_1 + c_2}{c_1 c_2} (y + c_1 + c_2)^{y -1}. 
\end{align}
 
\end{lemma}

We now state the general lemma which establishes the exponential decay of solutions to certain solitary wave equations in the Fourier (momentum) variable. \\

\begin{lemma} \label{lemma:expogeneral}
Let $a > d/2$, $K \geq 1$ be an integer, and $C_k \in \mathbb{R}, \ \tau_k \in \mathbb{R}^d$ be constants for $1 \leq k \leq K$. Let $A \subset \mathbb{R}^d$ and $M(q)$ be a continuous function which satisfy, for some constant $D_M > 0$, 
\begin{align}
& M(Q) \geq D_M |Q|^{2},  \qquad {\rm and} \qquad  |Q| \leq |Q + \tau_k |, \qquad {\rm for} \qquad Q \in A. 
\end{align}
Let $\Phi \in L^{2,a}(\mathbb{R}^d)$ be the solution to \eqref{eqn:generaldecayeqn2}. Then there exists a constant $\mu = \mu \left( \| \Phi \|_{L^{2,a}(\mathbb{R}^d)} \right) > 0$ such that
\begin{align}
\left\| e^{\mu | Q |} \ \Phi(Q) \right\|_{L^{2,a}(\mathbb{R}^d_Q)} \lesssim \left\|  \Phi \right\|_{L^{2,a}(\mathbb{R}^d)} . 
\end{align}
\end{lemma}

\noindent {\bf Proof of Lemma \ref{lemma:expogeneral}:} Our strategy is to estimate the moments $|Q|^j \ \Phi(Q)$, $j \geq 0$. Observe that by Taylor expansion, for any $\mu > 0$, 
\begin{align}
& \left\| e^{\mu |Q|} \ \Phi(Q) \right\|_{L^{2,a}(\mathbb{R}_Q^d)} = \left\| \sum_{j = 0}^{\infty} \frac{\mu^j \ |Q|^j}{j!} \ \Phi(Q) \right\|_{L^{2,a}(\mathbb{R}_Q^d)} \leq \sum_{j = 0}^{\infty} \frac{\mu^j}{j!} \  \left\| \ | Q |^j \ \Phi(Q) \ \right\|_{L^{2,a}(\mathbb{R}_Q^d)} . \label{eqn:exposeriesratio0}
\end{align}
We then have the following proposition. \\

\begin{proposition} \label{prop:momentsgeneral}
There exists a constant $b = b \left( \left\|  \Phi \right\|_{L^{2,a}(\mathbb{R}^d)} \right) >  0$ such that $\Phi \in L^{2,a}(\mathbb{R}^d)$ satisfies the estimates
\begin{align}
\left\| \ | Q |^j \ \Phi(Q) \ \right\|_{L^{2,a}(\mathbb{R}_Q^d)} \leq b^j \ (2j + 1)^{j - 1} \left\| \Phi \right\|_{L^{2,a}(\mathbb{R}^d)} , \quad j \in \mathbb{N}, \quad j \geq 0. \label{eqn:indhyp}
\end{align}
\end{proposition}

\noindent The proof of Proposition \ref{prop:momentsgeneral} is found below. By Proposition \ref{prop:momentsgeneral} and \eqref{eqn:exposeriesratio0}, for any $\mu > 0$, 
\begin{align}
& \left\| e^{\mu |Q|} \ \Phi(Q) \right\|_{L^{2,a}(\mathbb{R}_Q^d)} \leq  \ \left\| \Phi \right\|_{L^{2,a}(\mathbb{R}^d)}  \  \sum_{j = 0}^{\infty} A_j, \label{eqn:exposeriesratio}
\end{align}
where
$
A_j \equiv \frac{\mu^j}{j!} \ b ^j \ (2j + 1)^{j - 1} . 
$
and $\mu$ is to be determined so that the sum converges. 
The ratio test gives
\begin{align}
\underset{ j \to \infty}{\lim} \ \frac{A_{j+1}}{A_j} = \underset{ j \to \infty}{\lim} \ \frac{b  \ \mu  \ (2j + 3)^j}{ (j + 1) \ (2j + 1)^{j-1}} = 2 b \mu e. 
\end{align}
Thus, taking $\mu < 1/2 b e$ makes \eqref{eqn:exposeriesratio} convergent, completing the proof of Lemma \ref{lemma:expogeneral}. $\Box$ \\

\noindent {\bf Proof of Proposition \ref{prop:momentsgeneral}:}
Rewrite equation \eqref{eqn:generaldecayeqn2} as 
\begin{align}
\Phi(Q) = \frac{ \chi_{_{A}}(Q) }{1 + M(Q)} \  \sum_{k = 1}^K    \ C_k \ \Phi * \Phi * \Phi (Q + \tau_k )  .   \label{eqn:generaldecayeqn3}
\end{align}
We prove \eqref{eqn:indhyp} by induction. Define
\begin{align}
W(Q) \equiv \frac{1}{(2 \pi)^2} \ \Phi * \Phi(Q). \label{eqn:Wdef}
\end{align}
Let $m \geq 1$ and multiply equation \eqref{eqn:generaldecayeqn3} by $| Q |^m $ to get
\begin{align}
|Q|^m \ \Phi(Q) = \frac{ \chi_{_{A}}(Q) \ |Q|^m }{1 + M(Q)} \  \sum_{k = 1}^K    \ C_k \ W * \Phi (Q + \tau_k ) . \label{eqn:expo1}
\end{align}

Note that \eqref{eqn:indhyp} holds trivially for $j = 0$. We assume for all $0 \leq j \leq m - 1$ that there exists a constant $b > 0$ such that \eqref{eqn:indhyp} holds; this is the inductive hypothesis. 
We will prove that \eqref{eqn:indhyp} holds for $ j = m$. 

Observe that by the hypotheses of the lemma,
\begin{align}
 &  \frac{ \chi_{_{A}}(Q) \ |Q|^m}{1 + M(Q)} \leq \frac{ \chi_{_{A}}(Q) \  |Q|^m}{ 1 + D_M |Q|^{2} } = |Q|^{m - 1} \  \frac{ \chi_{_{A}}(Q) \  |Q|}{ 1 + D_M |Q|^{2} } \leq |Q|^{m -1} \ \frac{\chi_{_{A}}(Q) }{2 (D_M)^{1/2}}, \label{eqn:expo2} \\
{\rm and} \qquad & 
  \chi_{_{A}}(Q) \ |Q|^{m - 1}  \leq |Q + \tau_k |^{m-1}. \label{eqn:expo3}
\end{align}
Furthermore, the triangle inequality gives $
|Q|^{m-1} \leq \sum_{l = 0}^{m - 1} \left( \begin{array}{c} m-1 \\ l \end{array} \right) | Q - \xi|^l \ | \xi|^{m - l - 1}. $
Combining this with \eqref{eqn:expo1}, \eqref{eqn:expo2}, and \eqref{eqn:expo3}, we have
\begin{align}
|Q|^m \ | \Phi(Q) | & \leq  \frac{1}{2 (D_M)^{1/2}} \   \sum_{k = 1}^K    \ C_k  \ |Q + \tau_k |^{m - 1} \ \Big| \int_{\mathbb{R}^d} W(Q - \xi)   \Phi (\xi + \tau_k ) d\xi  \Big| \nonumber \\
& \leq  \frac{1}{2 (D_M)^{1/2}} \ \sum_{k = 1}^K    \ C_k \ \sum_{l = 0}^{m - 1}   \left( \begin{array}{c} m-1 \\ l \end{array} \right) \ \left( | \xi|^l \ |W| \right) *^{\xi} \left( |\xi|^{m - l - 1} \ |\Phi| \right) (Q + \tau_k ), 
\end{align}
 and therefore by the inductive hypothesis \eqref{eqn:indhyp} (which holds for $0 \leq m - l - 1 \leq m - 1$), we have {\small{
 \begin{align}
 & \left\| \ |Q|^m \   \Phi(Q)    \ \right\|_{L^{2,a}(\mathbb{R}_Q^d)} \nonumber \\
 & \leq  \frac{K \ D_a}{2 (D_M)^{1/2}} \ \underset{ k = 1, \dots, K} {\max} \{ C_k \} \  \sum_{l  = 0}^{m -1} \left( \begin{array}{c} m-1 \\ l \end{array} \right) \ \left\| | \xi|^l \ W( \xi) \ \right\|_{L^{2,a}(\mathbb{R}_{\xi}^d)} \ \left\| | \xi|^{m - l - 1} \ \Phi (\xi) \  \right\|_{L^{2,a}(\mathbb{R}_{\xi}^d)} \nonumber \\
 &  \leq  \frac{K \ D_a}{2 (D_M)^{1/2}} \ \underset{ k = 1, \dots, K} {\max} \{ C_k \} \  \sum_{l  = 0}^{m -1} \left( \begin{array}{c} m-1 \\ l \end{array} \right) \ b^{m - l - 1} \  [2 (m - l - 1) + 1]^{m - l - 2} \nonumber \\
 & \hspace{5cm} \cdot  \left\| | \xi|^l \ W( \xi) \ \right\|_{L^{2,a}(\mathbb{R}_{\xi}^d)}  \ \left\| \Phi \right\|_{L^{2,a}(\mathbb{R}^d)} . \label{eqn:fullbd}
 \end{align}
}}

Next, we must estimate $| \xi|^l \ W( \xi)$. Equation \eqref{eqn:Wdef} and the triangle inequality give
\begin{align}
& |  \xi |^l \ | W(\xi) | \leq \frac{1}{(2 \pi)^2} \ \sum_{k = 0}^l \left( \begin{array}{c} l \\ k \end{array} \right) \ \left( |\eta|^k | \Phi | \right) *^{\eta} \left( |\eta|^{l - k} \  | \Phi | \right) (\xi),
\end{align}
which in turn gives, by the inductive hypothesis \eqref{eqn:indhyp} (for $0 \leq k \leq l \leq m - 1$ and $0 \leq l - k \leq m - 1$) and Abel's identity \ref{lemma:abel},  {\small{
\begin{align}
 \left\|  |  \xi |^l \   W( \xi )  \right\|_{L^{2,a}(\mathbb{R}_{\xi}^d)}  
& \leq \frac{D_a}{(2 \pi)^2} \ \sum_{k = 0}^l \left( \begin{array}{c} l \\ k \end{array} \right) \ \left\| |\eta|^k \ \Phi(\eta)  \right\|_{L^{2,a}(\mathbb{R}_{\eta}^d)} \ \left\| |\eta|^{l - k} \ \Phi(\eta)  \right\|_{L^{2,a}(\mathbb{R}_{\eta}^d)} \nonumber \\
& \leq \frac{D_a}{(2 \pi)^2} \ \sum_{k = 0}^l \left( \begin{array}{c} l \\ k \end{array} \right) \ b^k \ b^{l - k} \ (2k + 1)^{k - 1} \ [ 2 (l - k) + 1 ]^{l - k - 1} \ \left\| \Phi \right\|^2_{L^{2,a}(\mathbb{R}^d)} \nonumber \\
& = \frac{ D_a \ \left\| \Phi \right\|^2_{L^{2,a}(\mathbb{R}^d)} }{(2 \pi)^2} \ b^l \ \sum_{k = 0}^l \left( \begin{array}{c} l \\ k \end{array} \right) \ (2k + 1)^{k - 1} \ [ 2 (l - k) + 1 ]^{l - k - 1} \nonumber \\
& = \frac{ D_a \ \left\| \Phi \right\|^2_{L^{2,a}(\mathbb{R}^d)} }{(2 \pi)^2} \ b^l \ 2^{l - 2} \sum_{k = 0}^l \left( \begin{array}{c} l \\ k \end{array} \right) \ (k + 1/2)^{k - 1} \ [   l - k + 1/2 ]^{l - k - 1} \nonumber \\
& =   \frac{ D_a \  \left\| \Phi \right\|^2_{L^{2,a}(\mathbb{R}^d)} }{(2 \pi)^2} \ b^l \ 2^l \ (l + 1)^{l - 1} =  \frac{ 2 \ D_a \ b^l \ \left\| \Phi \right\|^2_{L^{2,a}(\mathbb{R}^d)} }{(2 \pi)^2}  \ ( 2 l + 2 )^{l - 1}  . \label{eqn:Wbd}
\end{align}
}}

Combining \eqref{eqn:Wbd} with \eqref{eqn:fullbd}, along with Abel's identity again, gives {\small{
\begin{align}
 \left\| \ |Q|^m \   \Phi(Q)    \ \right\|_{L^{2,a}(\mathbb{R}_Q^d)} & \leq \frac{ K \ D_a^2}{(2 \pi)^2 (D_M)^{1/2}}  \ \underset{ k = 1, \dots, K} {\max} \{ C_k \} \ \left\| \Phi \right\|^3_{L^{2,a}(\mathbb{R}^d)}    \ \nonumber \\
 & \hspace{3.5cm} \cdot  \sum_{l  = 0}^{m -1} \left( \begin{array}{c} m-1 \\ l \end{array} \right)  b^{m - l - 1} \ b^l  \  (2l + 2)^{l -1} \  [2 (m - l - 1) + 1]^{m - l - 2}  \nonumber \\
& = \frac{K \ D_a^2}{(2 \pi)^2 (D_M)^{1/2}}  \ \underset{ k = 1, \dots, K} {\max} \{ C_k \} \ \left\| \Phi \right\|^3_{L^{2,a}(\mathbb{R}^d)}  \ b^{m - 1}  \ 2^{m - 3} \ \nonumber \\
 & \hspace{3.5cm} \cdot  \sum_{l  = 0}^{m -1} \left( \begin{array}{c} m-1 \\ l \end{array} \right) \ ( l + 1 )^{l -1} \  [  (m - l - 1) + 1 /2]^{m - l - 2}  \nonumber \\
 & = \frac{ K \ D_a^2}{(2 \pi)^2 (D_M)^{1/2}}  \ \underset{ k = 1, \dots, K} {\max} \{ C_k \} \ \left\| \Phi \right\|^3_{L^{2,a}(\mathbb{R}^d)}  \ b^{m - 1}  \ 2^{m - 3} \ 3 \ (m - 1 + 3/2)^{m - 2} \nonumber \\
 &  = \frac{3 K \ D_a^2}{8 \pi^2 (D_M)^{1/2}}  \ \underset{ k = 1, \dots, K} {\max} \{ C_k \} \ \left\| \Phi \right\|^3_{L^{2,a}(\mathbb{R}^d)}  \ b^{m - 1}  \ (2m + 1)^{m - 2}. 
\end{align}
}}To complete the proof that \eqref{eqn:indhyp} is satisfied for $j = m \geq 1$, we need to show that this quantity is bounded by
$ 
  b^m (2m + 1)^{m-1} \  \left\| \Phi \right\|_{L^{2,a}(\mathbb{R}^d)}. 
$
It will then follow that
\begin{align}
 \left\| |Q|^m \ \Phi(q) \right\|_{L^{2,a}(\mathbb{R}_Q^d)} \leq   b^m (2m + 1)^{m-1} \  \left\| \Phi \right\|_{L^{2,a}(\mathbb{R}^d)} . \label{eqn:hypfinal}
\end{align}
Condition \eqref{eqn:hypfinal} is satisfied for any $m \geq 1$ if 
\begin{align}
\frac{3 K \ D_a^2}{8 \pi^2 (D_M)^{1/2}}  \ \underset{ k = 1, \dots, K} {\max} \{ C_k \} \ \left\| \Phi \right\|^2_{L^{2,a}(\mathbb{R}^d)} \leq  b \ (2m + 1), \quad \forall m \geq 1. \label{eqn:conditionind}
\end{align}
It suffices to satisfy \eqref{eqn:conditionind} for $m = 1$, which can be done by choosing $b$ sufficiently large. 
This completes the proof of Proposition \ref{prop:momentsgeneral}. $\Box$

\medskip

\section{Proof of Proposition \ref{prop:off-on}} \label{subsection:details}
(a) Suppose that $ g = \{g_n \}_{n \in \mathbb{Z}}$ is on-site symmetric in the sense of Definition \ref{defn:onoff}. Then for all $n \in \mathbb{Z}: g_n = g_{-n}$ we have
\begin{align}
&  \widehat{g}(q) =  \mathcal{F}_{_D}[ g ](q) = \sum_{n \in \mathbb{Z}} e^{- i q n} g_n = g_0 + 2 \sum_{n = 1}^{\infty} \cos(q n) g_n. \label{eq:onrealsymproof1}
\end{align}
Therefore, $\widehat{g}(q)$ is real-valued and symmetric in $q$. The converse is easy to check. 

(b) Next, suppose that $g$  is off-site symmetric in the sense of Definition \ref{defn:onoff}. Then for all $n \in \mathbb{Z}: g_n = g_{-n + 1}$, we have
\begin{align}
&  \widehat{g}(q) =  \mathcal{F}_{_D}[ g ](q) = \sum_{n \in \mathbb{Z}} e^{- i q n} g_n =  2 e^{- i q/2} \sum_{n = 1}^{\infty} g_n \cos \left( q (n - 1/2) \right) \equiv  e^{- i q/2} \widehat{K}(q),
\end{align}
where $\widehat{K}(q)$ is real and symmetric in $q$. Conversely, suppose $\widehat{g}(q) = e^{- i q/2} \widehat{K}(q)$ with $\widehat{K}(q)$ real and symmetric. Then
\begin{align}
& g_n = \mathcal{F}^{-1}_{_D}[ \widehat{g} ]_n = \mathcal{F}^{-1}_{_D}[ e^{- i (\cdot) /2} \widehat{K} ]_n = (2 \pi)^{-1} \int_{\mathcal{B}} e^{i q (n - 1/2)} \widehat{K}(q) dq \nonumber \\
& = (2 \pi)^{-1} \int_{\mathcal{B}} e^{i q (n - 1/2)} \widehat{K}(-q) dq = - (2 \pi)^{-1} \int_{\pi}^{- \pi} e^{- i q (n - 1/2)} \widehat{K}(q) dq \nonumber \\
& = (2 \pi)^{-1} \int_{-\pi}^{\pi} e^{- i q/2} e^{i q (1 - n)} \widehat{K}(q) dq = g_{-n + 1} = \overline{g_n}. 
\end{align}
This completes the proof of Proposition \ref{prop:off-on}. $\Box$

\medskip

\section{Technical details in the proof of Proposition \ref{prop:ifthigh}} \label{section:technicalprops}
In this section we elaborate on the proofs of technical propositions used in Section \ref{subsect:lyapschm}. These technical points are used to prove Proposition \ref{prop:ifthigh}, which constructs the high-frequency component of the error: $(\alpha, \widehat{E_{_{\rm lo}}}) \mapsto \widehat{E_{_{\rm hi}}} \left[ \alpha,  \widehat{E_{_{\rm lo}}} \right]$. To construct this map, we solve equation \eqref{eqn:highrewrite} using the implicit function theorem. Propositions \ref{prop:R1estimate} and \ref{prop:Aprops} together assert that the hypotheses of the implicit function theorem hold. We discuss the most important elements here; all details may be found in \cite{J:15}.  \\

\noindent {\bf Proof of Proposition \ref{prop:R1estimate}:} We let $a > 1/2$ and make extensive use of \eqref{eqn:algebra0}: 
\begin{align}
\left\| \widetilde{f_1} * \widetilde{f_2} \right\|_{L^{2,a}(\mathbb{R}^d)} \lesssim \left\| \widetilde{f_1} \right\|_{L^{2,a}(\mathbb{R}^d)} \ \left\| \widetilde{f_2} \right\|_{L^{2,a}(\mathbb{R}^d)}. \label{eqn:algebra1}
\end{align}
 We also require the following lemma, stated generally since we use it in multiple sections of the proofs, which addresses the exponential smallness of ``shifted" ($m = \pm 1$ by our convention) convolutions of exponentially decaying functions.

\begin{lemma}
\label{lemma:expconvo}
Let $a > 1/2$ and let $ m = \pm 1$. Let $\widehat{f} \in L^{2,a}(\mathbb{R})$ and such that for $C > 0$,  $e^{\ C |q|} \widehat{f}(q) \in L^{2,a}(\mathbb{R})$. Then
\begin{align}
\bigg\| \ \chi_{_{\mathcal{B}_{\alpha}}} \  \widehat{f} * \widehat{f} * \widehat{f}(Q - 2 m \pi/\alpha) \ \bigg\|_{L^{2,a}(\mathbb{R}_Q)} \lesssim \  e^{- C \pi / \alpha} \ \left\|  \ e^{ C |Q | } \ \widehat{f} \ \right\|^3_{L^{2,a}(\mathbb{R}_Q)} . \label{eqn:expconvo2}
\end{align}
\end{lemma}
This follows by using $|Q - 2 m \pi / \alpha | \leq |Q - \xi - \zeta - 2 m \pi / \alpha| + |\xi | + | \zeta |$ to bound the exponent and \eqref{eqn:algebra1} twice. 

\begin{proposition}
\label{prop:M}
For any function $\widehat{f} \in L^{2,a}(\mathbb{R})$, we have
\begin{align}
\left\| \  \chi_{_{\rm hi}} \ \chi_{_{\mathcal{B}_{_{\alpha}}}} \  \left[ 1 + M_{\alpha}(Q) \right]^{-1} \  \widehat{f} \ \right\|_{L^{2,a}(\mathbb{R}_Q)} \lesssim \ \alpha^{2 - 2r} \left\| \ \widehat{f} \ \right\|_{L^{2,a}(\mathbb{R})}.
\end{align}
\end{proposition}
This follows from the lower bound $\chi_{_{\rm hi}}(Q) M_{\alpha}(Q) \geq C \ \chi_{_{\rm hi}}(Q) \ |Q|^2 \geq C \alpha^{2r - 2}$. 
 
Next, we bound the individual terms of $ \mathcal{R}_{J,1}^{\sigma} \left[ \alpha, \Gamma + \widehat{E_{_{\rm hi}}} \right] $.

\begin{proposition} [High frequency residual bound] \label{prop:RFbd}
Let $0 < r < 1$ and let the operator $\mathcal{D}^{\sigma,\alpha}$ be defined in \eqref{eqn:Phieqn}. We have
\begin{align}
\left\| \chi_{_{\rm hi}} \ \left[ 1 + M_{\alpha}(Q  ) \right]^{-1} \ \mathcal{D}^{\sigma,\alpha}[S_J^{\alpha}] \ \right\|_{L^{2,a}(\mathbb{R}_Q)} \lesssim  \ e^{- C / \alpha^{1 - r} }. 
\end{align}
for some constant $ C > 0$. 
\end{proposition}

This follows from the exponential decay bound $\left\|  \chi_{_{\rm hi}} S_J^{\alpha} \right\|_{L^{2,a}(\mathbb{R})} \lesssim e^{- C / \alpha^{1 - r} }  \left\| e^{C |Q|} \ S_J^{\alpha}(Q) \right\|_{L^{2,a}(\mathbb{R}_Q)}$,  Lemma \ref{lemma:expconvo} and Proposition \ref{prop:M}.

\begin{proposition} \label{prop:RLNLbd}
 Let $R_{_{\rm L}}  = \chi_{_{\mathcal{B}_{_{\alpha}}}}   R_{_{\rm  L}}  $ and  $R_{_{\rm NL}}  = \chi_{_{\mathcal{B}_{_{\alpha}}}}  R_{_{\rm NL}}  $ be as defined in \eqref{eqn:RLNLdef}. Then
 {\small{
\begin{align}
& \left\|  \  R_{_{\rm  L}} \left[ \alpha, \Gamma + \widehat{E_{_{\rm hi}}} \right] \  \right\|_{L^{2,a}(\mathbb{R})} \lesssim \  \left\| \ \Gamma \ \right\|_{L^{2,a}(\mathbb{R})} + \left\| \ \widehat{E_{_{\rm hi}}} \ \right\|_{L^{2,a}(\mathbb{R})}  , \nonumber \\
& \left\| \ R_{_{\rm NL}} \left[ \alpha, \Gamma + \widehat{E_{_{\rm hi}}} \right] \  \right\|_{L^{2,a}(\mathbb{R})} \lesssim  \ \left\| \ \Gamma \ \right\|^2_{L^{2,a}(\mathbb{R})} + \left\| \ \Gamma \ \right\|_{L^{2,a}(\mathbb{R})} \left\| \ \widehat{E_{_{\rm hi}}} \ \right\|_{L^{2,a}(\mathbb{R})} \nonumber \\
& \hspace{3cm} +   \left\| \ \widehat{E_{_{\rm hi}}} \ \right\|^2_{L^{2,a}(\mathbb{R})}  +  \left\| \ \Gamma \ \right\|^3_{L^{2,a}(\mathbb{R})} + \left\| \ \Gamma \ \right\|^2_{L^{2,a}(\mathbb{R})} \left\| \ \widehat{E_{_{\rm hi}}} \ \right\|_{L^{2,a}(\mathbb{R})}  \nonumber \\
& \hspace{3cm} +   \left\| \ \Gamma \ \right\|_{L^{2,a}(\mathbb{R})} \left\| \ \widehat{E_{_{\rm hi}}} \ \right\|^2_{L^{2,a} (\mathbb{R})} +  \left\| \ \widehat{E_{_{\rm hi}}} \ \right\|^3_{L^{2,a} (\mathbb{R})}  . 
\end{align}
}}
\end{proposition}
These bounds follow from \eqref{eqn:algebra1}. We now apply Propositions \ref{prop:RFbd}, \ref{prop:M}, and \ref{prop:RLNLbd} to $ \mathcal{R}_{J,1}^{\sigma} \left[ \alpha, \Gamma + \widehat{E_{_{\rm hi}}} \right] $ given in  \eqref{eqn:HJdef} to obtain \eqref{inv-bound}. This completes the proof of Proposition \ref{prop:R1estimate}. $\Box$

\noindent {\bf Proof of Proposition \ref{prop:Aprops}:} We apply Propositions \ref{prop:S} and \ref{prop:M} along with \eqref{eqn:algebra1} to \eqref{eqn:DfAhighproof} to get
\begin{align}
 \left\| D_{\widehat{E_{_{\rm hi}}}} \left( \mathcal{A}[\alpha, \Gamma, \widehat{E_{_{\rm hi}}} ] \right) - I \right\|_{L^{2,a}(\mathbb{R}) \longrightarrow L^{2,a}(\mathbb{R})} \lesssim \ \alpha^{2-2r} \ \bigg( 1 + \| \Gamma \|_{L^{2,a}(\mathbb{R})} + \| \widehat{E_{_{\rm hi}}} \|_{L^{2,a}(\mathbb{R})} \nonumber \\
 + \| \Gamma \|^2_{L^{2,a}(\mathbb{R})} + \| \Gamma \|_{L^{2,a}(\mathbb{R})} \| \widehat{E_{_{\rm hi}}} \|_{L^{2,a}(\mathbb{R})} + \| \widehat{E_{_{\rm hi}}} \|^2_{L^{2,a}(\mathbb{R})} \bigg). \qquad \Box
\end{align}

\noindent {\bf Proof of Proposition \ref{prop:highiftests}:} By Propositions \ref{prop:R1estimate} and \ref{prop:Aprops} , we have the map $\widehat{E_{_{\rm hi}}}: [0,\alpha_0) \times B_{_{\beta_0 }}(0) \rightarrow L^{2,a}(\mathbb{R}), \ (\alpha, \Gamma ) \mapsto \widehat{E_{_{\rm hi}}}[ \alpha, \Gamma ]$ which is the unique solution to \eqref{eqn:Asoln},
\begin{align}
 \widehat{E_{_{\rm hi}}} \left[ \alpha, \Gamma \right] (Q ) & =  \chi_{_{\rm hi}}( Q )  \ [ 1 + M_{\alpha}(Q) ]^{-1} \bigg( \chi_{_{\mathcal{B}_{_{\alpha}}}}(Q) \  \frac{3}{4 \pi^2}  \ \left[ \widetilde{\psi_1} * \widetilde{\psi_1} *  \Gamma (Q) + \widetilde{\psi_1} * \widetilde{\psi_1} * \widehat{E_{_{\rm hi}}} \left[ \alpha, \Gamma \right] (Q) \right]  \nonumber \\
& \hspace{5cm} - \mathcal{R}_{J,1}^{\sigma} \left[ \alpha, \Gamma + \widehat{E_{_{\rm hi}}} \left[ \alpha, \Gamma \right] \right]( Q ) \bigg) = 0.  \label{eqn:fhiequals}
\end{align}
We now estimate the mapping $\widehat{E_{_{\rm hi}}}[ \alpha, \Gamma ]$ in $L^{2,a}(\mathbb{R}^d)$. By \eqref{eqn:alpha0} and \eqref{eqn:fhiprops},
\begin{align}
\alpha \in [0, \alpha_0) \quad  \text{and}  \quad  \left\| \ \Gamma \ \right\|_{L^{2,a}(\mathbb{R})} \leq \beta_0 \qquad \Longrightarrow \qquad
\left\| \ \widehat{E_{_{\rm hi}}} \left[ \alpha, \Gamma \right] \right\|_{L^{2,a}(\mathbb{R})} \leq \kappa, \label{eqn:constantbound2}
\end{align}
for some constants $\alpha_0, \beta_0, \kappa > 0$. We apply Proposition \ref{prop:R1estimate} and \eqref{eqn:constantbound2} to (\ref{eqn:fhiequals}) to get
\begin{align}
\left\|  \ \widehat{E_{_{\rm hi}}} \left[ \alpha, \Gamma \right] \right\|_{L^{2,a}(\mathbb{R})} \lesssim \ e^{- C \alpha^{r-1}} \ + \alpha^{2 - 2r}  \left( \left\| \ \Gamma \ \right\|_{L^{2,a}(\mathbb{R})} + \left\| \ \widehat{E_{_{\rm hi}}} \left[ \alpha, \Gamma \right] \right\|_{L^{2,a}(\mathbb{R})} \right). \label{eqn:linearbound}
\end{align}
Since $\alpha^{2 - 2r} \to 0 $ when $\alpha \to 0$ when $0 < r < 1$, taking $\alpha$ sufficiently small in \eqref{eqn:linearbound} yields \eqref{eqn:highiftests1}.

Next, the differentiability of $\mathcal{A}$ stated in Proposition \ref{prop:Aprops} and the implicit function theorem guarantee that $D_{\Gamma} \widehat{E_{_{\rm hi}}} [\alpha, \Gamma] $ is well-defined and given by
\begin{align}
 \left( D_{\Gamma}   \widehat{E_{_{\rm hi}}} [ \alpha, \Gamma ] \right) \ \widehat{f} (Q ) & = \chi_{_{\rm hi}} \ \left[ 1 + M_{\alpha}(Q) \right]^{-1} \ \bigg[ \frac{3}{4 \pi^2} \  \chi_{_{\mathcal{B}_{_{\alpha}}}}  \ \bigg(    \widetilde{\psi_1} * \widetilde{\psi_1} * \widehat{f} (Q) \nonumber \\
+ & \widetilde{\psi_1} * \widetilde{\psi_1} * \left\{ D_{\Gamma} \left( \widehat{E_{_{\rm hi}}} [ \alpha, \Gamma ] \right) \ \widehat{f}  \right\} (Q) \bigg) + D_{\Gamma} \bigg( \mathcal{R}_{J,1}^{\sigma} \left[ \alpha, \Gamma +  \widehat{E_{_{\rm hi}}} \left[ \alpha, \Gamma \right] \right] \bigg)  \widehat{f} (Q) \bigg], \label{eqn:finalhigheqn}
\end{align}
for any $\widehat{f} \in L^{2,a}(\mathbb{R})$. We apply Propositions \ref{prop:S} and \ref{prop:M} along with equations \eqref{eqn:algebra1}, \eqref{eqn:constantbound2}, and  \eqref{eqn:constantbound2} to \eqref{eqn:finalhigheqn} to get \eqref{eqn:highiftests2}. This completes the proof of Proposition \ref{prop:highiftests}.  $\Box$

\section{Estimates on the low frequency equation} \label{section:lowests}  
We reduce the system \eqref{eqn:high} and \eqref{eqn:low} to equation \eqref{eqn:lowrescaled}, 
\begin{align}
\widetilde{L_+} \widehat{E_{_{\rm lo}}} (Q) = \mathcal{R}_{J,2}^{\sigma}[\alpha, \widehat{E_{_{\rm lo}}}](Q). \label{eqn:lowrescaled2}
\end{align}
for the low frequency components. Here, $ \mathcal{R}_{J,2}^{\sigma}[\alpha, \widehat{E_{_{\rm lo}}}] $ is given in \eqref{eqn:Htildedef} by
\begin{align}
\mathcal{R}_{J,2}^{\sigma} \left[ \alpha, \widehat{E_{_{\rm lo}}} \right](Q) & = \chi_{_{\rm lo}}( Q ) \ \mathcal{R}_{J,1}^{\sigma} \left[ \alpha, \widehat{E_{_{\rm lo}}} + \widehat{E_{_{\rm hi}}} [ \alpha, \widehat{E_{_{\rm lo}}} ] \right] ( Q)  + R_{_{\rm pert}} \left[ \alpha, \widehat{E_{_{\rm lo}}} \right] (Q)  \nonumber\\
& =  \chi_{_{\rm lo}}( Q )  \bigg( \mathcal{D}^{\sigma,\alpha}[S_J^{\alpha}](Q)  + R_{_{\rm L}}^{\sigma} \left[ \alpha, \  \chi_{_{\rm lo}} \widehat{E_{_{\rm lo}}}  \right] (Q)  + R_{_{\rm L}}^{\sigma} \left[ \alpha,  \widehat{E_{_{\rm hi}}} [ \alpha,  \widehat{E_{_{\rm lo}}} ]  \right] (Q)   \nonumber \\
& \qquad +  R_{_{\rm NL}}^{\sigma} \left[ \alpha, \widehat{E_{_{\rm lo}}} +  \widehat{E_{_{\rm hi}}} [ \alpha,  \widehat{E_{_{\rm lo}}} ]  \right] (Q) \bigg)  + R_{_{\rm pert}} \left[ \alpha, \widehat{E_{_{\rm lo}}} \right] (Q).  \label{eqn:Htildedef2}
\end{align}
where $R^{\sigma}_{_{L}}$ and $R^{\sigma}_{_{NL}}$ are defined in \eqref{eqn:RLNLdef} and $ R_{_{\rm pert}}$ is defined in \eqref{eqn:Rpertdef}. Here, we have used the linearity of $R^{\sigma}_{_{L}}$ in its second argument. We have also used that $\chi_{_{\rm hi}}  \widehat{E_{_{\rm lo}}} = 0$, which is a direct consequence of \eqref{eqn:lowrescaled2}.

In order to apply the implicit function theorem to \eqref{eqn:lowrescaled2}, we require the estimates on $\mathcal{R}_{J,2}^{\sigma}[\alpha, \widehat{E_{_{\rm lo}}}]$ which are given in \eqref{eqn:Hestimate} and \eqref{eqn:Hdiffestimate}. In sections \ref{subsection:toolslow} and \ref{section:Hanalysis}, we prove these bounds. We provide the essential steps; the full details are in \cite{J:15}.

\subsection{Some tools for estimates on the low frequency equation, \eqref{eqn:lowrescaled}} \label{subsection:toolslow}

To establish the properties of $\mathcal{R}_{J,2}^{\sigma}$ given in Proposition \ref{prop:rescale}, we require a number of general tools. . Recall that $\chi_{_{\rm hi}}(Q) = 1- \chi_{_{\rm lo}}(Q)$,
where $ \chi_{_{\rm lo}}(Q) \equiv \chi_{_{[ - \alpha^{r -1}, \alpha^{r-1} ]}}(Q)$ and $0 < r < 1$. The next four lemmata will be used to bound $\mathcal{R}_{J,2}^{\sigma}$.

\begin{lemma} \label{lemma:minus2}
For any function $\widehat{f} \in L^{2,a-2}(\mathbb{R})$ such that $ | Q |^{ - 2} \widehat{f}(Q) \in L^{2,a}(\mathbb{R}_Q)$  , we have
\begin{align}
\left\| \widehat{f}   \right\|_{L^{2,a-2 }(\mathbb{R})} \ \lesssim \ \left\| | Q |^{ - 2} \widehat{f} \ \right\|_{L^{2,a}(\mathbb{R}_Q)}.
\end{align}
\end{lemma}

\begin{lemma}
\label{lemma:ltwoasmall}
For any function $\widehat{f} \in L^{2,a}(\mathbb{R})$, we have
\begin{align}
\left\| \ \chi_{_{\rm hi}}   \widehat{f} \  \right\|_{L^{2,a-2}(\mathbb{R})} \lesssim \ \alpha^{2 - 2r} \ \| \widehat{f} \|_{L^{2,a}(\mathbb{R})}.
\end{align}
\end{lemma}
This follows from Lemma \ref{lemma:minus2} and $\chi_{_{\rm hi}} |Q|^{-2} \leq \alpha^{2 - 2r}$.

\begin{lemma} \label{lemma:symbolconvlo}
Let $0  < r < 1$. For any function $\widehat{g} \in L^{2,a}(\mathbb{R})$, we have
\begin{align}
\bigg\| \ \chi_{_{\rm lo}} \  \left[ \ | Q |^2 - M_{\alpha}( Q ) \ \right] \ \widehat{g} \ \bigg\|_{L^{2,a-2 }(\mathbb{R}_Q)} \ \lesssim \ \alpha^{2r} \ \left\| \ \widehat{g} \ \right\|_{L^{2,a}(\mathbb{R})}. 
\end{align}
\end{lemma} 
To prove Lemma \ref{lemma:symbolconvlo}, note that Taylor expansion of $M_{\alpha}(Q)$ gives
\begin{align}
 |Q|^2 - M_{\alpha}(Q) =  2 \sum_{j = 1}^{\infty} \frac{  \alpha^{2j}  \ (-1)^{j} \  |Q |^{2j + 2} }{ (2j + 2)!}. 
\end{align}
Since $\chi_{_{\rm lo}}(Q) \ |Q| \leq \alpha^{r - 1}$, it follows that
\begin{align}
\chi_{_{\rm lo}}(Q) \ |Q|^{-2} \ \big| \ |Q|^2 - M_{\alpha}(Q) \ \big|  =  2 \chi_{_{\rm lo}}(Q) \   \sum_{j = 1}^{\infty} \frac{  \alpha^{2j}  \  |Q |^{2j} }{ (2j + 2)!} \leq  2 \ \alpha^{2r} \left(    \sum_{j = 1}^{\infty} \frac{  \alpha^{2r (j - 1)} }{ (2j + 2)!} \right) \lesssim \alpha^{2r}. 
\end{align}
Lemma \ref{lemma:minus2} now gives
 \begin{align}
\left\| \ \chi_{_{\rm lo}}(Q) \  \left[ \ |Q|^2 - M_{\alpha}(Q) \ \right] \ \widehat{g}(Q)  \right\|_{L^{2,a-2 }(\mathbb{R}_Q)} \ & \lesssim \left\| \ \chi_{_{\rm lo}}(Q) \ |Q|^{-2} \ \left[ \ |Q|^2 - M_{\alpha}(Q) \ \right] \ \widehat{g}(Q) \right\|_{L^{2,a }(\mathbb{R}_Q)} \nonumber \\
& \lesssim \alpha^{2r} \ \left\| \ \widehat{g} \ \right\|_{L^{2,a}(\mathbb{R})}. 
\end{align}

\begin{lemma}
\label{lemma:expconvo2}
Suppose that $\widehat{f_1} \in L^{2,a}(\mathbb{R})$ and there exists $C > 0$ such that $e^{C |Q|} \widehat{f_1}(Q) \in L^{2,a}(\mathbb{R}_Q)$. Then for all $\widehat{f_2}, \widehat{f_3} \in L^{2,a}(\mathbb{R})$, we have
\begin{align}
& \left\| \ \overline{\chi}_{_{\mathcal{B}_{_{\alpha}}}} \ \widehat{f_1} \ \right\|_{L^{2,a}(\mathbb{R})} \lesssim e^{- C \pi / \alpha} \ \left\| \ e^{C |Q| } \  \widehat{f_1}(Q) \right\|_{L^{2,a}(\mathbb{R}_Q)}, \\
& \bigg\| \left( \overline{\chi}_{_{\mathcal{B}_{_{\alpha}}}} \widehat{f_1} \right) * \widehat{f_2} * \widehat{f_3} \bigg\|_{L^{2,a-2}(\mathbb{R})} \lesssim \  e^{- C \pi / \alpha} \ \left\| \ e^{C |Q |} \ \widehat{f_1}(Q) \right\|_{L^{2,a}(\mathbb{R}_Q)} \left\| \widehat{f_2} \right\|_{L^{2,a}(\mathbb{R})} \left\| \widehat{f_3} \right\|_{L^{2,a}(\mathbb{R})}.
\end{align}
where $  \overline{\chi}_{_{\mathcal{B}_{_{\alpha}}}} (Q) = \chi_{_{ \{ |Q| > \pi/\alpha \} }}$. 
\end{lemma}

\noindent Here, we used $\| \widehat{f} \|_{L^{2,a-2}(\mathbb{R})} \leq \| \widehat{f} \|_{L^{2,a}(\mathbb{R})}$.

\subsection{Analysis of $\mathcal{R}_{J,2}^{\sigma}$ and derivation of estimates \eqref{eqn:Hestimate} and \eqref{eqn:Hdiffestimate}} \label{section:Hanalysis}
\medskip
To bound  $\mathcal{R}_{J,2}^{\sigma} : \mathbb{R}  \times L^{2,a}(\mathbb{R}) \mapsto L^{2,a-2}(\mathbb{R})$, we use the following estimates from Proposition \ref{prop:ifthigh}, valid for $\alpha < \alpha_0 $:
\begin{align}
& \left\| \widehat{E_{_{\rm hi}}} \left[ \alpha, \widehat{E_{_{\rm lo}}} \right] \right\|_{L^{2,a}(\mathbb{R})} \lesssim  \ \alpha^{2 - 2r} \   \| \widehat{E_{_{\rm lo}}} \|_{L^{2,a}(\mathbb{R})} + e^{- C/\alpha^{1-r}}, \label{eqn:reschibound} \\
& \left\| \ D_{ \widehat{E_{_{\rm lo}}} } \   \widehat{E_{_{\rm hi}}} \left[ \alpha, \widehat{E_{_{\rm lo}}} \right] \  \right\|_{L^{2,a}(\mathbb{R})  \rightarrow L^{2,a}(\mathbb{R})} \lesssim  \ \alpha^{2 -2 r} \ . \label{eqn:reschibound2}
\end{align}

For any $\widehat{f} \in L^{2,a}(\mathbb{R})$, a direct computation using \eqref{eqn:Htildedef2} and the linearity of $R^{\sigma}_{_{\rm L}}$ and $R_{_{\rm pert}}$ in their second argument gives {\small{ 
\begin{align}
D_{\widehat{E_{_{\rm lo}}}} \mathcal{R}_{J,2}^{\sigma}[ \alpha, \widehat{E_{_{\rm lo}}} ] \ \widehat{f}(Q)  
  = & \ \chi_{_{\rm lo}}(Q) \ R_{_{\rm L}}^{\sigma} \left[ \alpha, \ \chi_{_{\rm lo}}  \widehat{f} \right] (Q) \nonumber \\
  & + \chi_{_{\rm lo}}(Q) \ R_{_{\rm L}}^{\sigma} \left[ \alpha, \left( D_{\widehat{E_{_{\rm lo}}}} \widehat{E_{_{\rm hi}}} [ \alpha, \widehat{E_{_{\rm lo}}} ] \right) \widehat{f} \right] (Q) + R_{_{\rm pert}} \left[ \alpha, \widehat{f} \right] (Q) \nonumber \\
& + D_{\widehat{E_{_{\rm lo}}}} \left( \chi_{_{\rm lo}} \ R^{\sigma}_{_{NL}} \left[ \alpha, \widehat{E_{_{\rm lo}}}  + \widehat{E_{_{\rm hi}}}  [ \alpha, \widehat{E_{_{\rm lo}}} ] \right] \right) \widehat{f}(Q)   . \label{eqn:DHtilde}
\end{align}
Here,
\begin{align}
D_{\widehat{E_{_{\rm lo}}}} \left( \chi_{_{\rm lo}} \ R^{\sigma}_{_{NL}} \left[ \alpha, \widehat{E_{_{\rm lo}}}  + \widehat{E_{_{\rm hi}}}  [ \alpha, \widehat{E_{_{\rm lo}}} ] \right] \right) \widehat{f}(Q)   = &  \chi_{_{\rm lo}}( Q)  \  \left( \frac{1}{2 \pi} \right)^2 \ \sum_{m = -1}^1 \ e^{2 m \pi i \sigma} \ \bigg[ 6  \ S_J^{\alpha} * \widehat{E_J^{\alpha, \sigma}} * \widehat{f} (Q - 2 m \pi / \alpha)  \nonumber \\
&   + 3 \  \widehat{E_{_{\rm lo}}} *  \widehat{E_{_{\rm lo}}} * \widehat{f} (Q - 2 m \pi / \alpha) \nonumber \\
&  +  6  \ S_J^{\alpha} * \widehat{E_{_{\rm lo}}} * \left( D_{\widehat{E_{_{\rm lo}}}} \widehat{E_{_{\rm hi}}} [ \alpha, \widehat{E_{_{\rm lo}}} ]  \cdot  \widehat{f} \right) (Q - 2 m \pi / \alpha)  \nonumber \\
&  + 3 \  \widehat{E_{_{\rm lo}}} *  \widehat{E_{_{\rm lo}}} * \left( D_{\widehat{E_{_{\rm lo}}}} \widehat{E_{_{\rm hi}}} [ \alpha, \widehat{E_{_{\rm lo}}} ]  \cdot  \widehat{f} \right) (Q - 2 m \pi / \alpha) \bigg]. \label{eqn:DNL}
\end{align}
}}
We proceed to bound the terms in \eqref{eqn:Htildedef2} and \eqref{eqn:DHtilde}. We will often use $\| \widehat{f} \|_{L^{2,a-2}(\mathbb{R})} \leq \| \widehat{f} \|_{L^{2,a}(\mathbb{R})}$ without explicitly stating it. 

\begin{proposition} \label{prop:Rpertbound}
Let $R_{_{\rm pert}} [\alpha, \widehat{f} ] $ be as defined in \eqref{eqn:Rpertdef}, $0 < r < 1$, and $0 < \alpha < \alpha_0$. Then for $\widehat{f} \in L^{2,a}(\mathbb{R})$, 
\begin{align}
& \left\| \ R_{_{\rm pert}} \left[ \alpha, \widehat{f} \right] \ \right\|_{L^{2,a-2 }(\mathbb{R})} \ \lesssim \  \left( \alpha^{2r} + \alpha^{2 - 2r} \right) \ \left\| \ \widehat{f} \ \right\|_{L^{2,a}(\mathbb{R})}. 
\end{align}
\end{proposition}
To prove this proposition, we apply Lemmata \ref{lemma:ltwoasmall} and \ref{lemma:symbolconvlo}. 
 
\begin{proposition} \label{prop:RLlow} Let $0 < r < 1$. Let $R^{\sigma}_{_{\rm L}} $ be defined in \eqref{eqn:RLNLdef} and let $\widehat{f} \in L^{2,a}(\mathbb{R})$. Then there exists some $\alpha_0$ such that for $0 < \alpha < \alpha_0$, we have for some $C > 0$ {\small{
\begin{align}
& \left\| \ \chi_{_{\rm lo}} \ R^{\sigma}_{_{\rm L}} \left[ \alpha, \ \chi_{_{\rm lo}} \widehat{f} \right] \ \right\|_{L^{2,a-2 }(\mathbb{R})} \ \lesssim \ \alpha^{2 -  2r} \ \left\| \ \widehat{f} \ \right\|_{L^{2,a}(\mathbb{R})}, \label{eqn:RLlowprop2}   \\
& \left\| \ \chi_{_{\rm lo}} \ R^{\sigma}_{_{\rm L}} \left[ \alpha, \widehat{f} \right] \right\|_{L^{2,a-2 }(\mathbb{R})} \ \lesssim   \left\| \ \widehat{f} \ \right\|_{L^{2,a}(\mathbb{R})} .  \label{eqn:RLlowprop3}
\end{align} }}
\end{proposition}
To obtain estimate \eqref{eqn:RLlowprop3}, we apply \eqref{eqn:algebra1}. To obtain estimate \eqref{eqn:RLlowprop2}, we first observe that
\begin{align*}
S_J^{\alpha} * S_J^{\alpha} * \widehat{f} (Q)  - \ \widetilde{\psi_1} * \widetilde{\psi_1}  * \widehat{f}(Q) \sim \alpha^2 
\end{align*}
 in $L^{2,a}(\mathbb{R})$,
since $S_J^{\alpha} = \chi_{_{\mathcal{B}_{_{\alpha}}}} \ \widetilde{\psi_1} + \mathcal{O}(\alpha^2)$ and using Lemma \ref{lemma:expconvo2}.

The essential difficulty in this estimate is due to terms of the form 
$
\chi_{_{\rm lo}}(Q) \  S_J^{\alpha}  *  S_J^{\alpha}  * \left( \chi_{_{ \rm lo}} \widehat{f} \right) (Q + 2 \pi / \alpha). 
$
 The difficulty lies in the fact that we do not have a uniform decay estimate on $\widehat{f} $, but require the bound \eqref{eqn:RLlowprop2} to be small in $\alpha$ in order to apply the implicit function theorem to equation \eqref{eqn:lowrescaled} (see the hypotheses of Lemma \ref{lemma:generallinearop}). We therefore use the localization of $Q$ and $\widehat{E_{_{\rm lo}}}$ along with the decay of $S_J^{\alpha}$. 

Observe that for any $Q,  \xi, \zeta \in \mathbb{R}$ satisfying 
$
 |Q| \leq \alpha^{r-1}$ and $ | \xi + \zeta - Q  - 2  \pi / \alpha| \leq \alpha^{r-1},
$
we have, for choice of $\alpha^r \leq \pi/2$ and any $C > 0$, 
\begin{align}
  \xi + \zeta - Q - \frac{2 \pi}{\alpha} \geq - \alpha^{r-1} \quad   &  \Longrightarrow \quad \xi + \zeta \geq \frac{ 1}{\alpha} \left(2 \pi - \alpha^r \right) + Q \geq \frac{2 }{\alpha} ( \pi - \alpha^r) \geq \frac{\pi}{\alpha} \nonumber \\
&  \Longrightarrow \quad 1 = e^{- C | \xi + \zeta |} e^{C | \xi + \zeta |} \leq e^{- C \pi / \alpha} \ e^{C |\xi|} \ e^{C |\zeta|}. 
\end{align}
We may therefore write 
\begin{align}
& \bigg|  \chi_{_{\rm lo}} (Q)  \   S_J^{\alpha} * S_J^{\alpha} * \left( \chi_{_{\rm lo}} \  \widehat{E_{_{\rm lo}}} \right) ( Q + 2 \pi / \alpha)  \ \bigg| \nonumber \\
& \leq  \int_{\mathbb{R}} \int_{\mathbb{R}} \ \chi_{_{\rm lo}} (Q) \   \chi_{_{\rm lo}}(Q - \xi - \zeta - 2 m \pi/\alpha) \ |S_J^{\alpha}(\xi)| \ |S_J^{\alpha}(\zeta) | \ | \widehat{E_{_{\rm lo}}}(Q - \xi - \zeta - 2 m \pi/\alpha) | \ d \xi \ d \zeta \nonumber \\
& \leq e^{- C \pi /   \alpha} \  \int_{\mathbb{R}} \int_{\mathbb{R}} e^{C |\xi|} \ |S_J^{\alpha}(\xi)| \ e^{C |\zeta|} \ |S_J^{\alpha}(\zeta) | \ | \widehat{E_{_{\rm lo}}}(Q - \xi - \zeta - 2 m \pi/\alpha) | \ d \xi \ d \zeta, 
\end{align}
and in turn, using Proposition \ref{prop:S} ( $ \left\| e^{C |Q| } \ S_J^{\alpha}(Q) \  \right\|_{L^{2,a}(\mathbb{R}_Q)} \lesssim 1$) and \eqref{eqn:algebra1},
\begin{align}
\left\| \  \chi_{_{\rm lo}}(Q)  \   S_J^{\alpha} * S_J^{\alpha} *  \widehat{E_{_{\rm lo}}} ( Q + 2 \pi / \alpha) \ \right\|_{L^{2,a -2 }(\mathbb{R}_Q)}  
 \lesssim e^{- C \pi / \alpha} \ \left\| \ \widehat{E_{_{\rm lo}}} \ \right\|_{L^{2,a}(\mathbb{R})} . \label{eqn:RLbd2}
\end{align}

\begin{proposition} \label{prop:RNLlow} Let  $0 < r < 1$. Let $R^{\sigma}_{_{\rm NL}} $ be defined in \eqref{eqn:RLNLdef} and recall its derivative given in \eqref{eqn:DNL}. Then there exists $\alpha_0, C > 0$ such that for $0 < \alpha < \alpha_0$, {\small{
\begin{align}
& \left\| \ \chi_{_{\rm lo}} \ R^{\sigma}_{_{\rm NL}} \left[ \alpha, \widehat{E_{_{\rm lo}}} + \widehat{E_{_{\rm hi}}}  [ \alpha, \widehat{E_{_{\rm lo}}} ] \right] \ \right\|_{L^{2,a-2 }(\mathbb{R})} \ \lesssim e^{- C/\alpha^{1-r}} +  e^{- C/\alpha^{1-r}} \  \left\| \  \widehat{E_{_{\rm lo}}} \ \right\|_{L^{2,a}(\mathbb{R})}  \nonumber \\
& \hspace{8cm} +  \ \left\| \  \widehat{E_{_{\rm lo}}} \ \right\|^2_{L^{2,a}(\mathbb{R})} +   \ \left\| \  \widehat{E_{_{\rm lo}}} \ \right\|^3_{L^{2,a}(\mathbb{R})},    \label{eqn:RNLlowprop} \\
& \left\| \ D_{\widehat{E_{_{\rm lo}}}} \left( \chi_{_{\rm lo}} \ R^{\sigma}_{_{NL}} \left[ \alpha, \widehat{E_{_{\rm lo}}}  + \widehat{E_{_{\rm hi}}}  [ \alpha, \widehat{E_{_{\rm lo}}} ] \right] \right) \ \right\|_{L^{2,a}(\mathbb{R}) \rightarrow L^{2,a -2 }(\mathbb{R})}  \lesssim e^{- C/ \alpha^{1-r}}   \nonumber \\
& \hspace{8cm} +  \left\| \  \widehat{E_{_{\rm lo}}} \ \right\|_{L^{2,a}(\mathbb{R})}  +  \ \left\| \  \widehat{E_{_{\rm lo}}} \ \right\|^2_{L^{2,a}(\mathbb{R})}  .  \label{eqn:RNLlowprop2}
\end{align}
}}
\end{proposition}
 This follows from \eqref{eqn:algebra1} and estimates \eqref{eqn:reschibound} and \eqref{eqn:reschibound2}.

\begin{proposition} \label{prop:RFlow} Let $\mathcal{D}^{\sigma,\alpha}$ be defined in \eqref{eqn:Phieqn}. Then there exists some $\alpha_0$ such that for $0 < r < 1$ and $0 < \alpha < \alpha_0$, we have
\begin{align}
& \left\| \ \chi_{_{\rm lo}} \ \mathcal{D}^{\sigma,\alpha}[S_J^{\alpha}] \ \right\|_{L^{2,a-2 }(\mathbb{R})} \ \lesssim   \alpha^{2J + 2}.    \label{eqn:RFlowprop}
\end{align}
\end{proposition}
To prove this proposition, we utilize the asymptotic construction of $F_j$ and therefore $S_J^{\alpha}$ given in Proposition \ref{prop:orderj}. Write
\begin{align}
S_J^{\alpha}(Q) & = \sum_{j = 0}^{J} \alpha^{2j} \  \chi_{_{\mathcal{B}_{_{\alpha}}}}( Q) \ F_j(Q) =  \sum_{j = 0}^{J} \alpha^{2j} \ \left( 1 -  \overline{\chi}_{_{\mathcal{B}_{_{\alpha}}}}( Q) \right) \  F_j(Q)  \equiv S_{_{\rm full}}^{\alpha} (Q) - S_{_{\rm tail}}^{\alpha} (Q). 
\end{align}
Since $ S_{_{\rm tail}}^{\alpha} (Q) =  \overline{\chi}_{_{\mathcal{B}_{_{\alpha}}}}( Q)  S_{_{\rm tail}}^{\alpha} (Q) $, all terms involving $S_{_{\rm tail}}^{\alpha} (Q) $ may be estimated using Lemma \ref{lemma:expconvo2}. We are then left with the terms only involving $S_{_{J, {\rm full}}}^{\alpha}$, to which we apply our formal expansion from Proposition \ref{prop:orderj}:
\begin{align}
R_{_{\rm full}} [ \alpha ] (Q) \equiv \chi_{_{\rm lo}}(Q)  \ \left( - \left[ 1 + M_{\alpha}( Q) \right] \ S_{_{J, {\rm full}}}^{\alpha}(Q) + \left( \frac{1}{2 \pi} \right)^2 \ S_{_{J, {\rm full}}}^{\alpha} * S_{_{J, {\rm full}}}^{\alpha}  * S_{_{J, {\rm full}}}^{\alpha}  (Q) \right). \label{eqn:Sfullterms}
\end{align} 
We expand
\begin{align}
 M_{\alpha} ( Q ) = \frac{4}{\alpha^2} \sin^2 \left( \frac{Q \alpha}{2} \right) = 2 \sum_{j = 0}^{\infty} \frac{  \alpha^{2j}  \ (-1)^{j} \  |Q |^{2j + 2} }{ (2j + 2)!}, \qquad {\rm and} \qquad  S_{_{J, {\rm full}}}^{\alpha} (Q) \equiv \sum_{j =0}^{J} \alpha^{2j} \ F_j(Q),
\end{align}
such that by the construction in Proposition \ref{prop:orderj},  \eqref{eqn:Sfullterms} becomes
\begin{align}
 R_{_{\rm full}} [ \alpha ] (Q) & =   \chi_{_{\rm lo}}(Q) \left(   2 \sum_{j = J + 1}^{\infty} \frac{  \alpha^{2j}  \ (-1)^{j + 1} \  |Q |^{2j + 2} }{ (2j + 2)!} \ S_{_{J, {\rm full}}}^{\alpha}(Q) +  \sum_{j = J + 1}^{3J} \frac{ \alpha^{2j} }{(2 \pi)^2} \  \sum_{ \substack{ k + l + m = j \\ 0 \leq k, l, m < j  }} \ \widehat{f_k} * \widehat{f_l} * \widehat{f_m} (Q)  \right) \nonumber \\
 & =  \chi_{_{\rm lo}}(Q) \ \alpha^{2J + 2} \  \bigg(   2 \sum_{j = J + 1}^{\infty} \frac{  \alpha^{2j - 2J - 2}  \ (-1)^{j + 1} \  |Q |^{2j + 2} }{ (2j + 2)!} \ S_{_{J, {\rm full}}}^{\alpha}(Q) \nonumber \\
 & \hspace{3cm} +  \frac{1}{( 2 \pi)^2} \ \sum_{j = J + 1}^{3J} \ \alpha^{2j - 2J - 2} \  \sum_{ \substack{ k + l + m = j \\ 0 \leq k, l, m \leq J }} \ \widehat{f_k} * \widehat{f_l} * \widehat{f_m} (Q)  \bigg). 
\end{align}
We use \eqref{eqn:algebra1} to estimate the convolutions to be $\lesssim 1$ in $L^{2,a}(\mathbb{R})$. Similarly, we write the remaining infinite summation term as
\begin{align}
&   2 \ \chi_{_{\rm lo}}(Q) \ \sum_{j = J + 1}^{\infty} \frac{  \alpha^{2j - 2J - 2}  \ (-1)^{j + 1} \  |Q |^{2j + 2} }{ (2j + 2)!} \ S_{_{J, {\rm full}}}^{\alpha}(Q) \nonumber \\
& = 2 \ \chi_{_{\rm lo}}(Q) \  \sum_{j = J + 1}^{\infty} \frac{  \alpha^{2j - 2J - 2}  \ (-1)^{j + 1} \  |Q |^{2j - 2J - 2} }{ (2j + 2)!} \ | Q|^{2J + 4}  \ S_{_{J, {\rm full}}}^{\alpha}(Q) .
\end{align}
which is we see is also $\lesssim 1$ in $L^{2,a}(\mathbb{R})$ since $ \left\| \ |Q|^{2J + 4} \ S_{_{J, {\rm full}}} (Q)  \right\|_{L^{2,a}(\mathbb{R}_Q)} \lesssim \left\| \ e^{C_S  |Q| } \  S_{_{J, {\rm full}}}(Q) \ \right\|_{L^{2,a}(\mathbb{R}_Q)} \lesssim 1$ and since $\chi_{_{\rm lo}}(Q) |Q| \leq \alpha^{r-1}$, 
\begin{align}
& \chi_{_{\rm lo}}(Q) \ \bigg| \sum_{j = J + 1}^{\infty} \frac{  \alpha^{2j - 2J - 2}  \ (-1)^{j + 1} \  |Q |^{2j - 2J - 2} }{ (2j + 2)!} \bigg| \leq   \sum_{j = 0}^{\infty} \frac{  \alpha^{2j}  \  \chi_{_{\rm lo}}(Q) \  |Q |^{2j} }{ (2J + 2j + 4)!}    \lesssim  \sum_{j = 0}^{\infty} \frac{  \alpha^{2jr} }{ (2J + 2j + 4)!} \lesssim 1. 
\end{align}
Therefore, $
\left\| R_{_{\rm full}} [ \alpha ]  \right\|_{L^{2,a}(\mathbb{R})} \lesssim \alpha^{2J + 2}. $

We now apply Propositions \ref{prop:Rpertbound}, \ref{prop:RLlow}, and \ref{prop:RNLlow} and estimates \eqref{eqn:reschibound} and \eqref{eqn:reschibound2} to \eqref{eqn:Htildedef2} and \eqref{eqn:DHtilde}.  This implies estimates \eqref{eqn:Hestimate} and \eqref{eqn:Hdiffestimate} in Proposition \ref{prop:rescale}.

\section{Proof of estimate \ref{eqn:sigmaest}} \label{subsection:details2}  We prove estimate \ref{eqn:sigmaest} in order to complete the proof of Lemma \ref{lemma:generallinearop}, where we follow a more detailed proof of the implicit function theorem as found in \cite{N:01}. We solve for $f = f[\alpha] \in L^{2,a}_{_{\rm even}}(\mathbb{R})$ in Lemma \ref{lemma:generallinearop} directly to get the estimate. We define the operators
\begin{align}
& \mathcal{J}[\alpha, f](q) \equiv \mathcal{L}f(q) - \mathcal{R}[\alpha, f](q), \\
& 
\mathcal{K}[\alpha, f](q) \equiv f(q) - \mathcal{L}^{-1} \mathcal{J}[\alpha, f](q) = \mathcal{L}^{-1} \left( \mathcal{L} f (q) -  \mathcal{J}[\alpha, f](q) \right).
\end{align}
First, observe that $\mathcal{K}[\alpha, f]$ is a contraction on $f \in B_{_{2 C \gamma_1(\alpha)}}(0)$, since for two functions $f_1, f_2 \in L^{2,a}_{_{\rm even}}(\mathbb{R})$, with
$ 
\| f_1 \| \leq 2 C \gamma_1(\alpha)$ and $ \| f_2 \| \leq 2 C \gamma_1(\alpha), 
$
we have {\small{
\begin{align}
\mathcal{K}[f_1,\alpha](q) - \mathcal{K}[f_2,\alpha](q) = \mathcal{L}^{-1} \bigg[ \mathcal{L} \bigg( f_1(q) - f_2(q) \bigg) -  \bigg( \mathcal{J}[f_1,\alpha](q) + \mathcal{J}[f_2, \alpha](q) \bigg) \bigg] \nonumber \\
= \mathcal{L}^{-1} \left[ \mathcal{L} \bigg( f_1(q) - f_2(q) \bigg) -  \left( \int_0^1 D_f \mathcal{J} \left[ t f_1 + (1 - t) f_2, \alpha \right] dt \right) \bigg( f_1(q) - f_2(q) \bigg) \right]  \nonumber \\ 
= \mathcal{L}^{-1} \left[ \int_0^1 \mathcal{L} - D_f \mathcal{J} \left[ t f_1 + (1 - t) f_2, \alpha \right] dt \right] \bigg( f_1(q) - f_2(q) \bigg) \nonumber \\
= \mathcal{L}^{-1} \left[ \int_0^1 D_f \mathcal{R} \left[ t f_1 + (1 - t) f_2, \alpha \right] dt \right] \bigg( f_1(q) - f_2(q) \bigg),
\end{align}
}}
and in turn by the hypothesis \eqref{eqn:HIFTestdiff},
\begin{align}
& \| \mathcal{K}[f_1,\alpha](q) - \mathcal{K}[f_2,\alpha](q) \|_{L^{2,a}(\mathbb{R})} \nonumber \\
& \leq \| \mathcal{L}^{-1} \|_{L^{2,a-2}(\mathbb{R}) \mapsto L^{2,a}(\mathbb{R})}     \| D_f \mathcal{R}[f_1 + f_2, \alpha] \|_{L^{2,a}(\mathbb{R}) \mapsto L^{2,a-2}(\mathbb{R})} \| f_1 - f_2 \|_{L^{2,a}(\mathbb{R})} \nonumber \\
& \leq C \left[ \gamma_2(\alpha) + \|f_1 \|_{L^{2,a}(\mathbb{R})} + \|f_2 \|_{L^{2,a}(\mathbb{R})} + \|f_1 \|^2_{L^{2,a}(\mathbb{R})} + \|f_2 \|^2_{L^{2,a}(\mathbb{R})} \right] \  \| f_1 - f_2 \|_{L^{2,a}(\mathbb{R})} \nonumber \\
& \leq C \left[ \gamma_2(\alpha) + 4  \gamma_1(\alpha) + 4 \gamma_1(\alpha)^2 \right] \ \| f_1 - f_2 \|_{L^{2,a}(\mathbb{R})}. \label{eqn:Kcont}
\end{align}
Thus, $\mathcal{K}$ is a contraction for $\alpha$ sufficiently small on the ball of radius $2 C \gamma_1(\alpha) $ in $L^{2,a}_{_{\rm even}}(\mathbb{R})$.  

Next, observe that by \eqref{eqn:Kcont} and \eqref{eqn:HIFTest},
\begin{align}
\| \mathcal{K}[\alpha, f] \|_{L^{2,a}(\mathbb{R})} & = \| \mathcal{K}[\alpha, f] - \mathcal{K}[0,\alpha] + \mathcal{K}[0,\alpha] \|_{L^{2,a}(\mathbb{R})} \nonumber \\
& \leq \| \mathcal{K}[\alpha, f] - \mathcal{K}[0,\alpha] \|_{L^{2,a}(\mathbb{R})} \ + \ \| \mathcal{K}[0,\alpha] \|_{L^{2,a}(\mathbb{R})} \nonumber \\
& = \| \mathcal{K}[\alpha, f] - \mathcal{K}[0,\alpha] \|_{L^{2,a}(\mathbb{R})} \ + \  \| \mathcal{R}[0,\alpha] \|_{L^{2,a}(\mathbb{R})}  \nonumber \\
& \leq C \gamma_1(\alpha) + C \left[  \gamma_2(\alpha) + 4 \gamma_1(\alpha) + 4 \gamma_1(\alpha)^2 \right] \| f \|_{L^{2,a}(\mathbb{R})} \nonumber \\
& \leq C \left[ \gamma_1(\alpha) + 2 \gamma_1(\alpha) \gamma_2(\alpha) + 8 \gamma_1(\alpha)^2 +  8 \gamma_1(\alpha)^3 \right] \leq 2 C \gamma_1(\alpha). 
\end{align}
for $\alpha$ sufficiently small, so $\mathcal{K}$ is a contraction which maps the ball of radius $2 C \gamma_1(\alpha)$ to itself. Thus, we guarantee by the contraction mapping principle that the unique solution $f$ to Lemma \ref{lemma:generallinearop} exists in the ball of size $2 C \gamma_1(\alpha)$ in $ L^{2,a}_{_{\rm even}}(\mathbb{R})$ such that
$ \ 
\| f \|_{L^{2,a}(\mathbb{R})} \leq 2 C \gamma_1(\alpha). \ $ $\Box$

\end{document}